\documentstyle[12pt]{article}
\begin{document}

\hfill DUKE-CGTP-99-07

\hfill hep-th/9909120

\vspace{1.25in}

\begin{center}

{\large\bf Discrete Torsion and Gerbes II}

\vspace{0.75in}

Eric R. Sharpe \\
Department of Physics \\
Box 90305 \\
Duke University \\
Durham, NC  27708 \\
{\tt ersharpe@cgtp.duke.edu} \\

 $\,$

\end{center}

In a previous paper we outlined how discrete torsion can be
understood geometrically, as an analogue of orbifold $U(1)$ Wilson lines.
In this paper we shall prove the remaining details.
More precisely, in this paper we describe gerbes in terms of
objects known as stacks (essentially, sheaves of categories),
and develop much of the basic theory of gerbes in such language.
Then, once the relevant technology has
been described, we give a first-principles geometric derivation
of discrete torsion.  In other words,
we define equivariant gerbes, and classify equivariant
structures on gerbes and on gerbes with connection.
We prove that in general, the set of equivariant structures on 
a gerbe with connection is a torsor under a group which includes
$H^2(\Gamma, U(1))$, where $\Gamma$ is the orbifold group.
In special cases, such as trivial gerbes, the set of equivariant
structures can furthermore be canonically identified with the
group.

\begin{flushleft}
September 1999
\end{flushleft}


\newpage

\tableofcontents

\newpage

\section{Introduction}

Historically discrete torsion has been a rather mysterious
aspect of string theory.  Discrete torsion was originally
discovered \cite{vafadt} as an ambiguity in the choice of
phases to assign to different sectors of string orbifold partition
functions.  Although other work has been done on the subject,
no work done to date
has succeeded in giving any sort of 
genuinely deep understanding
of discrete torsion.  In fact, discrete torsion has sometimes been
referred to has an inherently stringy degree of freedom, without
any geometric analogue.

In this paper we shall give a purely geometric understanding
of discrete torsion, as a precise analogue of orbifold
$U(1)$ Wilson lines, but for (two-form) $B$-fields rather than
vector fields.  In \cite{dt1} we outlined this description;
in this paper we shall prove the technical details omitted
in \cite{dt1}.  In an upcoming paper \cite{dt3} we shall rederive
these results in a more elementary fashion, and also describe
how this picture can be used to derive other physical manifestations
of discrete torsion.

More precisely, in this paper we shall argue that
discrete torsion should be understood as a (discrete) ambiguity
in lifting the action of an orbifold group $\Gamma$
on a space $X$ to a
1-gerbe with connection on $X$, just as orbifold Wilson lines
can be understood as an ambiguity in lifting the action of
$\Gamma$ to a bundle with connection.
This description makes no assumptions on the nature of $\Gamma$ -- 
it may or may not be freely-acting, it may or may not be abelian -- 
this description holds true regardless.

Our description of discrete torsion hinges on a deeper understanding
of type II $B$-fields than is common in the literature.
More specifically, just as vector fields are understood as connections
on bundles, we describe $B$-fields as connections on (1-)gerbes.
Although gerbes seem to be well-known in some circles, their
usefulness does not seem to be widely appreciated.
As accessible accounts of gerbes which provide 
the level of detail we need do not seem to exist, we
describe gerbes in considerable detail. 

We begin this paper by describing stacks (essentially,
sheaves of categories), in section~\ref{stacksec}.  
Next, we give a basic description
of gerbes in terms of stacks, in section~\ref{gerbesec}.
In order to define and study equivariant structures on gerbes,
we need some rather technical results, which we collect
in sections~\ref{advstacksec} and \ref{advgerbesec}, on
stacks and gerbes.  (Readers studying this paper for the first
time are advised to skip sections~\ref{advstacksec} and
\ref{advgerbesec}.)  Finally, we define equivariant structures
on gerbes, and classify equivariant structures on gerbes
and gerbes with connection, deriving $H^2(\Gamma, U(1))$ in
the process.

For additional information on gerbes,
the reader might consult 
\cite{hitchin,dcthesis,brylinski,brymcl1,giraud,sga1,breen,freed1}.

In passing, we should mention that
in addition to the description of gerbes in terms of stacks,
there exist alternative descriptions.
For example, (equivalence classes of) gerbes can be described 
in terms of objects
on loop spaces, as described in (for example) \cite{brylinski}.
(We briefly review this description, using it to check our results,
in section~\ref{loopspaces}.)
One noteworthy description not in \cite{brylinski}
is known as ``bundle gerbes''
\cite{murray1,murray2}.
Because of certain technical difficulties with the description
of gerbes in terms of bundle gerbes, notably the difficulty in
determining whether two bundle gerbes are isomorphic,
we shall not refer to them any further in this paper.
In general, which description of gerbes is most
useful clearly depends upon both the application in mind and 
personal preference.

In an earlier version of this paper, we argued that the
difference between any two equivariant structures on a 1-gerbe
with connection is an element
of $H^2(\Gamma, U(1))$.  We have since corrected a minor error
in that calculation, and weakened the result to only claim
that the difference between equivariant structures 
is an element of a group which includes $H^2(\Gamma, U(1))$.
In other words, there are additional degrees of freedom beyond
just those encoded in $H^2(\Gamma, U(1))$, which we missed 
previously.

\section{Stacks}    \label{stacksec}

The reader might well ask what object should be
associated to a gerbe -- after all, in \cite{dt1} we 
really only referred to gerbes in terms of sheaf cohomology groups.
The answer is that gerbes can be understood in terms of sheaves
of categories, also known as stacks, which we shall review in this
section.

Our presentation of stacks closely follows 
\cite[section 5]{brylinski}, \cite[section 3]{brymcl1},
and \cite[chapter 1]{breen}.

\subsection{Presheaf of categories}

Before defining a sheaf of categories, we shall first define
a presheaf of categories, which are sometimes also called
prestacks.  We shall closely follow the definitions
of \cite{breen}.  In passing we shall note that stacks and related
ideas are often defined with respect to Grothendieck topologies
and sites, i.e.,
in the language of descent,
rather than the topologies that most physicists are acquainted with.
We feel that such definitions add an essentially irrelevant (for
our purposes) layer
of technical abstraction, and so have circumvented them.
(For definitions of stacks in terms of sites,
see for example \cite{brylinski}, \cite[expos\'e VI]{sga1},
or \cite{groth1}; for a 
basic introduction to the 
ideas of Grothendieck topologies, see for example \cite{artin,mumford}.)

Before we define a presheaf of categories, let us take a moment
to review the notion of a presheaf of sets, following 
\cite[section 1.1]{brylinski}.  A presheaf of sets ${\cal S}$ on
a space $X$ is
an assignment of a set ${\cal S}(U)$ to every open set $U \subseteq X$,
together with a map $\rho_{VU}^*: {\cal S}(U) \rightarrow {\cal S}(V)$
associated to each inclusion $\rho_{VU}: V \hookrightarrow U$,
such that if $W \subseteq V \subseteq U$ are open sets, then
\begin{displaymath}
\rho_{VU}^* \circ \rho_{WV}^* \: = \: \rho_{WU}^*
\end{displaymath}
and such that $\rho_{UU}^*$, the map associated to the trivial inclusion
$U \hookrightarrow U$, is the identity.
To define a presheaf of categories we shall follow a similar pattern:
we shall associate a category to each open set, and a functor to
each inclusion of open sets, with certain constraints on the functors. 

A presheaf of categories ${\cal C}$ on a topological space $X$
associates, to any open set $U \subseteq X$, a category ${\cal C}(U)$,
and for every inclusion of open sets $\rho: U_2 \hookrightarrow U_1$,
there is a restriction functor $\rho^*: {\cal C}(U_1) \rightarrow
{\cal C}(U_2)$, which may be taken to be the identity whenever
$U_1 = U_2 = U$ and $\rho = 1_U$.  (A word on notation:  we shall sometimes
use the notation ${\cal C}(\rho)$ or $|_{U_2}$ instead of 
$\rho^*$ in this paper.)

In passing, we should point out that not every category ${\cal C}(U)$ need
be nonempty -- for example, we shall see later that
a nontrivial 1-gerbe over a space $X$ necessarily has
${\cal C}(X)$ empty.

The restriction functors are required to satisfy two conditions.
The first is that for every pair of composable inclusions
of open sets $\rho_1: U_2 \hookrightarrow U_1$ and $\rho_2:
U_3 \hookrightarrow U_2$, one is given an invertible natural 
transformation\footnote{Readers closely watching
related references will note that in \cite[chapter 5]{brylinski}, these
natural transformations are defined in the opposite direction.}
\begin{displaymath}
\varphi_{\rho_1, \rho_2}: \: (\rho_1 \rho_2)^* \: \Longrightarrow \:
\rho_2^* \circ
\rho_1^*
\end{displaymath}

The second condition on restriction functors is that if one has
three composable inclusions $\rho_1: U_2 \hookrightarrow U_1$,
$\rho_2: U_3 \hookrightarrow U_2$, and $\rho_3: U_4 \hookrightarrow U_3$,
then the following diagram of natural transformations
is required to commute:
\begin{equation}   \label{trans}
\begin{array}{ccc}
(\rho_1 \rho_2 \rho_3)^* & \Longrightarrow & \rho_3^* \circ 
( \rho_1 \rho_2)^* \\
\Downarrow & & \Downarrow \\
(\rho_2 \rho_3)^* \circ \rho_1^* & \Longrightarrow &
\rho_3^* \circ \rho_2^* \circ \rho_1^*
\end{array}
\end{equation}

An assignment of categories to open sets, together with inverse
image functors satisfying the two conditions above, defines
a presheaf of categories.

In order to get a better grasp of this material, the reader
is encouraged to compare the presheaf of categories defined above with the
general definition of presheaf.

At this point an example might help the reader.
It is possible to describe a presheaf of sets as a special
kind of presheaf of categories.  To do this, consider a set
as a discrete category -- a category whose objects are the elements
of the set, and whose only morphisms are the identity morphisms
mapping any object back to itself.
It can then be shown fairly easily that if we think of a set
as a discrete category, then a presheaf of sets is a special
kind of presheaf of categories.

Before going on, we should also describe how any presheaf of
categories contains within it, a number of presheaves of sets.
More specifically, given a presheaf of categories ${\cal C}$ on 
a space $X$, we shall define, for any open $U \subseteq X$
and any two objects $P_a, P_b \in \mbox{Ob } {\cal C}(U)$,
a presheaf (of sets) $\underline{\mbox{Hom}}_U(P_a, P_b)$
of local morphisms from $P_a$ to $P_b$.
This presheaf of sets is defined as follows.
For any open $V \subseteq U$, the set of sections of the presheaf
is given by $\mbox{Hom}_{{\cal C}(V)}(P_a|_V, P_b|_V)$.
For any inclusion $\rho: U_2 \hookrightarrow U_1$ of open sets
$U_1, U_2 \subseteq U$, define the restriction map $\rho^*$ by,
\begin{eqnarray*}
\rho^*: \: \beta & \mapsto & \varphi_{1,2}^{-1} \circ \beta|_{U_2}
\circ \varphi_{1,2} \\
 & & \: \in \mbox{Hom}\left(P_a|_{U_2}, P_b |_{U_2} \right)
\end{eqnarray*}
for any $\beta \in \mbox{Hom}(P_a|_{U_1}, P_b|_{U_1})$.
It is straightforward to check that the restriction map satisfies
the axioms for a presheaf of sets, and so $\underline{ \mbox{Hom} }_U
(P_a, P_b)$ is a presheaf of sets.
In order to distinguish this presheaf of morphisms from a set
of morphisms, we shall always denote the presheaf by
$\underline{\mbox{Hom}}$ and the set by $\mbox{Hom}$.

In passing, we should mention that we shall sometimes use
the notation $\mbox{Aut}_{{\cal C}(U)}(P)$ to denote
the set $\mbox{Hom}_{{\cal C}(U)}(P,P)$,
and $\underline{\mbox{Aut}}_U(P)$ to denote
the presheaf $\underline{\mbox{Hom}}_U(P,P)$.

\subsection{Sheaf of categories}

We shall use the terms ``sheaf of categories'' and ``stack'' 
synonymously in this paper, though not all authors quite agree
\cite{breen}.  In the conventions of \cite{breen}, where the
two concepts are distinguished, what we shall define in this
section will technically be a ``stack,'' rather than a ``sheaf
of categories'' (which is required to satisfy a  
stronger gluing condition on objects in \cite{breen}).

Before giving the technical definitions, let us review the
gluing conditions for a sheaf of sets, on which the gluing conditions
below shall be modelled.  Following \cite[section 1.1]{brylinski},
given a presheaf ${\cal S}$ of sets over a space $X$,
we say the presheaf is a sheaf of sets if for any open set $U \subseteq X$
and every open cover $\{ U_{\alpha} \}$ of $U$, if $\{ s_{\alpha} \in
{\cal S}(U_{\alpha}) \}$ is a family of elements such that
\begin{displaymath}
\rho_{U_{\alpha \beta} U_{\alpha}}^* s_{\alpha} \: = \:
\rho_{U_{\alpha \beta} U_{\beta}}^* s_{\beta}
\end{displaymath}
then there exists a unique $s \in {\cal S}(U)$ such that
$\rho_{U_{\alpha} U}^* s = s_{\alpha}$ for all $\alpha$.
The constraints for a presheaf of categories to be a sheaf of
categories are closely related; one must give a rather similar
gluing condition for both the objects and the morphisms of the
categories.

\subsubsection{Gluing law for objects}

First, we shall define the gluing condition for objects.
Let $\{ U_{\alpha} \}$ be an open cover of some open set $U \subseteq X$,
and suppose we are given a family of objects $x_{\alpha} \in
\mbox{Ob }{\cal C}(U_{\alpha})$, and a family of isomorphisms $\phi_{\alpha
\beta}$:
\begin{displaymath}
\phi_{\alpha \beta}: x_{\beta} |_{U_{\alpha \beta}} \stackrel{\sim}
{\longrightarrow} x_{\alpha} |_{U_{\alpha \beta}}
\end{displaymath}
satisfying the compatibility condition $\phi_{\alpha \beta} \circ
\phi_{\beta \gamma} = \phi_{\alpha \gamma}$ in ${\cal C}(U_{\alpha
\beta \gamma})$, and such that $\phi_{\alpha \alpha} = 1_{\alpha}$
in ${\cal C}(U_{\alpha})$.

Gluing holds for objects (more technically, the descent condition
is effective) if there exists an object $x \in \mbox{Ob }{\cal C}(U)$,
together with a family of isomorphisms $\psi_{\alpha}: x |_{U_{\alpha}}
\stackrel{\sim}{\longrightarrow} x_{\alpha}$,
such that the following diagram commutes:
\begin{equation}  \label{gluobj1}
\begin{array}{ccccc}
x |_{U_{\beta}} |_{U_{\alpha \beta}}
 & \stackrel{ \varphi_{\beta, \alpha \beta} }
{\longleftarrow} & x |_{U_{\alpha \beta}} & \stackrel{
\varphi_{\alpha, \alpha \beta} }{ \longrightarrow } &
x |_{U_{\alpha}} |_{U_{\alpha \beta}}  \\
\makebox[0pt][r]{ $\scriptstyle{\psi_{\beta} |_{U_{\alpha \beta}}}$ } 
\downarrow 
& & & & 
\downarrow
\makebox[0pt][l]{ $\scriptstyle{\psi_{\alpha} |_{U_{\alpha \beta}}}$ }  \\
x_{\beta} |_{U_{\alpha \beta}} & & \stackrel{ \phi_{\alpha \beta} }
{\longrightarrow} & & x_{\alpha} |_{U_{\alpha \beta}}  
\end{array}
\end{equation}
where we have used the notation $|_{U}$ to indicate the restriction 
functor, and where $\varphi_{\alpha, \alpha \beta}$
indicates the natural transformation $|_{U_{\alpha \beta}} \Rightarrow
|_{U_{\alpha}} |_{U_{\alpha \beta}}$ appearing in the definition of
presheaf of categories.

Before we proceed to the gluing law for morphisms, we shall take
a moment to reflect on the gluing law for objects given above.
First, note that one can not assume that the objects 
$x |_{U_{\alpha}} |_{U_{\alpha
\beta}}$, $x |_{U_{\beta}} |_{U_{\alpha \beta}}$, and
$x |_{U_{\alpha \beta}}$ in $\mbox{Ob } {\cal C}(U_{\alpha \beta})$
are the same object.  Rather, they are related by invertible
natural transformations between the restriction functors, nothing more.

We should also mention that we were sloppy in part of the gluing law
given above.  Strictly speaking, the relation $\phi_{\alpha \beta} \circ
\phi_{\beta \gamma} = \phi_{\alpha \gamma}$ does not make sense,
even after naively using the restriction functors, as the morphisms in
question act on distinct objects.  In order to properly make sense
out of this relation, one must make use of the natural transformations
between restriction functors appearing in the definition of presheaf.
More concretely, the relation $\phi_{\alpha \beta} \circ \phi_{\beta
\gamma} = \phi_{\alpha \gamma}$ should be replaced with the
constraint that the following diagram commutes:
\begin{equation}  \label{transgluobj}
\begin{array}{ccccc}
 x_{\gamma} |_{U_{\beta \gamma}} |_{U_{\alpha \beta \gamma}} 
& \stackrel{ \phi_{\beta \gamma} |_{U_{\alpha \beta \gamma}} }
{\longrightarrow} & x_{\beta} |_{U_{\beta \gamma}} |_{U_{\alpha
\beta \gamma}} & \stackrel{ \varphi_{\beta \gamma, \alpha \beta
\gamma}^{-1} }{ \longrightarrow } &
x_{\beta} |_{U_{\alpha \beta \gamma}}  \\
\makebox[0pt][r]{$\scriptstyle{\varphi_{\beta \gamma, \alpha \beta \gamma}^{-1}}$}
\downarrow & & & & 
\downarrow \makebox[0pt][l]{$\scriptstyle{\varphi_{\alpha \beta, \alpha \beta \gamma}}$} \\
x_{\gamma} |_{U_{\alpha \beta \gamma}} & & & & x_{\beta} |_{U_{\alpha \beta}}
|_{U_{\alpha \beta \gamma}}  \\
\makebox[0pt][r]{$\scriptstyle{\varphi_{\alpha \gamma, \alpha \beta \gamma}}$} \downarrow
& & & & \downarrow
\makebox[0pt][l]{$\scriptstyle{\phi_{\alpha \beta} |_{U_{\alpha \beta \gamma}}}$} \\
x_{\gamma} |_{U_{\alpha \gamma}} |_{U_{\alpha \beta \gamma}} & & & &
x_{\alpha} |_{U_{\alpha \beta}} |_{U_{\alpha \beta \gamma}} \\
\makebox[0pt][r]{$\scriptstyle{\phi_{\alpha \gamma} |_{U_{\alpha \beta \gamma}}}$}
\downarrow &
& & & \downarrow 
\makebox[0pt][l]{$\scriptstyle{\varphi_{\alpha \beta, \alpha \beta \gamma}^{-1}}$} \\
x_{\alpha} |_{U_{\alpha \gamma}} |_{U_{\alpha \beta \gamma}} &
& \stackrel{ \varphi_{\alpha \gamma, \alpha \beta \gamma}^{-1} }
{\longrightarrow}  & & x_{\alpha} |_{U_{\alpha \beta \gamma}}  
\end{array}
\end{equation}

Finally, the reader might ask to what extent the object $x$ constructed
in the gluing law above is unique.  Certainly in the definition of
sheaves of sets, the gluing law specified unique objects, but uniqueness
was not mentioned in the definition above.  In fact, it can be shown
using the gluing law for morphisms defined in the next section that the
object $x$ is unique up to unique isomorphism commuting with the 
$\psi_{\alpha}$.  We shall not work through
the details, but the basic idea is as follows.  If $x' \in \mbox{Ob }
{\cal C}(U)$ is another object with isomorphisms $\psi'_{\alpha}:
x'|_{U_{\alpha}} \stackrel{ \sim }{ \longrightarrow } x_{\alpha}$
that make diagram~(\ref{gluobj1}) commute, then define
a set of morphisms $f_{\alpha}: x|_{U_{\alpha}} \rightarrow
x'|_{U_{\alpha}}$ by, $f_{\alpha} = \psi'^{-1}_{\alpha} \circ
\psi_{\alpha}$.  Using the gluing law for morphisms these can be
glued together to form a unique morphism $f: x \rightarrow x'$ whose
restriction to each $U_{\alpha}$ commutes with $\psi_{\alpha}$ and
$\psi'_{\alpha}$, i.e,
\begin{displaymath}
\begin{array}{ccc}
x|_{U_{\alpha}} & \stackrel{ f|_{U_{\alpha}} }{ \longrightarrow } &
x'|_{U_{\alpha}} \\
\makebox[0pt][r]{ $\scriptstyle{ \psi_{\alpha} }$ } \downarrow & &
\downarrow \makebox[0pt][l]{ $\scriptstyle{ \psi'_{\alpha} }$ } \\
x_{\alpha} & = & x_{\alpha}
\end{array}
\end{displaymath}
commutes,
and with further work it can be shown that $f$ is an isomorphism.
Note that if we dropped the constraint that the restriction of $f$
commute with the $\psi_{\alpha}$, then we would lose uniqueness -- 
given any $f$, we could compose with any automorphism of either
$x$ or $x'$ to obtain another morphism $x \rightarrow x'$.

\subsubsection{Gluing law for morphisms}

The gluing condition on morphisms can be stated as follows.
Let $U \subseteq X$ be an open set, $x, y \in \mbox{Ob } {\cal C}(U)$,
and $\{ U_{\alpha} \}$ be an open cover of $U$.
If $\{ f_{\alpha}: x |_{U_{\alpha}} \rightarrow y |_{U_{\alpha}}
\}$ is a set of maps such that $f_{\alpha} |_{U_{\alpha \beta}} 
= f_{\beta} |_{U_{\alpha \beta}}$, then there exists a unique
$f: x \rightarrow y$ such that $f_{\alpha} = f |_{U_{\alpha}}$.

Unfortunately this phrasing is slightly sloppy. 
Strictly speaking, the relation $f_{\alpha} |_{U_{\alpha \beta}}
= f_{\beta} |_{U_{\alpha \beta}}$ does not make sense:  
the objects $x |_{U_{\alpha}} |_{U_{\alpha \beta}}$
and $x|_{U_{\beta}} |_{U_{\alpha \beta}}$ are (in general)
distinct objects of $\mbox{Ob }{\cal C}(U_{\alpha \beta})$, so the
morphisms $f_{\alpha} |_{U_{\alpha \beta}}$ and $f_{\beta} |_{U_{\alpha
\beta}}$ are morphisms between (in general) distinct 
objects of ${\cal C}(U_{\alpha \beta})$, and so cannot be immediately
compared.  Thus, we must replace the condition that $f_{\alpha} |_{U_{\alpha
\beta}} = f_{\beta} |_{U_{\alpha \beta}}$ with the condition that the
following diagram commutes:
\begin{equation}   \label{glumorph}
\begin{array}{ccccc}
x |_{U_{\beta}} |_{U_{\alpha \beta}} &
\stackrel{ \varphi_{\beta, \alpha \beta} }{ \longleftarrow } &
x |_{U_{\alpha \beta}} & \stackrel{ \varphi_{\alpha, \alpha \beta} }
{\longrightarrow} & x |_{U_{\alpha}} |_{U_{\alpha \beta}} \\
\makebox[0pt][r]{$\scriptstyle{ f_{\beta} |_{U_{\alpha \beta}} }$}
\downarrow & & & & \downarrow \makebox[0pt][l]{$\scriptstyle{
f_{\alpha} |_{U_{\alpha \beta}} }$} \\
y |_{U_{\beta}} |_{U_{\alpha \beta}} &
\stackrel{ \varphi_{\beta, \alpha \beta}^{-1} }{ \longrightarrow } &
y |_{U_{\alpha \beta}} & \stackrel{ \varphi_{\alpha, \alpha \beta}^{-1} }
{\longleftarrow} & y |_{U_{\alpha}} |_{U_{\alpha \beta}}
\end{array}
\end{equation}

Finally, note that we can rephrase the gluing condition for morphisms
somewhat more elegantly by saying that morphisms satisfy the gluing law
if for any pair of objects $x, y \in \mbox{Ob } {\cal C}(U)$ and any
open cover $\{ U_{\alpha} \}$ of $U$, the ordinary sheaf (of sets)
axiom for gluing in the presheaf of morphisms $\underline{ \mbox{Hom} }_U
(x,y)$ is satisfied.  Put another way, satisfying the gluing condition
for morphisms is equivalent to the presheaf of sets $\underline{ \mbox{Hom} }$
being a sheaf of sets.  Phrased yet another way, in a sheaf of categories,
each presheaf of sets of morphisms $\underline{ \mbox{Hom} }$ is a sheaf
of sets, not just a presheaf.

\subsubsection{Examples}

One easy example of a stack is the sheaf of discrete categories
associated to a sheaf of sets.  Recall we pointed out earlier
that a presheaf of sets can be understood as a presheaf of
discrete categories.  (Identify the elements of each set with
objects in each category.  By definition, the only morphisms in
a discrete category are the identity morphisms, so we have
completely characterized the categories.)  It is easy to check
that if the presheaf of sets is actually a sheaf of sets,
then the corresponding presheaf of (discrete) categories
is actually a stack.

A trivial example of a stack is
the stack of all principal $G$-bundles on $X$, for some 
Lie group $G$.
More precisely, define a stack ${\cal C}$ by
associating to each open set $U \subseteq X$, the category
${\cal C}(U)$ whose objects are all principal $G$ bundles over $U$,
with morphisms all bundle isomorphisms\footnote{In fact, any morphism
of principal $G$-bundles for fixed $G$ over a fixed space $X$ is
necessarily an isomorphism \cite[section 4.3]{husemoller}.}.  
It is straightforward to check that this defines a presheaf
of categories (with the restriction functors defined naturally,
and the natural transformations trivial), 
and furthermore this presheaf of
categories is a stack.  We shall denote this example of
a stack by $\mbox{Tors}(G)$.

Now, let us describe an example of a presheaf of
categories that is not a stack.  Fix some principal $G$
bundle $P$ on $X$, and define a presheaf of categories as
follows.  To each open set $U$, associate a category
with one object, equal to $P |_U$, and let the morphisms in this
category be the automorphisms of $P|_U$.  It is easy to check
that this defines a presheaf of categories, with the
restriction functors defined naturally, and the natural
transformations trivial.  Denote this presheaf of categories
by ${\cal P}$.

We shall now argue that the presheaf of categories ${\cal P}$ is not a stack
in general,
by observing that it does not always\footnote{In special cases, such as
$X$ contractible, ${\cal P}$ may be a stack.  However, we shall consider
more general $X$, for which ${\cal P}$ will not be a stack.}
satisfy the gluing law for objects.
Let $U \subseteq X$ be an open subset of $X$ such that
there exists a principal $G$ bundle $Q$ over $U$ such that
$P |_U \otimes Q$ is not 
topologically equivalent to $P |_U$.  Let $\{ U_{\alpha} \}$ be
a good cover of $U$.  
Let $g_{\alpha \beta}: U_{\alpha \beta} \rightarrow G$ denote the
transition functions for $Q$, defined with respect to the cover
$\{ U_{\alpha} \}$.
Define a family of isomorphisms $\phi_{\alpha \beta}: P|_{U_{\beta}}
|_{U_{\alpha \beta}} \rightarrow P|_{U_{\alpha}} |_{U_{\alpha \beta}}$
by
\begin{displaymath}
\phi_{\alpha \beta} \: = \: g_{\alpha \beta} \circ
\varphi_{\alpha, \alpha \beta} \circ 
\varphi_{\beta, \alpha \beta}^{-1}
\end{displaymath}
where the $\varphi$ are the (trivial) natural transformations
appearing in the definition of presheaf of categories ${\cal P}$.
(We have made them explicit for completeness.)
It is easy to check these isomorphisms satisfy the gluing condition,
and so by the gluing condition for objects we should find a corresponding
object in ${\cal P}(U)$.  This new object was essentially created
by tensoring local sections of $P |_U$ with local sections of $Q$,
and so in general should be local sections\footnote{In writing
this slightly loose statement, we are assuming that $G$
is abelian.} of $P |_U \otimes Q$.
(This argument is somewhat weak; rigorous versions can be found
in section~\ref{gtap}, in the discussion of gauge transformations
for gerbes.)
By assumption, however,
no such object
exists in ${\cal P}(U)$.  Thus, the gluing law is not satisfied,
and so the presheaf of categories ${\cal P}$ cannot be a stack.

\subsection{Cartesian functors}

Let ${\cal C}$ and ${\cal D}$ denote two presheaves of categories
over a space $X$.
A map $F: {\cal C} \rightarrow {\cal D}$ is defined to be
\cite[section 1]{breen} a family of functors $F(U):
{\cal C}(U) \rightarrow {\cal D}(U)$, together with, for every
inclusion $\rho: U_2 \hookrightarrow U_1$, an invertible natural transformation
$\chi_{\rho}: \rho_{{\cal D}}^* \circ F(U_1) \Rightarrow F(U_2)
\circ \rho_{{\cal C}}^*$.  These invertible natural transformations
are required to have the property that for any pair
of composable inclusions $\rho_2: U_3 \hookrightarrow U_2$,
$\rho_1: U_2 \hookrightarrow U_1$, the following diagram commutes:
\begin{equation}   \label{cartdef}
\begin{array}{ccccc}
\rho_{2 {\cal D}}^* \circ \rho_{1 {\cal D}}^* \circ F(U_1)
& \stackrel{ \chi_{\rho_1} }{ \Longrightarrow } &
\rho_{2 {\cal D}}^* \circ F(U_2) \circ \rho_{1 {\cal C}}^* &
\stackrel{ \chi_{\rho_2} }{ \Longrightarrow } &
F(U_3) \circ \rho_{2 {\cal C}}^* \circ \rho_{1 {\cal C}}^* \\
\makebox[0pt][r]{ $\scriptstyle{ \varphi_{\rho_1, \rho_2}^{{\cal D}} }$ } 
\Uparrow &
& & & \Uparrow \makebox[0pt][l]{ $\scriptstyle{ 
\varphi_{\rho_1, \rho_2}^{{\cal C}} }$ } \\
(\rho_1 \rho_2)_{{\cal D}}^* \circ F(U_1) & &
\stackrel{ \chi_{ \rho_1 \rho_2 } }{ \Longrightarrow } &
& F(U_3) \circ ( \rho_1 \rho_2 )_{{\cal C}}^*
\end{array}
\end{equation}
where the $\varphi$ are the invertible natural transformations appearing
in the definition of presheaf of categories.

Such maps between presheaves of categories are called
Cartesian functors.  Note that a Cartesian functor is not precisely a functor,
in the sense that it is not a map between categories, but rather
a map between presheaves of categories.

A morphism between stacks is precisely a morphism between the 
underlying presheaves of categories, that is, a morphism between
stacks is precisely a Cartesian functor.

Cartesian functors can be composed.  That is, if ${\cal C}$, ${\cal D}$,
and ${\cal E}$ are three presheaves of categories on $X$, and
$F: {\cal C} \rightarrow {\cal D}$ and $G: {\cal D} \rightarrow
{\cal E}$ are two Cartesian functors, then one can define
a Cartesian functor $G \circ F: {\cal C} \rightarrow {\cal E}$.
We shall outline the definition.  For any open set $U$,
the functors $F(U): {\cal C}(U) \rightarrow {\cal D}(U)$
and $G(U): {\cal D}(U) \rightarrow {\cal E}(U)$ can certainly be composed.
Given an inclusion $\rho: U_2 \hookrightarrow U_1$,
we can define the invertible natural transformation $\chi_{\rho}^{GF}$
as the composition of the natural transformations associated to $F$
and $G$.  In other words, $\chi_{\rho}^{GF}: \rho_{{\cal E}}^* \circ (GF)(U_1)
\Rightarrow (GF)(U_2) \circ \rho_{{\cal C}}^*$ is defined by
\begin{displaymath}
\chi_{\rho}^{GF}: \rho_{{\cal E}}^* \circ G(U_1) \circ F(U_1) 
\stackrel{ \chi_{\rho}^G }{ \Longrightarrow } G(U_2) \circ
\rho_{{\cal D}}^* \circ F(U_1) \stackrel{ \chi_{\rho}^{F} }{ \Longrightarrow }
G(U_2) \circ F(U_2) \circ \rho_{{\cal C}}^*
\end{displaymath}
It can be shown that $\chi_{\rho}^{GF}$ satisfies the pentagonal 
identity~(\ref{cartdef}).

Now, what does it mean for two stacks to be equivalent?
We say that two stacks ${\cal C}$, ${\cal D}$ are equivalent if
there exists a Cartesian functor $F: {\cal C} \rightarrow {\cal D}$
such that the functor $F(U): {\cal C}(U) \rightarrow {\cal D}(U)$
associated to any open set $U$ is an equivalence of
categories\footnote{Recall that a functor $F: {\cal E} \rightarrow
{\cal F}$ is said to be an equivalence of the categories ${\cal E}$,
${\cal F}$ if there is a functor $G: {\cal F} \rightarrow {\cal E}$
and there are invertible natural transformations
$\mbox{Id}_{ {\cal E} } \Rightarrow G F$ and
$\mbox{Id}_{ {\cal F} } \Rightarrow F G$, i.e., $\mbox{Id}_{ {\cal E} } \cong
G F$ and $\mbox{Id}_{ {\cal F} } \cong F G$.}.


We shall now show that
a Cartesian functor $F: {\cal C} \rightarrow {\cal D}$ between
presheaves of categories ${\cal C}$, ${\cal D}$ induces
a morphism of presheaves
\begin{displaymath}
\underline{\mbox{Hom}}_{ {\cal C}(U) }( P_a, P_b ) \: \longrightarrow
\: \underline{\mbox{Hom}}_{ {\cal D}(U) }( F(U)(P_a), F(U)(P_b) )
\end{displaymath}
for any open set $U$ and any two objects $P_a, P_b \in
\mbox{Ob } {\cal C}(U)$.
For any open $V \subseteq U$, define a map of sets
\begin{displaymath}
\lambda(V): \: \mbox{Hom}_{ {\cal C}(V) }(P_a|_V, P_b|_V)
\: \longrightarrow \: \mbox{Hom}_{ {\cal D}(V) }(
F(U)(P_a)|_V, F(U)(P_b)|_V)
\end{displaymath}
by,
\begin{displaymath}
\lambda(V)(\beta) \: \equiv \: \left( \chi^F_{U, V}(P_b) \right)^{-1}
\circ F(V)(\beta) \circ \chi^F_{U,V}(P_a)
\end{displaymath}
for all $\beta: P_a|_V \rightarrow P_b|_V$, where $\chi^F$ denotes
the invertible natural transformation defining $F$ as a Cartesian functor.
It is straightforward to check that $\lambda$ defines a morphism of
presheaves of sets, in other words, that for every inclusion
$\rho: U_2 \hookrightarrow U_1$ of open $U_1, U_2 \subseteq U$,
the following diagram commutes:
\begin{equation}
\begin{array}{ccc}
\underline{ \mbox{Hom} }_U( P_a, P_b )(U_1) &
\stackrel{ \lambda(U_1) }{ \longrightarrow } &
\underline{ \mbox{Hom} }_U( F(U)(P_a), F(U)(P_b) )(U_1) \\
\makebox[0pt][r]{ $\scriptstyle{ \rho^* }$ } \downarrow & &
\downarrow \makebox[0pt][l]{ $\scriptstyle{ \rho^* }$ } \\
\underline{ \mbox{Hom} }_U(P_a, P_b)(U_2) &
\stackrel{ \lambda(U_2) }{ \longrightarrow } &
\underline{ \mbox{Hom} }_U( F(U)(P_a), F(U)(P_b) )(U_2)
\end{array}
\end{equation}

\subsection{2-arrows}

Now that we have defined analogues of functors for presheaves
of categories (namely, Cartesian functors), we shall define
analogues of natural transformations between Cartesian functors.
These analogues of natural transformations are known as 2-arrows
\cite[section 1.1]{breen}.  We shall also define sheaves of sets
describing local natural transformations between Cartesian functors.

\subsubsection{2-arrows}

Let $F, G: {\cal C} \rightarrow {\cal D}$ be Cartesian functors between
a pair of presheaves of categories ${\cal C}$, ${\cal D}$ over a space
$X$.
A 2-arrow $\Psi: F \Rightarrow G$ is a family of natural transformations
$\Psi(U): F(U) \Rightarrow G(U)$ (one for each open $U \subseteq X$),
such that, for any inclusion $\rho: U_2 \hookrightarrow U_1$,
the following diagram commutes:
\begin{equation}
\begin{array}{ccc}
\rho_{{\cal D}}^* \circ F(U_1) & \stackrel{ \Psi(U_1) }{ \Longrightarrow }
& \rho_{{\cal D}}^* \circ G(U_1) \\
\makebox[0pt][r]{ $\scriptstyle{ \chi_{\rho}^F }$ } \Downarrow & & 
\Downarrow \makebox[0pt][l]{ $\scriptstyle{ \chi_{\rho}^G }$ } \\
F(U_2) \circ \rho_{{\cal C}}^* & \stackrel{ \Psi(U_2) }{ \Longrightarrow }
& G(U_2) \circ \rho_{{\cal C}}^*
\end{array}
\end{equation}
where the $\chi_{\rho}$ are natural transformations appearing in 
the definition of Cartesian functors.

In passing, note that 2-arrows can be composed.  In other words,
if $F, G, H: {\cal C} \rightarrow {\cal D}$ are Cartesian functors
and $\Psi_1: F \Rightarrow G$, $\Psi_2: G \Rightarrow H$ are a pair of
2-arrows, then the composition $\Psi_2 \circ \Psi_1: F \Rightarrow H$ 
is well-defined
as a 2-arrow.

An invertible 2-arrow is a 2-arrow $\Psi$ such that $\Psi(U)$ is an
invertible natural transformation for all open $U \subseteq X$.

\subsubsection{Sheaves of natural transformations}

Let $F, G: {\cal C} \rightarrow {\cal D}$ be Cartesian functors
between a pair of presheaves of categories ${\cal C}$ and ${\cal D}$,
over a space $X$.  Fix some open set $U \subseteq X$.
We shall define a presheaf of sets $\underline{2R}_U(F,G)$ which
shall describe local 2-arrows between $F$ and $G$,
as well as a presheaf of sets $\underline{NT}_U(F,G)$, a related
sheaf which shall describe
local natural transformations between $F$ and $G$.

We shall define a presheaf of 2-arrows from $F|_U$ to $G|_U$,
which we shall denote $\underline{2R}_U(F,G)$.
To any open $V \subseteq U$, define the set $\underline{2R}_U(F,G)(V)$
to be the set of all 2-arrows $F |_V \Rightarrow G |_V$.
In other words, an element $\psi \in \underline{2R}_U(F,G)(V)$
is a collection of natural transformations 
\begin{displaymath}
\psi(W): \: F(W) \: \Longrightarrow \:
G(W) 
\end{displaymath}
(one such for each open $W \subseteq V$), such that
for any inclusion $\rho: W_2 \hookrightarrow W_1$ of open sets
$W_1, W_2 \subseteq V$, the following diagram commutes:
\begin{equation}   \label{2Rdef}
\begin{array}{ccc}
\rho^* \circ F(W_1)  & \stackrel{ \psi(W_1) }{ \Longrightarrow }
& \rho^* \circ G(W_1)  \\
\makebox[0pt][r]{ $\scriptstyle{ \chi^F_{\rho} }$ } \Downarrow & &
\Downarrow \makebox[0pt][l]{ $\scriptstyle{ \chi^G_{\rho} }$ } \\
F(W_2) \circ \rho^* & \stackrel{ \psi(W_2) }{ \Longrightarrow }
& G(W_2) \circ \rho^* 
\end{array}
\end{equation}
where the $\chi$ are the natural transformations defining $F$, $G$
as Cartesian functors.

Restriction maps in the presheaf $\underline{2R}_U(F,G)$ are defined
as follows.  Let $\rho: V_2 \hookrightarrow V_1$ denote an inclusion
of open sets $V_1, V_2 \subseteq U$.
An element $\psi \in \underline{2R}_U(F,G)(V_1)$ is a collection
of natural transformations $\psi(W): F(W) \Rightarrow
G(W)$, for $W \subseteq V_1$,
as above.  We define $\rho^* \psi$ to be the new collection of natural
transformations obtained from the collection $\psi$
by removing all elements corresponding
to open $W$ such that $V_2 \subset W \subseteq V_1$. 

It should be clear that these definitions yield a presheaf of sets.
For example, if $\rho_2: V_3 \hookrightarrow V_2$ and $\rho_1:
V_2 \hookrightarrow V_1$ are a pair of composable inclusions of open
sets, then for all $\psi \in \underline{2R}_U(F,G)(V_1)$,
$\rho_2^* \rho_1^* \psi = (\rho_1 \rho_2)^* \psi$.

In the special case that ${\cal D}$ is a stack, not just a presheaf
of categories, it can be shown that $\underline{2R}_U(F,G)$ is
a sheaf of sets, not just a presheaf.  We shall outline the
details here.
Let $V \subseteq U$ be an open set, and let $\{ U_{\alpha} \}$ be
an open cover of $V$.  Let $\{ \psi_{\alpha} \in 
\underline{2R}_U(F,G)(U_{\alpha})
\}$ be a set of elements such that $\rho_{\alpha, \alpha \beta}^*
\psi_{\alpha} = \rho_{\beta, \alpha \beta}^* \psi_{\beta}$
(where the $\rho$ are the natural inclusions from $U_{\alpha}
\cap U_{\beta}$ into $U_{\alpha}$, $U_{\beta}$).
We need to show that there exists a unique $\psi \in
\underline{2R}_U(F,G)(V)$ such that $\rho_{\alpha}^* \psi = \psi_{
\alpha}$ (where $\rho_{\alpha}: U_{\alpha} \hookrightarrow V$ is
inclusion).

Finding a 2-arrow $\psi$ means
finding a set of natural transformations
$\psi(W): F(W) \Rightarrow G(W)$, one for each open $W \subseteq V$,
obeying the usual compatibility condition.  Let $P \in \mbox{Ob }
{\cal C}(W)$, then we can define $\psi(W)(P)$ to be the unique
morphism generated by gluing together morphisms
\begin{displaymath}
(\chi_{\alpha}^G)^{-1} \circ \psi_{\alpha}( W \cap U_{\alpha} )(P|_{
W \cap U_{\alpha} }) \circ \chi_{\alpha}^F: \:
F(W)(P)|_{ W \cap U_{\alpha} } \: \longrightarrow \:
G(W)(P)|_{ W \cap U_{\alpha} }
\end{displaymath}
It is straightforward to check that this defines a natural transformation
$\psi(W): F(W) \Rightarrow G(W)$ for all open $W \subseteq V$,
and also that these natural transformations satisfy
diagram~(\ref{2Rdef}).  Thus, we have defined a unique 2-arrow
$\psi$ such that $\rho^*_{\alpha} \psi = \psi_{\alpha}$,
and so we shown the gluing law for sheaves of sets is satisfied.
Thus, when ${\cal D}$ is a stack, the presheaf of sets
$\underline{2R}_U(F,G)$ is a sheaf of sets.

In addition to the presheaf of sets $\underline{2R}_U(F,G)$, one can
also define another presheaf of natural transformations,
which we shall label $\underline{NT}_U(F,G)$.
Our use of this presheaf will be very limited; we mention it
solely for completeness.
For any open $V \subseteq U$, define the set $\underline{NT}_U(F,G)(V)$
to be the set of all natural transformations
\begin{displaymath}
F(V) \circ |_{V} \: \Longrightarrow \: G(V) \circ |_{V}
\end{displaymath}
where $|_{V}$ indicates the restriction functor from ${\cal C}(U)$ to
${\cal C}(V)$.  For any inclusion $\rho: U_2 \hookrightarrow U_1$
between open sets $U_1, U_2 \subseteq U$, define the restriction
map
\begin{displaymath}
\rho^*: \: \underline{NT}_U(F,G)(U_1) \: \longrightarrow \:
\underline{NT}_U(F,G)(U_2)
\end{displaymath}
by,
\begin{displaymath}
\rho^* \eta \: \equiv \: \varphi_{1,2}^{{\cal C} -1} \circ \chi^G_{\rho} \circ
\eta \circ ( \chi^F_{\rho} )^{-1} \circ \varphi_{1,2}^{{\cal C}}
\end{displaymath}
for $\eta \in \underline{NT}_U(F,G)(U_1)$ (i.e., $\eta: F(U_1) \circ |_{U_1}
\Rightarrow G(U_1) \circ |_{U_1}$), where the $\varphi^{{\cal C}}$
are the natural
transformations defining ${\cal C}$ as a presheaf
of categories, and where the $\chi$ are the natural transformations
defining $F$, $G$ as Cartesian functors.

It is straightforward to check that this defines a presheaf of sets.
For example, for two composable inclusions $\rho_1: U_2 \hookrightarrow U_1$,
$\rho_2: U_3 \hookrightarrow U_2$, $U_1, U_2, U_3 \subseteq U$,
the restriction maps obey $\rho_2^* \rho_1^* \eta = (\rho_1 \rho_2)^* \eta$
for all $\eta: F(U_1) \circ |_{U_1} \Rightarrow G(U_1) \circ |_{U_1}$.

In the special case that ${\cal D}$ is a stack, not just a presheaf of
categories, it is straightforward to check that $\underline{NT}_U(F,G)$
is a sheaf of sets, not just a presheaf, for all open $U \subseteq X$.

In passing, note that any 2-arrow $\psi: F|_U \Rightarrow G|_U$
defines an element
of $\underline{NT}_U(F,G)(V)$ for all open $V \subseteq U$ for which
${\cal D}(V) \neq \emptyset$, and moreover if $\rho: U_2 \hookrightarrow
U_1$ is inclusion of open sets, then $\rho^* \psi(U_1) =
\varphi^{{\cal C} -1}_{1,2} \circ \psi(U_2) \circ \varphi^{{\cal C}}_{1,2}$,
i.e., the restriction functor relates these elements in a natural way.

\section{Gerbes and stacks}   \label{gerbesec}

For a brief but readable discussion of gerbes in terms of
stacks, see for example \cite[section 3]{brymcl1}.  More detailed
information is available in \cite{brylinski,giraud,breen}.
This section reviews material that can be
found in \cite[chapter 5]{brylinski},
\cite[section 3]{brymcl1},
and \cite[chapter 1]{breen}.

\subsection{Definitions and examples}

A stack ${\cal C}$ is called a (1-)gerbe if the following three
conditions are satisfied:
\begin{enumerate}
\item For every open set $U \subseteq X$, every
morphism in the category ${\cal C}(U)$ is invertible.  (In more
technical language, this means ${\cal C}(U)$ is a groupoid.)
\item Each point $x \in X$ has a neighborhood $U_x$ for which
${\cal C}(U_x)$ is nonempty.
\item Any two objects $P_1$, $P_2$ of ${\cal C}(U)$ are locally
isomorphic.  In other words, each $x \in U$ has a neighborhood
$V_x$ such that the restrictions of $P_1$ and $P_2$ to
$V_x$ are isomorphic.
\end{enumerate}

One says that a gerbe ${\cal C}$ is bound by a sheaf of
abelian groups ${\cal A}$ (or, that the gerbe has band ${\cal A}$)
if for any open set $U$ and object $P \in \mbox{Ob } {\cal C}(U)$,
there exists an isomorphism of sheaves of groups $\alpha_U(P):
{\cal A}|_U \cong \underline{\mbox{Aut}}_U(P)$.  These isomorphisms
are required to satisfy a constraint which we shall describe shortly.
In the rest of this paper, when we speak
of gerbes, we will implicitly refer to gerbes bound by some sheaf of
abelian groups.

In passing, we should point out that since $\alpha(U)(P)$ is a morphism
of sheaves of groups, it is compatible with restriction, i.e.,
the following diagram commutes:
\begin{equation}
\begin{array}{ccc}
{\cal A}(U_1) & \stackrel{ \alpha_U(U_1)(P) }{ \longrightarrow } &
\mbox{Hom}_{ {\cal C}(U_1) }( P, P ) \\
\makebox[0pt][r]{ $\scriptstyle{ \rho^* }$ } \downarrow & &
\downarrow \makebox[0pt][l]{ $\scriptstyle{ \rho^* }$ } \\
{\cal A}(U_2) & \stackrel{ \alpha_U(U_2)(P|_{U_2}) }{ \longrightarrow } &
\mbox{Hom}_{{\cal C}(U_2) }(P|_{U_2}, P|_{U_2})
\end{array}
\end{equation}
where the $\rho^*$ are the restriction maps defining the sheaves of sets,
$P \in \mbox{Ob } {\cal C}(U_1)$,
and $\rho: U_2 \hookrightarrow U_1$ is an inclusion between open sets
$U_1, U_2 \subseteq U$.

We mentioned above that the isomorphisms $\alpha_U(P): {\cal A}|_U
\rightarrow \underline{\mbox{Aut}}_U(P)$ are required to satisfy a
condition, which we shall now explain.
Let $U$ be an open set,
let $P_1, P_2 \in \mbox{Ob } {\cal C}(U)$, let 
$\beta \in \mbox{Hom}_{{\cal C}(U)}
\left( P_1, P_2 \right)$, and let $g \in \mbox{Aut}(P_1)$.
The morphism $\beta \circ g \circ \beta^{-1} \in \mbox{Aut}(P_2)$,
and in fact inner automorphisms by $\beta$ of the form above clearly define
a group homomorphism (in fact, an isomorphism)
from $\mbox{Aut}(P_1)$ to $\mbox{Aut}(P_2)$.
Put another way, for any $g \in \mbox{Aut}(P_1)$,
there exists $g' \in \mbox{Aut}(P_2)$ such that $\beta \circ
g = g' \circ \beta$, and $g$ and $g'$ are isomorphic
to the same element of the group ${\cal A}(U)$.
More generally, the morphism $\beta$ defines a morphism of
sheaves
\begin{displaymath}
\beta^*: \: \underline{\mbox{Aut}}_U(P_1) \: \stackrel{
\sim}{ \longrightarrow } \: \underline{\mbox{Aut}}_U(P_2)
\end{displaymath}
We can now finally state the constraint on the isomorphisms
$\alpha_U(P)$.
The $\alpha_U(P)$ are required to be such that the
following diagram commutes:
\begin{equation}
\begin{array}{ccc}
{\cal A}|_U & \stackrel{ \alpha_U(P_1) }{ \longrightarrow } &
\underline{\mbox{Aut}}_U(P_1) \\
\parallel & & \downarrow \makebox[0pt][l]{ $\scriptstyle{ \beta^* }$ } \\
{\cal A}|_U & \stackrel{ \alpha_U(P_2) }{ \longrightarrow } &
\underline{\mbox{Aut}}_U(P_2)
\end{array}
\end{equation}

The 1-gerbes discussed in \cite{dt1} all describe gerbes with
band $C^{\infty}(U(1))$.  Here we see that more general gerbes can be
defined.

A trivial example of a 1-gerbe with band $C^{\infty}(G)$ (for $G$
an abelian Lie group) is the
stack $\mbox{Tors}(G)$, the stack of all principal $G$-bundles,
introduced earlier.  In fact, we shall see later that all gerbes
with band $C^{\infty}(G)$ look locally like $\mbox{Tors}(G)$, in the
same sense that all fiber bundles look locally like the trivial bundle.

A nontrivial example of a 1-gerbe with band $C^{\infty}(U(1))$
is the stack describing $\mbox{Spin}^c(n)$ lifts of a principal $\mbox{SO}(n)$
bundle, which we shall now describe. 
(Relevant information can be found in \cite[appendix D]{lm}.) 
Let $P$ be a 
principal $\mbox{SO}(n)$-bundle on $X$.  We shall describe
a 1-gerbe, call it ${\cal C}$, that implicitly describes obstructions to lifting
the structure group of $P$ from $\mbox{SO}(n)$ to $\mbox{Spin}^c(n)$.
(We shall roughly follow \cite[section 5.2]{brylinski}.)
To any open set $U \subset X$, define the objects of ${\cal C}(U)$
to be pairs $(Q, \phi)$,
where $Q$ is a principal $\mbox{Spin}^c(n)$ bundle on $U$
which is a lift of $P |_U$,
and where $\phi: Q \rightarrow P|_U$ is a morphism of principal
bundles.  (Note that
if $P$ does not admit a global lift, then ${\cal C}(X) = \emptyset$,
for example.)  Morphisms between objects of ${\cal C}(U)$ are defined
as follows. 
Let $(Q, \phi)$, $(Q', \phi')$ be two objects of ${\cal C}(U)$.
A morphism $u: (Q, \phi) \rightarrow (Q', \phi')$ is defined to
be a morphism\footnote{Recall \cite[section 4.3]{husemoller} that
if $u: Q_1 \rightarrow Q_2$ is a morphism of principal bundles over 
the same space
with the same fiber, then it is necessarily an isomorphism.  Thus,
the morphisms we define in ${\cal C}(U)$ are all necessarily
isomorphisms.} 
$u: Q \rightarrow Q'$ of principal $\mbox{Spin}^c(n)$
bundles, such that the following diagram commutes:
\begin{equation}
\begin{array}{ccc}
Q & \stackrel{u}{\longrightarrow} & Q' \\
\makebox[0pt][r]{ $\scriptstyle{ \phi }$ } \downarrow & &
\downarrow \makebox[0pt][l]{ $\scriptstyle{ \phi' }$ } \\
P |_U & = & P |_U
\end{array}
\end{equation}
It is straightforward to check that this structure
defines a stack, and furthermore is a 1-gerbe.  The element of
$H^3(X, {\bf Z})$ associated to this 1-gerbe precisely classifies
the obstruction to lifting the structure group of $P$ to $\mbox{Spin}^c(n)$.
If the category ${\cal C}(X)$ is nonempty, then its objects are 
principal $\mbox{Spin}^c(n)$ bundles on $X$, globally defined lifts
of the principal $\mbox{SO}(n)$ bundle $P$.

More generally, we shall see later that a nontrivial gerbe on a space
$X$ can be
distinguished from a trivial gerbe on $X$ by the category ${\cal C}(X)$.
If this category is nonempty, then the gerbe is trivial -- objects
in ${\cal C}(X)$ define trivializations of the gerbe, just as global
sections of a principal bundle trivialize the bundle.  We shall return
to this matter later.

\subsection{Equivalences of gerbes}

Let ${\cal C}$ and ${\cal D}$ be 1-gerbes, both with band
${\cal A}$ (a sheaf of abelian groups).  Under what circumstances
can we say that ${\cal C}$ is equivalent to ${\cal D}$ ?

A map between two gerbes ${\cal C}$ and ${\cal D}$
with specified band ${\cal A}$ is defined to be a Cartesian functor
$F: {\cal C} \rightarrow {\cal D}$
such that for all open sets $U$ and for all $P \in \mbox{Ob }
{\cal C}(U)$, the following diagram of sheaves of groups commutes
(by definition of morphism of sheaves):
\begin{equation}  \label{gerbmap}
\begin{array}{ccc}
\underline{\mbox{Aut}}_{{\cal C}(U)}(P) & \stackrel{ F(U) }{ \longrightarrow } &
\underline{\mbox{Aut}}_{{\cal D}(U)}(F(U)(P)) \\
\makebox[0pt][r]{ $\scriptstyle{ \alpha_{{\cal C}(U)}(P) }$ } \downarrow
\makebox[0pt][l]{ $\scriptstyle{ \sim }$ }
& & \makebox[0pt][r]{ $\scriptstyle{ \sim }$ }
\downarrow \makebox[0pt][l]{ $\scriptstyle{ \alpha_{{\cal D}(U)}(P) }$ } \\
{\cal A} |_U & = &  {\cal A} |_U
\end{array}
\end{equation}
where the $\alpha$ are the isomorphisms between the band and the
automorphisms of an object (given in the definition of band),
and we have used $F(U)$ to denote the induced morphism of sheaves
(here, sheaves of abelian groups) discussed in the section on Cartesian
functors.  
More intuitively, this condition means
that the action of the band on the gerbe commutes with the Cartesian
functor.

An equivalence of two gerbes ${\cal C}$, ${\cal D}$ with band ${\cal A}$
is defined to be a Cartesian functor $F: {\cal C} \rightarrow
{\cal D}$ obeying the constraint~(\ref{gerbmap}), such that
the Cartesian functor defines an equivalence of stacks.

We shall show in section~\ref{gmapequiv} that any map between two
gerbes with the same band, over the same space, is necessarily
an equivalence of gerbes.  This is closely analogous to the
result that any morphism of principal $G$-bundles, for fixed
$G$, over the same space, is necessarily an isomorphism
\cite[section 4.3]{husemoller}.

\subsection{Sheaf cohomology and gerbes}

In \cite{dt1} we claimed that gerbes were classified by
elements of sheaf cohomology groups.
How can we derive an element of $H^2(X, {\cal A} )$
from the description of gerbes given above? 
We shall work through the details in this subsection.
More precisely, we shall show how to obtain a \v{C}ech representative
of the relevant sheaf cohomology group, associated to some fixed
open cover.
For convenience, we shall assume the band ${\cal A} = C^{\infty}(U(1))$,
though the reader should be able to easily extend to more general cases.

Before describing how to associate sheaf (and also, more usually, \v{C}ech)
cohomology elements to gerbes, we shall take a moment to review
how this procedure works for sheaves of local sections of bundles.
Let $I$ be a sheaf of local sections of some principal 
$G$-bundle on a space $X$, 
and let $\{ U_{\alpha} \}$
be a good open cover of $X$.  Let $\{ s_{\alpha} \}$ be any choice
of local sections of $I$ with respect to the cover $\{ U_{\alpha} \}$
(i.e., $s_{\alpha} \in I(U_{\alpha})$ for all $\alpha$).
Then on each overlap $U_{\alpha \beta} = U_{\alpha} \cap U_{\beta}$, 
the sections 
$s_{\alpha} |_{U_{\alpha \beta}}$ and $s_{\beta}|_{U_{\alpha \beta}}$
will differ by some element of $C^{\infty}(G)$ over $U_{\alpha \beta}$.
Denote each such element by $g_{\alpha \beta}$.
It is straightforward to check that the $g_{\alpha \beta}$ define a
cocycle representative of an element of $H^1(X, C^{\infty}(G))$
(or, rather, the corresponding \v{C}ech cohomology group associated to
the cover $\{ U_{\alpha} \}$).  Picking different local sections
corresponds to changing the cocycle by a coboundary.
Thus, we have derived an element of $H^1(X, C^{\infty}(G))$ classifying
the sheaf $I$.  Finally, note that $I$ admits a global section if and only
if there exist sections $s_{\alpha}$ such that $s_{\alpha}|_{U_{\alpha \beta}}$
and $s_{\beta}|_{U_{\alpha \beta}}$ agree on overlaps,
i.e., $g_{\alpha \beta}$ is
the identity on each overlap, i.e., the corresponding element of sheaf and
\v{C}ech cohomology is trivial.

Now that we have described how to associate \v{C}ech cohomology elements
to any given sheaf of local sections of a bundle, we shall discuss
how to associate cohomology elements to gerbes.  
We shall see that the details are closely analogous to the case above.

Let $\{ U_{\alpha} \}$ be a good cover
of $X$, i.e., a cover such that every element and every intersection
of elements is contractible.  We shall assume this is sufficient
for every object of any category ${\cal C}(U_{\alpha})$ to be isomorphic.
(If not, pick a suitable refinement of $\{ U_{\alpha} \}$.)
Then, let $P_{\alpha}$ denote an object of ${\cal C}(U_{\alpha})$.
(Since all objects in ${\cal C}(U_{\alpha})$ are isomorphic,
the precise choice of $P_{\alpha}$ is irrelevant.)
Let $u_{\alpha \beta}$ denote the isomorphisms
\begin{displaymath}
u_{\alpha \beta}: P_{\alpha} |_{U_{\alpha \beta}} 
\stackrel{\sim}{\longrightarrow} P_{\beta} |_{U_{\alpha \beta}}
\end{displaymath}
where $U_{\alpha \beta} = U_{\alpha} \cap U_{\beta}$,
and we implicitly assume $u_{\alpha \beta} = u_{\beta \alpha}^{-1}$.
Suppose $U_{\alpha \beta \gamma} = U_{\alpha} \cap U_{\beta} \cap
U_{\gamma} \neq \emptyset$.  Define $h_{\alpha \beta \gamma}:
U_{\alpha \beta \gamma} \rightarrow U(1)$
by
\begin{eqnarray*}
h_{\alpha \beta \gamma} & = &
u_{\gamma \alpha} \circ u_{\beta \gamma} \circ u_{\alpha \beta} \\
& \in & \mbox{Aut}( P_{\alpha} |_{U_{\alpha \beta \gamma}} )
\end{eqnarray*}

In fact, we have been slightly sloppy about the distinction between
objects $x |_{U_{\alpha}} |_{U_{\alpha \beta}}$ and $x |_{U_{\alpha \beta}}$,
for example.  To rigorously define $h_{\alpha \beta \gamma}$ we must
introduce the invertible natural transformations $\varphi$ defining
${\cal C}$ as a presheaf of categories.  Define maps
$\overline{u}_{\alpha \beta, \alpha \beta \gamma}: P_{\alpha} |_{U_{\alpha
\beta \gamma}} \rightarrow P_{\beta} |_{U_{\alpha \beta \gamma}}$
by,
\begin{displaymath}
\overline{u}_{\alpha \beta, \alpha \beta \gamma} \: = \:
\varphi_{\alpha \beta, \alpha \beta \gamma}^{-1} \circ u_{\alpha \beta}
 |_{\alpha
\beta \gamma} \circ \varphi_{\alpha \beta, \alpha \beta \gamma}
\end{displaymath}
then the rigorous definition of $h_{\alpha \beta \gamma}$ is as
\begin{eqnarray*}
h_{\alpha \beta \gamma} & = & \overline{u}_{\gamma \alpha, \alpha \beta \gamma}
\circ \overline{u}_{\beta \gamma, \alpha \beta \gamma} \circ 
\overline{u}_{\alpha \beta,
\alpha \beta \gamma} \\
 & \in & \mbox{Aut}(P_{\alpha} |_{U_{\alpha \beta \gamma}})
\end{eqnarray*}
It is straightforward to check that $h_{\alpha \beta \gamma}$ defines
a \v{C}ech 2-cocycle.

Naively, our description of $h_{\alpha \beta \gamma}$ above
might appear to always be a coboundary, as $h_{\alpha \beta \gamma}$
naively appears to be the coboundary of a 1-cocycle defined by the
$u_{\alpha \beta}$.  However, there is an important distinction at work
here.  The $\overline{u}_{\alpha \beta, \alpha \beta \gamma}$ are maps between
(in general)
distinct objects, not automorphisms of a single object.
Since they are maps between distinct objects, they will not
(in general) be valued in the band of the gerbe.  Thus, $h_{\alpha
\beta \gamma}$ will not be a trivial \v{C}ech cocycle in general.

It can also be shown that if $\{ P_{\alpha} , u_{\alpha \beta} \}$
and $\{ P'_{\alpha} , u'_{\alpha \beta} \}$ are two choices
of objects and isomorphisms, then the 2-cocycles $h_{\alpha \beta
\gamma}$, $h'_{\alpha \beta \gamma}$ defined by either differ by
a coboundary.  We shall leave the details of this verification to
the reader, but in passing we will mention three important points that
come up.  First, one needs the fact that elements of the band commute
with morphisms, as demonstrated earlier.  Second, if we let $\psi_{\alpha}:
P'_{\alpha} \stackrel{\sim}{\longrightarrow} P_{\alpha}$ be a set
of isomorphisms, then it is a useful fact that the $\psi_{\alpha}$
commute with the natural transformations used to define the presheaf
of categories, by definition of natural transformation.
Finally, note that $u_{\alpha \beta}$ and 
$\psi_{\beta} |_{U_{\alpha \beta}} \circ u'_{\alpha \beta} \circ
\psi_{\alpha}^{-1} |_{U_{\alpha \beta}}$ may differ by an element
of the band, in general.

Thus, any set of choices $\{ P_{\alpha}, u_{\alpha \beta} \}$
will define the same element of cohomology. 

Furthermore, any two equivalent gerbes define the same element of
cohomology.  Let $F: {\cal C} \rightarrow {\cal D}$ be a map 
between two gerbes on $X$ with band ${\cal A}$, which also defines
an equivalence of gerbes.  Let $\{ U_{\alpha} \}$ be a good
open cover of $X$, let $\{ P_{\alpha} \in \mbox{Ob }{\cal C}(U_{\alpha}) \}$
be a set of objects, and $\{ u_{\alpha \beta}: P_{\alpha} |_{U_{\alpha \beta}}
\stackrel{\sim}{\longrightarrow} P_{\beta} |_{U_{\alpha \beta}} \}$ be
a set of isomorphisms.  Then $\{ F(U_{\alpha})(P_{\alpha}) \in
\mbox{Ob } {\cal D}(U_{\alpha}) \}$ is a set of objects in ${\cal D}$,
and 
\begin{displaymath}
\left\{ F(U_{\alpha \beta})( u_{\alpha \beta} ):
F(U_{\alpha \beta})( P_{\alpha} |_{U_{\alpha \beta}} )
\stackrel{\sim}{\longrightarrow} F(U_{\alpha \beta})( P_{\beta} |_{U_{\alpha
\beta}} ) \right\}
\end{displaymath}
is a set of isomorphisms between objects in ${\cal D}$.
These objects and isomorphisms in ${\cal D}$ define a cocycle $h^{\cal D}_{
\alpha \beta \gamma}$, but as noted above, the cohomology class of the
cocycle is independent of the choice of objects and isomorphisms.
Thus, any two equivalent gerbes define the same element of cohomology.

An astute reader may be slightly confused by the paragraph above.
Nowhere in our discussion did we seem to use the fact that
$F: {\cal C} \rightarrow {\cal D}$ is an equivalence of gerbes;
we only used the fact that it is a map of gerbes.  
In particular, the reader might be concerned that if we could derive
a contradiction:  if $F$ were not an equivalence of gerbes, then
we should not get the same cohomology element.  However, we shall
show in section~\ref{gmapequiv} that any map of gerbes with the same
band and over the same space is necessarily an equivalence of gerbes.
Thus, it is not possible for $F$ to not be an equivalence of gerbes,
and so any potential contradiction is averted.

So far we have described how a sheaf cohomology element can be associated
to a given gerbe.  The converse is also possible -- given a sheaf 
cohomology element, we can construct an associated gerbe.
This construction is carried out in, for example, 
\cite[section 5.2]{brylinski}.  We shall not repeat the construction here,
but in passing we shall mention that it uses techniques closely akin
to the descent categories we describe in section~\ref{desccat}.

It should now be clear that equivalence classes of gerbes with band
${\cal A}$ on a space $X$ are in one-to-one correspondence with elements of
$H^2(X, {\cal A})$.

Let us take a moment to try to gain some intuition for the meaning
of this sheaf cohomology description of gerbes.
Suppose a gerbe ${\cal C}$ on a space $X$ 
is described by a cohomologically-trivial
cocycle $h_{\alpha \beta \gamma}$; what does this imply about
the gerbe?  If $h_{\alpha \beta \gamma}$ is a coboundary,
then by slight redefinitions of the isomorphisms $u_{\alpha \beta}:
P_{\alpha} |_{U_{\alpha \beta}} \stackrel{\sim}{\longrightarrow}
P_{\beta} |_{U_{\alpha \beta}}$ we can arrange for $h_{\alpha \beta 
\gamma} = 1$.  (Simply compose each isomorphism with an automorphism
dictated by the cochain defining the coboundary; use the fact that
elements of the band commute with morphisms.)  Then, we can use the
gluing law for objects to construct an object of ${\cal C}(X)$.
Thus, a gerbe described by a cohomologically-trivial cocycle
has ${\cal C}(X) \neq \emptyset$.
Conversely, it is straightforward to check that if ${\cal C}(X) \neq
\emptyset$, then the corresponding element of cohomology is trivial.
This is closely analogous to the fact that a global section of
a principal $G$-bundle is a trivialization of the bundle;
here, an object of ${\cal C}(X)$ is a trivialization of the gerbe.

In particular, let us consider the example of the gerbe $\mbox{Tors}(G)$,
for $G$ an abelian Lie group, on a space $X$.  In this case,
$\mbox{Tors}(G)(X) \neq \emptyset$, that is, there exists a globally
defined object (in fact, several globally defined objects, in general), 
so it should be clear that on a good open cover $\{ U_{\alpha} \}$
one can pick objects $\{ P_{\alpha} \}$ and isomorphisms
$\{ u_{\alpha \beta} \}$ such that $h_{\alpha \beta \gamma} = 1$.
In other words, $\mbox{Tors}(G)$ is an example of a gerbe with
a globally defined object, which implies that the corresponding
element of cohomology is trivial.
$\mbox{Tors}(G)$ is an example of a trivial gerbe.

The $\mbox{Spin}^c$ gerbes, discussed earlier,
are somewhat more interesting.  From the general discussion above,
as this class of gerbes has band $C^{\infty}(U(1))$, they should
be topologically classified by elements of 
\begin{displaymath}
H^2(X, C^{\infty}(U(1)) ) \: \cong \: H^3(X, {\bf Z})
\end{displaymath}
Indeed, in discussing whether an $SO(n)$-bundle can be lifted to
a $\mbox{Spin}^c(n)$-bundle, there arises an element of
$H^3(X,{\bf Z})$, usually labeled $W_3$, which is the image under
a Bockstein homomorphism of the second Stiefel-Whitney class of the
$SO(n)$-bundle in question.  The characteristic class $W_3$ is precisely
the integral characteristic class classifying the $\mbox{Spin}^c$
gerbe.  An $SO(n)$-bundle admits a $\mbox{Spin}^c$ lift
if and only if $W_3$ vanishes, in precise accord with the
general framework above -- a 1-gerbe admits a global trivialization
if and only if the classifying integral characteristic class vanishes.
A global trivialization of the $\mbox{Spin}^c$ gerbe associated to
some $SO(n)$-bundle is precisely a $\mbox{Spin}^c$ lift of the bundle.

\subsection{Gauge transformations of gerbes}    \label{gerbegt}

Just as a gauge transformation of a principal $G$-bundle on a space $X$
is defined
by a map $X \rightarrow G$, it turns out that a gauge transformation of
a gerbe with band\footnote{More generally, a gauge transformation
on a 1-gerbe with band ${\cal A}$ is defined by an ${\cal A}$-torsor.
We are specifically trying to avoid using the language of torsors
in this subsection.} ${\cal A} = C^{\infty}(G)$ is a principal $G$-bundle.
(In terminology introduced in \cite{dt1}, a gauge transformation of
an $n$-gerbe is defined by an $(n-1)$-gerbe.)
(Note that we are implicitly assuming $G$ is an abelian Lie group.)

Strictly speaking, only equivalence classes of principal $G$-bundles
will have distinct actions, but we shall defer discussion of this
technicality until later.

How precisely does a principal $G$-bundle act on a gerbe with
band $C^{\infty}(G)$?
A complete discussion of the technical details is beyond the
intended scope of this section -- see instead section~\ref{gtap}
for a complete discussion.  However, we can give some general
intuition.  Suppose, for example, the objects of ${\cal C}(U)$ are
line bundles on $U$.  In this case, the action of $I$ amounts
to tensoring each object with $I$.  This yields, for any object
$P$, a map $P \mapsto P \times I$.  (We use the notation $P \times I$
instead of something like $P \otimes I$ because at the end of the
day, we need not be manipulating bundles.  Again,
see section~\ref{gtap} for details.)

Not only does a principal $G$-bundle yield an action on the
objects of ${\cal C}(U)$, but it can also be used to define
a self-equivalence of the category ${\cal C}(U)$.

In section~\ref{gtap} we will give a number of results related 
gauge transformations of gerbes.  One result, as noted
above, is that a gauge transformation of a gerbe is defined
by a bundle (a 0-gerbe), just as a gauge transformation of
a bundle is given by a function (a (-1)-gerbe).  We also argue
that gauge transformations of gerbes commute with gerbe maps,
just as gauge transformations of principal bundles commute
with principal bundle maps, and that any map between two gerbes
of the same band, over the same space, is necessarily an isomorphism,
just as any morphism of principal $G$-bundles, for fixed $G$
and over a fixed space, is necessarily an isomorphism.
Finally, we shall argue that any gerbe with band $C^{\infty}(G)$
looks locally
like the trivial gerbe $\mbox{Tors}(G)$, just as any principal
bundle looks locally like the trivial principal bundle.

\subsection{Connections on gerbes}    \label{gerbeconnecintro}

In this subsection we shall restrict to gerbes with band
$C^{\infty}(U(1))$, for convenience.

Now, how does one define a connection on a 1-gerbe, defined in terms
of stacks as above?  It is tempting to proceed as follows.
(This description will be wrong, but useful pedagogically.)
Let $\{ U_{\alpha} \}$ be an open cover.  Identify ${\cal C}(U_{\alpha})$
with $\mbox{Tors}( U(1) )(U_{\alpha})$, i.e., identify objects
of ${\cal C}(U_{\alpha})$ with principal $U(1)$-bundles on $U_{\alpha}$.
To each open set $U_{\alpha}$ and principal $U(1)$ bundle ${\cal L}_{\alpha}$,
associate a connection $\nabla^{\alpha}$.  Note that since
${\cal L}_{\alpha}$ need not be trivial, and since $U_{\alpha}$ need
not be contractible, there is no reason why $\nabla^{\alpha}$ should
be expressible in terms of a single 1-form definable over all of
$U_{\alpha}$.  Let $F^{\alpha}$ denote the curvature of $\nabla^{\alpha}$
on $U_{\alpha}$.  It is very tempting (and incorrect) to then identify 
the $B$-field
associated to $U_{\alpha}$ with $F^{\alpha}$.  Then, on overlaps,
$F^{\alpha} - F^{\beta} = 0$ in cohomology, so there exists
1-forms $A^{\alpha \beta}$ such that $F^{\alpha} - F^{\beta} = d A^{\alpha
\beta}$.  We could then clearly build up the \v{C}ech-de Rham complex describing
a connection on a gerbe, as described in \cite{dt1}.

Unfortunately, this natural-looking idea will not work in general.
The essential problem is that this would define a 3-form $H$ that
was always zero in cohomology, as $H |_{U_{\alpha}} = d F^{\alpha} = 0$.
Put more simply, the $B$-field associated to any open set need not
be closed, whereas the idea described in the paragraph above would
always necessarily associate a closed 2-form to each $U_{\alpha}$.

The correct way to associate a connection to a 1-gerbe is somewhat
more complicated to explain.

In this section we shall make frequent use of the idea of a torsor.
We have strenuously avoided speaking of torsors in previous sections,
but at this point their use becomes unavoidable.
Torsors are defined in section~\ref{torsdef}; we shall assume
henceforward that the reader is acquainted with the material in that
section.

Connections on gerbes are defined in \cite{brylinski} in terms
of ``connective structures'' and ``curvings'' on the gerbe.
In order to get some intuition for the meaning of these concepts,
we shall take a moment to define analogues of ``connective structure''
and ``curving'' for a bundle.
On a fixed principal $U(1)$-bundle on a space $X$, a ``connective structure''
is defined to be an $\Omega^1(X)$-torsor consisting of all
the connections on the principal $U(1)$-bundle.  This connective
structure is required to obey the constraint that any gauge
transformation $\phi: X \rightarrow U(1)$ defines an automorphism
$\phi_*$ of the connective structure, such that any section
$\nabla$ of the connective structure (a single connection on the
bundle) transforms under $\phi_*$ as, $\phi_*(\nabla) = \nabla - d \, \mbox{ln }
\phi$.  One then defines a ``curving,'' which is a map that assigns
a closed 2-form $K(\nabla)$ to any section $\nabla$ of the connective
structure, such that 
\begin{enumerate}
\item if $\phi: X \rightarrow U(1)$ is a gauge transformation,
then $K(\phi_*(\nabla)) = K(\nabla)$ for any $\nabla$
\item for any $\alpha \in \Omega^1(X)$, 
$K(\nabla + \alpha) = K(\nabla) + d \alpha$
\end{enumerate}
Clearly the curving corresponds to the curvature of the connection $\nabla$. 

These notions of connective structure and curving for a bundle 
seem quite clumsy, however they are more useful when discussing gerbes.

Now that we have given some basic intuitions, we shall give
the rigorous definition of a connection on a 1-gerbe.
This is formally described as assigning a connective structure
and curving to a gerbe, call it ${\cal C}$.
We shall closely follow the presentation of \cite[section 5.3]{brylinski}.

Let $\Omega^1(U)$ denote the sheaf of 1-forms on an open set $U$.
(Note that $\Omega^1(U)$ is a sheaf of abelian groups on $U$.)

A connective structure on a gerbe ${\cal C}$ is defined to be
a Cartesian functor $\mbox{Co}: {\cal C} \rightarrow \mbox{Tors}(
\Omega^1 )$ between the underlying stacks, subject to the following constraint.
Let $U$ be an open set and $P \in \mbox{Ob } {\cal C}(U)$,
let $\nabla$ be a section of the $\Omega^1(U)$-torsor
$\mbox{Co}(U)(P)$, and let $g: P \rightarrow P$ be an automorphism
of $P$, which we shall identify with an element of the band.
Then we demand
$\mbox{Co}(U)(g) (\nabla) = \nabla - d \ln g$.

At this point we shall introduce some notation.
For any isomorphism $\phi: P_1 \stackrel{ \sim }{ \longrightarrow }
P_2$ of objects of ${\cal C}(U)$, its image under the
functor $\mbox{Co}(U)$ is denoted $\phi_*$.  In other words,
$\phi_* = \mbox{Co}(U)(\phi)$.
For any inclusion $\rho: U_1 \hookrightarrow U$, let $\chi_{\rho}$
denote the invertible natural transformation $\rho^* \circ \mbox{Co}(U)
\Rightarrow \mbox{Co}(U_1) \circ \rho^*$ appearing in the definition
of $\mbox{Co}$ as a Cartesian functor.  
(Readers also studying \cite[section 5.3]{brylinski} will note that
in that reference, $\mbox{Co}(U)(P)$ is abbreviated to
$\mbox{Co}(P)$, and the natural transformation we denote by $\chi_{\rho}$
is there denoted $\alpha_{\rho}$.)

An example of a gerbe with connective structure is in order at this
point.  Consider the (trivial) gerbe $\mbox{Tors}(G)$ of all principal
$G$-bundles, where we assume $G$ is an abelian Lie group.
An example of a connective structure on $\mbox{Tors}(G)$ is the one
obtained by assigning, to each principal $G$-bundle $P$ over any
open set $U$, the $\Omega^1(U)$-torsor of all connections on $P$.
In other words, define $\mbox{Co}(U)(P)$ to be the $\Omega^1(U)$-torsor
of all connections on $P$, and for any isomorphism $f: P_1 \rightarrow P_2$
define $\mbox{Co}(U)(f)$ to be the morphism such that for any
$\nabla \in \Gamma(U, \mbox{Co}(U)(P_1))$,
the morphism
\begin{displaymath}
f: \: \left( P_1, \nabla \right) \: \longrightarrow \:
\left( P_2, \mbox{Co}(U)(f)(\nabla) \right)
\end{displaymath}
is an equivalence of bundles with connection.
One can then define the rest of the structure of
a Cartesian functor in the obvious way.

It can be shown that any gerbe with band $C^{\infty}(U(1))$ admits a
connective structure.  We shall not work through the details of
this argument here; see instead, for example, \cite[section 5.3]{brylinski}.

Note that \cite[section 5.3]{brylinski} if $\Phi: {\cal G}
\rightarrow {\cal G}'$ defines an equivalence of two gerbes
${\cal G}$, ${\cal G}'$ on a space $X$, both with band $C^{\infty}(
U(1) )$, and the gerbe ${\cal G}'$ has a connective structure,
call it $\mbox{Co}'$, then the connective structure $\mbox{Co}'$ 
on ${\cal G}'$ can
be pulled back to form a connective structure $\mbox{Co}$ on ${\cal G}$.
More specifically, $\mbox{Co} = \mbox{Co}' \circ \Phi$.

It should now be clear that there is a natural notion of equivalence
of gerbes with connection structure.  Let $( {\cal G}_1, \mbox{Co}_1 )$
and $( {\cal G}_2, \mbox{Co}_2 )$ be a pair of gerbes with connective
structure.  We say that a Cartesian functor $\Phi: {\cal G}_1
\rightarrow {\cal G}_2$ defines an equivalence of gerbes with connective
structure if 
\begin{enumerate}
\item $\Phi$ defines an equivalence of gerbes, and
\item there exists an invertible 2-arrow $\Psi: \mbox{Co}_1 \Rightarrow
\mbox{Co}_2 \circ \Phi$ between the Cartesian functors
$\mbox{Co}_1$ and $\mbox{Co}_2 \circ \Phi$.
\end{enumerate}
For example, we shall see later that a principal $U(1)$ bundle $I$
defines a gerbe automorphism $I_{{\cal C}}: {\cal C} \rightarrow 
{\cal C}$, and in such a case, the 2-arrow $\Psi$ defined above
is equivalent to a choice of connection on the bundle.

Just as the difference between any two connections on a principal
$U(1)$ bundle on $X$ is a 1-form on $X$, i.e., an element of
$\Omega^1(X)$, it can be shown \cite[prop. 5.3.6]{brylinski} that
the difference between any two connective structures on the same
gerbe on $X$ is given by an $\Omega^1(X)$-torsor.

So far we have yet to describe precisely how to associate a $B$ field to
a 1-gerbe.  $B$ fields are described as a ``curving'' of the connective
structure introduced above.

More precisely, given some gerbe ${\cal C}$ on $X$ with connective
structure $\mbox{Co}$, we can define a curving of the
connective structure as follows.  A curving of the connective structure
is a rule that assigns to any object $P \in \mbox{Ob } {\cal C}(U)$
and to any section $\nabla$ of the $\Omega^1(U)$-torsor $\mbox{Co}(U)(P)$,
a (${\bf R}$-valued) 2-form $K(\nabla)$ on $U$, called the 
curvature\footnote{The reader should note that the 2-form which we here
denote the ``curvature'' of a connection, is not the curvature
in the usual sense.  Rather, it is merely some 2-form -- not necessarily
closed -- associated to the connection.  The nomenclature is unfortunate,
but seems to be standard.} of
$\nabla$, such that the following three properties are satisfied:
\begin{enumerate}
\item Given an inclusion $\rho: U_1 \hookrightarrow U$, the curvature
$K( \chi_{\rho}(\rho^* \nabla) )$ of the section $\chi_{\rho}(
\rho^* \nabla)$ of $\mbox{Co}(U_1)(\rho^* P)$ is equal to 
$\rho^* K(\nabla)$, where $\chi_{\rho}$ denotes the natural transformation
defining $\mbox{Co}$ as a Cartesian functor.
\item Let $\phi: P \stackrel{ \sim }{ \longrightarrow } P'$ be an isomorphism
with another object $P' \in \mbox{Ob } {\cal C}(U)$.
Let $\phi_*(\nabla)$ be the corresponding section of $\mbox{Co}(U)(P')$.
Then $K(\nabla) = K(\phi_*(\nabla))$.
\item Let $\omega \in \Omega^1(U)$.  Then $K(\nabla + \omega) =
K(\nabla) + d \omega$.
\end{enumerate}

From the last two conditions on the curving, we see that
$K$ associates to any isomorphism class of objects in ${\cal C}(U)$,
a 2-form (an element of $\Omega^2(U)$, not necessarily closed)
modulo exact 2-forms.

It can be shown \cite[section 5.3]{brylinski} that, 
given a connective structure $\mbox{Co}$ on
a gerbe, there always exist curvings.

In passing, we should mention that if $\Phi_1, \Phi_2: {\cal G}_1
\rightarrow {\cal G}_2$ are any pair of gerbe maps between the
gerbes ${\cal G}_1$, ${\cal G}_2$, and $\psi: \Phi_1 \Rightarrow \Phi_2$
is any 2-arrow, then for any curving $K$ on $({\cal G}_2, \mbox{Co})$,
we have as an immediate consequence of the definition of curving that
\begin{displaymath}
K(\nabla) \: = \: K( \, \psi_* \nabla) \, )
\end{displaymath}
where $\nabla \in \Gamma(U, ( \mbox{Co}\circ \Phi_1)(U)(P) )$,
for any open $U$ and any object $P \in \mbox{Ob } {\cal G}_1(U)$.
Intuitively, this means that if $(I, \nabla)$, $(I', \nabla')$ are
two bundles with connection which are isomorphic (as bundles with connection),
then the associated 2-forms $K(\nabla)$ and $K(\nabla')$ should
be identical -- the 2-forms $K$ should be the same on equivalence
classes of bundles with connection.

Suppose that two gerbes with connective structure $( {\cal G}_1,
\mbox{Co}_1 )$ and $( {\cal G}_2, \mbox{Co}_2 )$ come with specified
curvings $K_1$, $K_2$, respectively.  Then we say that $\Phi$ defines
an equivalence of gerbes with connective structure and curving if
\begin{enumerate}
\item $(\Phi, \Psi)$ defines an equivalence of gerbes with connective structure,
where $\Psi: \mbox{Co}_1 \Rightarrow \mbox{Co}_2 \circ \Phi$ is the 
associated 2-arrow between connective structures, and
\item for all open $U$, for all objects $P \in \mbox{Ob }{\cal G}_1(U)$,
and for all $\nabla \in \Gamma(U, \mbox{Co}_1(U)(P))$, we have that
\begin{displaymath}
K_1( \, \nabla \,) \: = \: K_2( \, \Psi(\nabla) \, )
\end{displaymath}
\end{enumerate}

Earlier, we mentioned that for a gerbe automorphism $I_{{\cal C}}$
defined by a principal bundle $I$, specifying a 2-arrow $\Psi:
\mbox{Co} \Rightarrow \mbox{Co} \circ I_{{\cal C}}$ is equivalent
to specifying a connection on the bundle.  In order for
such an automorphism of a gerbe with connective structure
to be an automorphism of a gerbe with connective structure and curving,
the constraint on the $K$'s implies that the connection on $I$
must be flat.

We define an ``equivalence of gerbes with connection'' to be
an equivalence of gerbes with connective structure and curving.
We shall usually use the former notation rather than the latter, as it
is briefer.

How can we make contact with the description of connection given
earlier in section~\ref{gerbesec}?  
Let $\{ U_{\alpha} \}$ be a good cover of $X$,
such that the objects in any one category ${\cal C}(U_{\alpha_1 \cdots
\alpha_n})$ are all isomorphic.  Then to each open set $U_{\alpha}$,
the curving $K$ associates a 2-form (not necessarily closed),
defined up to the addition of an exact 2-form.
These 2-forms are precisely the 2-forms appearing in the
earlier definition of connections on 1-gerbes.
The rest of the earlier description -- connections on principal
$U(1)$ bundles on overlaps -- can be understood directly in terms
of the transition functions.

Earlier we argued that principal $G$-bundles define gauge
transformations on gerbes with band $C^{\infty}(G)$; how does
such a gauge transformation act on the connective structure and
curving?  We shall examine this in detail in section~\ref{gtconnec};
we shall outline the results here.
Let $I$ denote a principal $G$-bundle 
defining a gauge transformation
of a 1-gerbe.  Let $P \in \mbox{Ob } {\cal C}(U)$.
We argued earlier that $I$ defines a map $P \mapsto P \times (I|_U)$.
Now, a precise specification of how a particular section
$\nabla \in \Gamma(U, Co(U)(P))$ is mapped by $I$ is equivalent
to a specification of a connection on $I$ 
\cite[section 5.3, equ'n~(5-11)]{brylinski}.
(Note in passing that this statement dovetails with the earlier observation
that any two connective structures differ by an $\Omega^1$-torsor.)

More explicitly, let $\{ A^{\alpha} \}$ be a connection on $I |_U$,
defined with respect to an open cover $\{ U_{\alpha} \}$ of $U$.
In other words, each $A^{\alpha}$ is a 1-form on $U_{\alpha}$.
Then if $\nabla_{\alpha} \in \Gamma( U_{\alpha}, Co(U_{\alpha})(P|_{\alpha}) )$,
then under the action of $I$ on the gerbe, $\nabla_{\alpha} \mapsto
\nabla_{\alpha} + A^{\alpha}$.

So far we have described how the bundle $I$ defining a gauge transformation
acts on the connective structure.  How does $I$ act on the curving?
It is clear from the definition of curving that
$K(\nabla_{\alpha} + A^{\alpha} ) = K(\nabla_{\alpha}) + d A^{\alpha}$.

Thus, we have recovered the description of gauge transformations
on gerbe connections outlined in \cite{dt1}.

\subsection{Transition functions for gerbes}

In section~\ref{morphtors}, we show that if ${\cal C}$ is a gerbe
with band\footnote{More generally, for any open $U$ such that
${\cal C}(U)$ is nonempty, the category ${\cal C}(U)$ is equivalent
to the category $\mbox{Tors}({\cal A})(U)$ of ${\cal A}$-torsors
on $U$.}
${\cal A} = C^{\infty}(G)$, then for any open $U$
such that ${\cal C}(U)$ is nonempty, the category ${\cal C}(U)$
is equivalent to the category $\mbox{Tors}(G)(U)$.

We can use this fact to define transition functions for gerbes.
Now, such a term should be explained -- we have described
gerbes in terms of sheaves of categories, analogously to
describing bundles in terms of sheaves of sections.
Transition functions are not a necessary component of such
a description.  However, we can certainly recover transition
functions, if we choose to do so.  (For a discussion of
sheaves of sets in terms of transition-function-like language,
see for example \cite[ch. I.A.iii, cor. I-11]{eisenhar}.)

The transition functions for a gerbe should be clear.
Given some open cover $\{ U_{\alpha} \}$ of $X$ such that
${\cal C}(U_{\alpha}) \neq \emptyset$ for all $\alpha$,
and a set of equivalences of categories from ${\cal C}(U_{\alpha})$
into $\mbox{Tors}(G)(U_{\alpha})$, it should be clear that
we could describe the gerbe in terms of transition functions
between the categories $\mbox{Tors}(G)(U_{\alpha})$.
In other words, we can specify the gerbe by specifying
principal $G$-bundles on overlaps $U_{\alpha \beta} = U_{\alpha}
\cap U_{\beta}$.  Each such bundle determines a functor,
and so determines how $\mbox{Tors}(G)(U_{\alpha})$ is
mapped into $\mbox{Tors}(G)(U_{\beta})$.  Elements of \v{C}ech
cohomology classifying the gerbe are determined from
tensor products of the bundles on triple overlaps;
conversely, if one wishes to describe a gerbe with fixed
\v{C}ech cohomology, one can demand that tensor products
of bundles on triple overlaps have appropriate
canonical trivializations.  This description
is precisely analogous to describing bundles in terms of
gauge transformations on overlaps, describing how the local  
trivializations are mapped into one another.  
Note furthermore that this description is precisely the description
of 1-gerbes given in \cite{hitchin,dcthesis}.

It should be clear that in order to describe
a 1-gerbe with connection, one would specify a principal $G$-bundle
with connection on each overlap $U_{\alpha \beta}$.

Morphisms of principal $G$-bundles (of same structure
group, over same space) can be described in local
trivializations as a set of gauge transformations
\cite[section 5.5]{husemoller}
\begin{displaymath}
\phi_{\alpha}: \: U_{\alpha} \: \longrightarrow \: G
\end{displaymath}
one for each element $U_{\alpha}$ of an open cover $\{ U_{\alpha} \}$.
It should be clear that one can describe a map of gerbes
in a similar fashion:  for each $U_{\alpha}$, associate
a principal $G$-bundle $T_{\alpha}$.  This bundle determines
an automorphism of the category $\mbox{Tors}(G)(U_{\alpha})$,
which describes how the gerbes are mapped into one another at
the level of local trivializations.

It should also be clear that a map of gerbes with connection
can be described in terms of a set of principal $G$-bundles
$T_{\alpha}$ with connection.

\subsection{Nonabelian gerbes}

In the physics literature it is sometimes claimed\footnote{For
example, see \cite{dijk1}.} that
certain physical theories have an understanding in terms of
nonabelian gerbes.  In the rest of this paper we have specialized
to abelian gerbes, that is, gerbes with abelian band.

Just as an abelian gerbe can be described in terms of a stack
that locally looks like a stack of principal $G$-bundles for
abelian $G$, it is presumably the case that a nonabelian gerbe
can be described in terms of a stack that locally looks like a stack
of principal $G$-bundles for nonabelian $G$.

Unfortunately, more than this is difficult to say within the
present framework.  Much of our discussion of gerbes has hinged,
either implicitly or explicitly, on the assumption that the band
is a sheaf of abelian groups.  We have not made a thorough study
of how matters would be altered if the band became nonabelian.

Considerably more information on nonabelian gerbes can be found
in \cite{giraud}.  Related material can be found in \cite{dmos}.

\section{Technical notes on stacks}   \label{advstacksec}

In this section we give some highly technical material on stacks.
A reader perusing this paper for the first (or even the second
or third) time is strongly encouraged to skip this section entirely.

In particular, we shall describe sheafification of presheaves of
categories, pullbacks of stacks, and stalks of stacks.
The basics of these topics are outlined in \cite[section 5]{brylinski}.

\subsection{Sheafification}

Given any presheaf of sets, it is possible to construct a sheaf
of sets through a process sometimes called sheafification.
Similarly, given any presheaf of categories, it is possible
to construct a sheaf of categories.  In this section we shall
describe this procedure.

We should warn the reader that this section is extremely technical
in nature.  A reader visiting this material for the first time is
urged to skip ahead to the next section.

First, we shall review sheafification for presheaves of sets,
then we shall describe the process for presheaves of categories.  
In sheafification of presheaves of categories, one defines
a family of ``descent categories,'' then takes a direct limit
of descent categories.  We shall devote subsections to both of
these notions.  Finally, we shall describe how to lift a Cartesian
functor between presheaves of categories to a Cartesian functor
between their sheafifications.

The bulk of our discussion will be based on material
in \cite[section 5.2]{brylinski}.

\subsubsection{Sheafification of presheaves of sets}

First, let us take a moment to review how sheafification works
for presheaves of sets, following \cite[section 5.1]{brylinski}. 
(For a less technical review, see for example \cite[section 0.3]{gh}.) 
Let $F$ be a presheaf of sets on a space $X$,
and pick some open set $U \subseteq X$.  We want to build a sheaf of
sets $\tilde{F}$, i.e., a presheaf of sets $\tilde{F}$ such that
there is a bijective correspondence between elements of $\tilde{F}(U)$
and collections $\{ s_{\alpha} \in F(U_{\alpha}) \}$, $\{ U_{\alpha} \}$
an open cover of $U$, such that $s_{\alpha} |_{U_{\alpha \beta}}
= s_{\beta} |_{U_{\alpha \beta}}$.

The basic idea will be to build $\tilde{F}(U)$ by taking as elements,
collections $\{ s_{\alpha} \in F(U_{\alpha}) \}$ for $\{ U_{\alpha} \}$
an open cover, such that $s_{\alpha} |_{U_{\alpha \beta}}
= s_{\beta} |_{U_{\alpha \beta}}$.  In other words,
we will build a set $\tilde{F}(U)$ in which the bijective correspondence
between elements of the set and local elements satisfying compatibility
conditions is built in from the definition.

More formally, each set $\tilde{F}(U)$ is constructed as follows.
Take $\tilde{F}(U)$ to be the disjoint union over all open covers
$\{ U_{\alpha} \}$ of $U$ of collections $\{ s_{\alpha} \in
F(U_{\alpha}) \}$ such that $s_{\alpha} |_{U_{\alpha \beta}} = 
s_{\beta} |_{U_{\alpha \beta}}$, modulo an equivalence relation
$\sim$.  In other words,
\begin{equation}  \label{sheafifyset1}
\tilde{F}(U) \: = \: \coprod_{ \{ U_{\alpha} \} }
\left\{ \, \{ s_{\alpha} \in F(U_{\alpha}) \} \: | \: s_{\alpha} 
|_{U_{\alpha \beta}}
= s_{\beta} |_{U_{\alpha \beta}} \, \right\} / \sim
\end{equation}

The equivalence relation $\sim$ is defined as follows.
Identify two collections $\{ U_{\alpha}, s_{\alpha} \}$
and $\{ U'_{\alpha}, s'_{\alpha} \}$ if there exists another open cover
$\{ U''_{\alpha} \}$, a refinement of both 
$\{ U_{\alpha} \}$ and $\{ U'_{\alpha} \}$,
such that whenever $U''_{\alpha} \subseteq U_{\beta} \cap U'_{\gamma}$,
we have $s_{\beta} |_{U''_{\alpha}} = s'_{\gamma} |_{U''_{\alpha}}$.


We can rewrite equation~(\ref{sheafifyset1}) in terms of a direct
limit over open covers, as
\begin{displaymath}
\tilde{F}(U) \: = \: \lim_{ \stackrel{ \longrightarrow }{ \{ U_{\alpha} \} } }
\, \left\{ \, \{ s_{\alpha} \in F(U_{\alpha}) \} \: | \: s_{\alpha} |_{U_{\alpha
\beta}} = s_{\beta} |_{U_{\alpha \beta}} \, \right\}
\end{displaymath}
where the partial ordering on open covers $\{ U_{\alpha} \}$ is provided by
the notion of refinement.

Note that if $F$ is a sheaf, not just a presheaf, then $\tilde{F} = F$,
i.e., for all open sets $U$, $\tilde{F}(U) = F(U)$.

\subsubsection{Descent categories}     \label{desccat}

Let ${\cal C}$ denote a presheaf of categories on a space $X$,
and let $U$ denote an open set in $X$.
In order to define the category $\tilde{{\cal C}}(U)$ in the
associated sheaf $\tilde{{\cal C}}$, we shall take a direct limit
over ``descent categories'' defined for the open set $U$.
In this subsection we shall describe descent categories.
In subsequent subsections we shall describe the process of taking
direct limits, and the resulting sheafification.

To each open cover $\{ U_{\alpha} \}$ of $U$,
we shall define a category, which we shall denote by
\begin{displaymath}
\mbox{Desc}_U( {\cal C},
\{ U_{\alpha} \} )
\end{displaymath}
and which is called a descent category.

The objects of $\mbox{Desc}( {\cal C}, \{ U_{\alpha} \} )$ are 
collections $\{ ( x_{\alpha} ), ( \phi_{\alpha \beta} ) \}$,
where each $x_{\alpha} \in \mbox{Ob } {\cal C}(U_{\alpha})$,
and where $\phi_{\alpha \beta}: x_{\beta} |_{U_{\alpha \beta}}
\stackrel{ \sim }{ \longrightarrow } x_{\alpha} |_{U_{\alpha \beta}}$
are isomorphisms,
such that 
(schematically\footnote{The correct statement
is commutativity of diagram~(\ref{transgluobj}).}) 
\begin{displaymath}
\phi_{\alpha \beta} \circ \phi_{\beta \gamma} \: = \:
\phi_{\alpha \gamma}
\end{displaymath}
in ${\cal C}(U_{\alpha
\beta \gamma})$, and $\phi_{\alpha \alpha} = 1_{\alpha}$ in
${\cal C}(U_{\alpha})$.

Morphisms $\{ (x_{\alpha}), ( \phi_{\alpha \beta} ) \}
\rightarrow \{ (y_{\alpha}), ( \phi'_{\alpha \beta} ) \}$
in $\mbox{Desc}({\cal C}, \{ U_{\alpha} \} )$ are
collections of morphisms $\{ f_{\alpha}: x_{\alpha} 
\rightarrow y_{\alpha} \}$ in $\{ {\cal C}(U_{\alpha}) \}$
such that the following diagram commutes:
\begin{equation} \label{descmorph}
\begin{array}{ccc}
x_{\beta} |_{U_{\alpha \beta}} & \stackrel{ \phi_{\alpha \beta} }
{ \longrightarrow } & x_{\alpha} |_{U_{\alpha \beta}} \\
\makebox[0pt][r]{ $\scriptstyle{ f_{\beta} |_{U_{\alpha \beta}} }$ }
\downarrow & & \downarrow
\makebox[0pt][l]{ $\scriptstyle{ f_{\alpha} |_{U_{\alpha \beta}} }$ } \\
y_{\beta} |_{U_{\alpha \beta}} & \stackrel{ \phi'_{\alpha \beta} }
{ \longrightarrow } & y_{\alpha} |_{U_{\alpha \beta}} 
\end{array}
\end{equation}

Note that if ${\cal C}$ is a sheaf of categories, not just a presheaf,
then for any open cover $\{ U_{\alpha} \}$ of $U$, 
the category $\mbox{Desc}( {\cal C}, \{ U_{\alpha} \} )$ is
equivalent to ${\cal C}(U)$.

Suppose that $\{ U'_i \}$ is a refinement of $\{ U_{\alpha} \}$
and $\rho: \{ U'_i \} \rightarrow \{ U_{\alpha} \}$ is the inclusion,
i.e., $\rho$ is a collection of maps $\{ \rho_i: U'_i \hookrightarrow
U_{ \alpha(i) } \}$.  Then $\rho$ induces a functor
\begin{displaymath}
\rho^*:  \mbox{Desc}( {\cal C}, \{ U_{\alpha} \} )
\rightarrow \mbox{Desc}( {\cal C}, \{ U'_i \} )
\end{displaymath}
The action of $\rho^*$ on an object $\{ (x_{\alpha}), (\phi_{\alpha
\beta}) \} \in \mbox{Ob } \mbox{Desc}( {\cal C}, \{ U_{\alpha} \} )$
is given by
\begin{displaymath}
\begin{array}{cl}
\rho^*: & (x_{\alpha}) \mapsto (x_i \equiv x_{\alpha(i)} |_{U'_i} ) \\
\rho^*: & (\phi_{\alpha \beta}) \mapsto
\left( \phi_{ij} \equiv \varphi_{i, ij} \circ 
\varphi^{-1}_{\alpha(i) \alpha(j), ij} \circ
\phi_{\alpha(i) \alpha(j)} |_{U'_{ij}} \circ 
\varphi_{\alpha(i) \alpha(j), ij} \circ
\varphi^{-1}_{j, ij} \right)
\end{array}
\end{displaymath}
where the $\varphi$ are the invertible natural transformations appearing
in the definition of the presheaf of categories ${\cal C}$.
One can verify (after an extremely lengthy diagram chase) that
this functor is well-defined on objects, i.e., that the image under $\rho^*$
of the set of morphisms $(\phi_{\alpha \beta})$ satisfies the compatibility
relation~(\ref{transgluobj}).

We still need to define the action of the proposed functor $\rho^*$
on morphisms.  Given objects 
\begin{displaymath}
\{ (x_{\alpha}, (\phi_{\alpha \beta}) \}, \:
\{ (y_{\alpha}), (\phi'_{\alpha \beta}) \} \: \in \: \mbox{Ob }
\mbox{Desc}({\cal C}, \{ U_{\alpha} \} )
\end{displaymath}
and a morphism
\begin{displaymath}
\{ f_{\alpha}: x_{\alpha} \rightarrow y_{\alpha} \}: \:
\{ (x_{\alpha}), (\phi_{\alpha \beta}) \} \: \longrightarrow \:
\{ (y_{\alpha}), (\phi'_{\alpha \beta}) \}
\end{displaymath}
define
\begin{displaymath}
\rho^* \{ f_{\alpha} \} \: = \: \left\{ f_i \equiv f_{\alpha(i)} |_{U'_i}:
x_{\alpha(i)} |_{U'_i} \rightarrow y_{\alpha(i)} |_{U'_i} \right\}
\end{displaymath}
It can be shown that this map is well-defined, i.e.,
the appropriate version of diagram~(\ref{descmorph}) commutes.

It is now straightforward to check that $\rho^*$ does indeed define
a functor 
\begin{displaymath}
\rho^*: \: \mbox{Desc}( {\cal C}, \{ U_{\alpha} \} ) 
\: \longrightarrow \: \mbox{Desc} ( {\cal C}, \{ U'_i \} )
\end{displaymath}

Now, suppose $\rho_1: \{ U'_i \} \hookrightarrow \{ U_{\alpha} \}$
and $\rho_2: \{ U''_a \} \hookrightarrow \{ U'_i \}$
are a pair of families of inclusions, associated to the refinement
$\{ U'_i \}$ of $\{ U_{\alpha} \}$, and to the refinement
$\{ U''_a \}$ of $\{ U'_i \}$.  Then there exists an invertible
natural transformation $\lambda: (\rho_1 \rho_2)^* \Rightarrow
\rho_2^* \circ \rho_1^*$.

This invertible natural transformation is defined as follows.
To each object 
\begin{displaymath}
\{ (x_{\alpha}), (\phi_{\alpha \beta}) \}
\: \in \: \mbox{Ob } \mbox{Desc}( {\cal C}, \{ U_{\alpha} \})
\end{displaymath}
$\lambda$ associates a morphism
\begin{displaymath}
\begin{array}{cl}
\lambda( \{ (x_{\alpha}), (\phi_{\alpha \beta}) \} ): &
(\rho_1 \rho_2)^* \{ (x_{\alpha}), (\phi_{\alpha \beta}) \} 
\, \rightarrow \, (\rho_2^* \circ \rho_1^*) \{ (x_{\alpha}), (\phi_{\alpha
\beta}) \} 
\end{array}
\end{displaymath}
given by
\begin{displaymath}
\lambda( \{ (x_{\alpha}), (\phi_{\alpha \beta}) \} ) \: = \:
\left\{ \, \varphi_{i(a), a}(x_{\alpha(a)} |_{U''_a}): \, 
x_{\alpha(a)} |_{U''_a} \, \rightarrow \, x_{\alpha(a)} |_{U'_{i(a)}} |_{U''_a}
\, \right\} 
\end{displaymath}
where the $\varphi$ are the invertible natural transformations appearing
in the definition of ${\cal C}$ as a presheaf of categories.
It is straightforward to check that this indeed is a morphism, i.e.,
the appropriate version of diagram~(\ref{descmorph}) commutes,
and moreover $\lambda$ is an invertible natural transformation.

Finally, it is straightforward to check that if we are given three
composable inclusions of open covers $\rho_1$, $\rho_2$, $\rho_3$,
then the following diagram (corresponding to diagram~(\ref{trans}))
of natural transformations commutes:
\begin{displaymath}
\begin{array}{ccc}
(\rho_1 \rho_2 \rho_3)^* & \Longrightarrow & \rho_3^* \circ (\rho_1 \rho_2)^* \\
\Downarrow & & \Downarrow \\
(\rho_2 \rho_3)^* \circ \rho_1^* & \Longrightarrow & \rho_3^* \circ
\rho_2^* \circ \rho_1^*
\end{array}
\end{displaymath}

So far we have discussed functors associated to sets of inclusion maps
between two open covers of a fixed open set $U$.
Other functors can also be constructed, as we shall now discuss.
Suppose $\rho: V \hookrightarrow U$ is an inclusion of open sets.
Then, for any open cover $\{ U_{\alpha} \}$ of $U$,
$\rho$ induces a functor $\rho_{V, \{ U_{\alpha} \} }^*$ between 
descent categories associated
to $U$ and $V$:
\begin{displaymath}
\rho_{V, \{ U_{\alpha} \} }^*: 
\: \mbox{Desc}_U( {\cal C}, \{ U_{\alpha} \} ) \: \rightarrow \:
\mbox{Desc}_V( {\cal C}, \{ U_{\alpha}|_V \} )
\end{displaymath}
where we have used the fact that $\{ U_{\alpha} |_V \}$ is an open
cover of $V$.  This functor can be defined in precise analogy with
the previously-described functor between descent categories between
open covers of a single open set $U$; we shall not repeat the details,
as they are virtually identical.  Similarly, given two inclusions
$\rho_2: U_3 \hookrightarrow U_2$ and $\rho_1: U_2 \hookrightarrow U_1$,
one can define (as above) an invertible natural transformation
$\kappa_{12}: (\rho_1 \rho_2)_{U_3, \{ U_{\alpha} \} }^*
\Rightarrow \rho_{2 \: U_3, \{ U_{\alpha} |_{U_2} \} }^* \circ
\rho_{1 \: U_2, \{ U_{\alpha} \} }^*$, such that analogues
of diagram~(\ref{trans}) commute. 

Finally, we can also construct natural transformations between
compositions of the two types of functors listed above.
Let $\rho_{12}: \{ U^2_{\alpha} \} \rightarrow \{ U^1_{\alpha} \}$
be a set of inclusions between covers of an open set $U \subseteq X$,
and let $\rho: V \hookrightarrow U$ be an inclusion map between
open sets, inducing functors $\rho_{V, \{ U^1_{\alpha} \} }^*$
and $\rho_{V, \{ U^2_{\alpha} \} }^*$.
Let $\rho_{V, 12}: \{ U^2_{\alpha} |_V \} \rightarrow \{
U^1_{\alpha} |_V \}$ be the set of inclusion maps between covers
of $V$ that makes the following diagram commute:
\begin{equation}   
\begin{array}{ccc}
\{ U^1_{\alpha} \} & \stackrel{ \rho_{V, \{ U^1_{\alpha} \}} }
{ \longleftarrow } & \{ U^1_{\alpha} |_V \} \\
\makebox[0pt][r]{ $\scriptstyle{ \rho_{12} }$ } \uparrow & &
\uparrow \makebox[0pt][l]{ $\scriptstyle{ \rho_{V, 12} }$ } \\
\{ U^2_{\alpha} \} & \stackrel{ \rho_{V, \{ U^2_{\alpha} \} } }
{ \longleftarrow } & \{ U^2_{\alpha} |_V \}
\end{array}
\end{equation}
Then we can define the following two invertible natural transformations:
\begin{displaymath}
\begin{array}{cl}
\delta_{V, 12}: & ( \rho_{12} \circ \rho_{ V, \{ U^2_{\alpha} \} } )^*
\: \Rightarrow \: \rho_{V, \{ U^2_{\alpha} \} }^* \circ \rho_{12}^* \\
\delta_{12, V}: & ( \rho_{V, \{ U^1_{\alpha} \} } \circ \rho_{V, 12} )^*
\: \Rightarrow \: \rho_{V, 12}^* \circ \rho_{V, \{ U^1_{\alpha} \} }^*
\end{array}
\end{displaymath}
These natural transformations also make analogues of diagram~(\ref{trans})
commute.

\subsubsection{Direct limits}

Given our presheaf of categories ${\cal C}$ and for any open $U$,
the descent categories $\mbox{Desc}({\cal C}, \{ U_{\alpha} \})$,
we can now construct the associated sheaf of categories.

The sheaf of categories $\tilde{{\cal C}}$ associated to the
presheaf of categories ${\cal C}$ is defined by
\begin{equation}   \label{sheafifydef}
\tilde{{\cal C}}(U) \: = \: \lim_{ \stackrel{ \longrightarrow }
{ \{ U_{\alpha} \} }
} \mbox{Desc}( {\cal C}, \{ U_{\alpha} \})
\end{equation}
In words, $\tilde{{\cal C}}(U)$ is defined to be the direct limit
over open covers of $U$ of descent categories.

We shall now give the definition of the direct limit used above.

The objects of the direct limit~(\ref{sheafifydef}) are the disjoint
union of the objects of all the descent categories associated to
open covers of $U$.  In other words,
\begin{equation}   \label{shifyobdef}
\mbox{Ob } \lim_{ \stackrel{ \longrightarrow }{ \{ U_{\alpha} \} } }
\mbox{Desc}( {\cal C}, \{ U_{\alpha} \} ) \: = \:
\coprod_{ \{ U_{\alpha} \} } \mbox{Ob } \mbox{Desc}( {\cal C},
\{ U_{\alpha} \} )
\end{equation}

Now, we shall describe morphisms of the direct limit category.
Let $P_1$, $P_2$ be objects:
\begin{eqnarray*}
P_1 & \in &
\mbox{Ob } \mbox{Desc}({\cal C}, \{ U^1_{\alpha} \} ) \\
P_2 & \in & \mbox{Ob }
\mbox{Desc}( {\cal C}, \{ U^2_{\alpha} \} )
\end{eqnarray*}
Let $\{ U^3_{\alpha} \}$ be any (open cover) refinement of both
$\{ U^1_{\alpha} \}$ and $\{ U^2_{\alpha} \}$, let $\rho_{13}: \{
U^3_{\alpha} \} \rightarrow \{ U^1_{\alpha} \}$
and $\rho_{23}: \{ U^3_{\alpha} \} \rightarrow \{ U^2_{\alpha} \}$ 
be the two sets of inclusions, and  
define $S_{ \{ U^3_{\alpha} \} }$ to be the set of all morphisms
$\beta: \rho_{13}^* P_1 \rightarrow \rho_{23}^* P_2$ in
$\mbox{Desc}( {\cal C}, \{ U^3_{\alpha} \})$.

Define the set of all morphisms $\mbox{Hom}(P_1, P_2)$ to be
the disjoint union of all the $S_{ \{ U'_{\alpha} \} }$
(for $\{ U'_{\alpha} \}$ an (open cover) refinement of 
both $\{ U^1_{\alpha} \}$ and $\{ U^2_{\alpha} \}$),
modulo an equivalence relation $\sim$ to be defined momentarily.
In other words,
\begin{displaymath}
\mbox{Hom}( P_1, P_2 ) \: = \: \coprod_{ \{ U'_{\alpha} \} }
S_{ \{ U'_{\alpha} \} } / \sim
\end{displaymath}

The equivalence relation $\sim$ is defined as follows.
If $\{ U_{\alpha} \}$ and $\{ U'_{\alpha} \}$ are two open covers
which both refine both $\{ U^1_{\alpha} \}$ and $\{ U^2_{\alpha} \}$,
and 
\begin{displaymath}
\begin{array}{cl}
\rho_i: & \{ U_{\alpha} \} \rightarrow \{ U^i_{\alpha} \} \\
\rho'_i: & \{ U'_{\alpha} \} \rightarrow \{ U^i_{\alpha} \} 
\end{array}
\end{displaymath}
are the sets of inclusion maps ($i \in \{1, 2 \}$), then we say
$\beta: \rho_1^* P_1 \rightarrow \rho_2^* P_2$ in $S_{ \{ U_{\alpha} \} }$
is equivalent to $\beta': \rho'^*_1 P_1 \rightarrow \rho'^*_2 P_2$
in $S_{ \{ U'_{\alpha} \} }$ if and only if there exists an (open
cover) refinement $\{ U''_{\alpha} \}$ of both $\{ U_{\alpha} \}$
and $\{ U'_{\alpha} \}$, with sets of inclusion maps
\begin{displaymath}
\begin{array}{cl}
\gamma: & \{ U''_{\alpha} \} \rightarrow \{ U_{\alpha} \} \\
\gamma': & \{ U''_{\alpha} \} \rightarrow \{ U'_{\alpha} \}
\end{array}
\end{displaymath}
such that 
\begin{enumerate}
\item $\rho_i \gamma = \rho'_i \gamma'$ for $i \in \{ 1, 2 \}$,
\item the following diagram commutes:
\begin{equation}   \label{dodescmor}
\begin{array}{ccccc}
(\rho_1 \gamma)^* (P_1) & \stackrel{ \lambda_{1} }{ \longrightarrow } &
\gamma^* \circ \rho_1^* (P_1) & \stackrel{ \gamma^*(\beta) }
{ \longrightarrow } &
\gamma^* \circ \rho_2^* (P_2) \\
\makebox[0pt][r]{ $\scriptstyle{ \lambda'_{1} }$ } \downarrow &
& & & \uparrow \makebox[0pt][l]{ $\scriptstyle{ \lambda_2 }$ } \\
\gamma'^* \circ \rho'^*_1 (P_1) & \stackrel{ \gamma'^*(\beta') }
{ \longrightarrow } &
\gamma'^* \circ \rho'^*_2 (P_2) & \stackrel{ \lambda'_2 }{ \longleftarrow }
& ( \rho_2 \gamma )^* (P_2)
\end{array}
\end{equation}
\end{enumerate}
where the $\lambda$ are the natural transformations defined in
the previous subsection.

We have defined objects and morphisms in the categories $\tilde{{\cal C}}(U)$.
We shall now take a moment to discuss composition of morphisms in this 
category, as the correct definition might not be completely obvious to
the reader.  Let $P_1, P_2, P_3 \in \mbox{Ob } \tilde{{\cal C}}(U)$,
i.e., $P_i \in \mbox{Ob } \mbox{Desc}_U({\cal C}, \{ U^i_{\alpha} \})$
for some open covers $\{ U^i_{\alpha} \}$ of $U$, $i \in \{ 1, 2, 3 \}$.
Let $\beta \in \mbox{Hom}_{ \tilde{{\cal C}}(U) }(P_1, P_2)$,
$\alpha \in \mbox{Hom}_{ \tilde{{\cal C}}(U) }(P_2, P_3)$.
In other words, there exist refinements $\{ U^4_{\alpha} \}$ of
$\{ U^1_{\alpha} \}$ and $\{ U^2_{\alpha} \}$,
and $\{ U^5_{\alpha} \}$ of $\{ U^2_{\alpha} \}$ and $\{ U^3_{\alpha} \}$,
such that
\begin{eqnarray*}
\beta & \in & \mbox{Hom}_{ {\it Desc}({\cal C}, \{ U^4_{\alpha} \}) }
\left( \rho_{14}^* P_1, \rho_{24}^* P_2 \right) \\
\alpha & \in & \mbox{Hom}_{ {\it Desc}( {\cal C}, \{ U^5_{\alpha} \} }
\left( \rho_{25}^* P_2, \rho_{35}^* P_3 \right)
\end{eqnarray*}
We define the composition $\alpha \circ \beta$ as follows.
Let $\{ U^6_{\alpha} \}$ be a refinement of both $\{ U^4_{\alpha} \}$
and $\{ U^5_{\alpha} \}$, such that $\rho_{24} \rho_{46} = \rho_{25}
\rho_{56}$.  Define $\alpha \circ \beta$ to be
\begin{eqnarray*}
\alpha \circ \beta & \equiv & \lambda_{356}^{-1} \circ \rho_{56}^* \alpha
\circ \lambda_{256} \circ \lambda_{246}^{-1} \circ \rho_{46}^* \beta
\circ \lambda_{146} \\
& \in & \mbox{Hom}_{ {\it Desc}({\cal C}, \{ U^6_{\alpha} \} ) }
\left( \rho_{16}^* P_1, \rho_{36}^* P_3 \right)
\end{eqnarray*}
where the $\lambda$ are the natural transformations defined in the previous 
subsection.  It is straightforward to check that this composition is
well-defined, i.e., $\alpha \sim \alpha'$ and $\beta \sim \beta'$
implies $\alpha \circ \beta \sim \alpha' \circ \beta'$.

So far we have discussed how to construct the categories $\tilde{{\cal C}}(U)$
appearing in the sheaf associated to the presheaf of categories ${\cal C}$.
To completely define the sheaf we must also specify restriction functors
and natural transformations, which we shall now do.

Let $\rho: V \hookrightarrow U$ be an inclusion of open sets.
We shall now show that $\rho$ induces a functor
\begin{displaymath}
\rho^*: \: \tilde{{\cal C}}(U) = 
\lim_{ \stackrel{ \longrightarrow }{ \{ U_{\alpha} \} } }
\mbox{Desc}_U ( {\cal C}, \{ U_{\alpha} \} ) \: \longrightarrow \:
\tilde{{\cal C}}(V) = \lim_{ \stackrel{ \longrightarrow }{ \{ U'_{\alpha} \} } }
\mbox{Desc}_V ( {\cal C}, \{ U'_{\alpha} \} )
\end{displaymath}

First we shall describe how $\rho^*$ acts on objects.
Recall that an object of 
$\tilde{{\cal C}}(U)$
is an object, call it $P$, of $\mbox{Desc}_U( {\cal C},
\{ U_{\alpha} \} )$ for some open cover $\{ U_{\alpha} \}$ of $U$.
The functor $\rho^*$ acts on the object $P$ as,
\begin{displaymath}
P \: \mapsto \: \rho_{V, \{ U_{\alpha} \} }^* (P) \: \in \:
\mbox{Ob } \mbox{Desc}_V( {\cal C}, \{ U_{\alpha} |_V \} )
\end{displaymath}

Now we shall describe how $\rho^*$ acts on morphisms.
Let $P_1, P_2 \in \mbox{Ob }
\tilde{{\cal C}}(U)$.
In other words,
\begin{eqnarray*}
P_1 \: \in \: \mbox{Ob } \mbox{Desc}_U( {\cal C}, \{ U^1_{\alpha} \} ) \\
P_2 \: \in \: \mbox{Ob } \mbox{Desc}_U( {\cal C}, \{ U^2_{\alpha} \} )
\end{eqnarray*}
for some open covers $\{ U^1_{\alpha} \}$, $\{ U^2_{\alpha} \}$ of
$U$.  Recall
\begin{displaymath}
\mbox{Hom}(P_1, P_2) \: = \: \coprod_{ \{ U^3_{\alpha} \} } \, 
\mbox{Hom}( \rho_{13}^* P_1, \rho_{23}^* P_2 ) / \sim
\end{displaymath}
Let $\beta \in \mbox{Hom}( \rho_{13}^* P_1, \rho_{23}^* P_2 )$ for some
refinement $\{ U^3_{\alpha} \}$ of $\{ U^1_{\alpha} \}$ and
$\{ U^2_{\alpha} \}$.
Define $\rho^*(\beta)$ as,
\begin{displaymath}
\rho^*(\beta) \: = \: \delta_{23, V} \circ \delta_{V, 23}^{-1} \circ
\rho_{V, \{ U^3_{\alpha} \} }^* (\beta) \circ \delta_{V, 13} \circ
\delta_{13, V}^{-1}
\end{displaymath}
so that 
\begin{displaymath}
\rho^*(\beta) \in \mbox{Hom}_{ \mbox{Desc}_V( {\cal C},
\{ U^3_{\alpha} |_V \} ) } \left( 
\rho_{V, 13 }^* \circ \rho_{V, \{ U^1_{\alpha} \} }^* (P_1),
\rho_{V, 23 }^* \circ \rho_{V, \{ U^2_{\alpha} \} }^* (P_2) \right)
\end{displaymath}
It is straightforward to check that this functor is well-defined -- 
for example, $\beta \sim \beta'$ implies $\rho^*(\beta) \sim \rho^*(\beta')$.

Let $\rho_V: V \hookrightarrow U$ and $\rho_W: W \hookrightarrow V$
be inclusions between open sets.  It is easy to check that
the natural transformations $\kappa$ defined in the previous
subsection give a set of invertible natural transformations 
$\tilde{\varphi}_{VW}: (\rho_V \rho_W)^*
\Rightarrow \rho_W^* \circ \rho_V^*$ between the restriction
functors acting on the direct limit categories.
Moreover, these natural transformations make analogues of
diagram~(\ref{trans}) commute.  (Note that we have chosen to denote
these natural transformations by $\tilde{\varphi}$ instead of $\kappa$,
in keeping with our general tendency to denote sheafified objects
with a tilde.)

We have now given $\tilde{{\cal C}}$ the structure of a presheaf
of categories -- we have associated categories (direct limits of
descent categories) to each open set, and given restriction functors
and appropriate natural transformations.

Furthermore, it can be shown that $\tilde{{\cal C}}$ is a sheaf
of categories, not just a presheaf.  

Note that if ${\cal C}$ is a sheaf of categories, not just a 
presheaf, then $\tilde{{\cal C}}$ is equivalent to ${\cal C}$ as
a stack.  In other words, if you sheafify a sheaf of categories,
then you recover the original sheaf.

\subsubsection{Lifts of Cartesian functors}

Suppose $\Phi: {\cal C} \rightarrow {\cal D}$ is a Cartesian
functor between presheaves of categories ${\cal C}$, ${\cal D}$
on a space $X$.
In this section we shall show that $\Phi$ lifts to a
Cartesian functor $\tilde{\Phi}: \tilde{{\cal C}} \rightarrow
{{\cal D}}$ between the sheafifications of ${\cal C}$ and
${\cal D}$.  

Let $U$ be an open set, and $\{ U_{\alpha} \}$ an open cover of $U$.
We shall first define a functor
\begin{displaymath}
\Phi(U, \{ U_{\alpha} \}): \: \mbox{Desc}_U( {\cal C}, \{ U_{\alpha} \} )
\: \longrightarrow \: \mbox{Desc}_U( {\cal D}, \{ U_{\alpha} \})
\end{displaymath}

We shall define the functor $\Phi(U, \{ U_{\alpha} \})$ on
objects as follows.  Let $\{ (x_{\alpha}), (\phi_{\alpha \beta}) \} \in
\mbox{Ob } \mbox{Desc}_U( {\cal C}, \{ U_{\alpha} \} )$.
Define the action of $\Phi(U, \{ U_{\alpha} \} )$ by,
\begin{eqnarray*}
x_{\alpha} & \mapsto & \Phi(U_{\alpha})(x_{\alpha}) \\
\phi_{\alpha \beta} & \mapsto & \chi_{\alpha, \alpha \beta}^{-1}
\circ \Phi(U_{\alpha \beta})( \phi_{\alpha \beta}) \circ
\chi_{\beta, \alpha \beta} \\
& & \in \mbox{Hom}_{{\cal D}(U_{\alpha \beta})}\left( \Phi(U_{\beta})
(x_{\beta}) |_{U_{\alpha \beta}} , \Phi(U_{\alpha})(x_{\alpha}) 
|_{U_{\alpha \beta}} \right)
\end{eqnarray*}
where the $\chi$ are the natural transformations appearing in the
definition of $\Phi$ as a Cartesian functor.
It is straightforward to check that this action is well-defined,
i.e., that the images of the $\phi_{\alpha \beta}$ make
the analogue of diagram~(\ref{transgluobj}) commute.

Now we shall define the functor $\Phi(U, \{ U_{\alpha} \})$ on
morphisms.  Given objects
\begin{displaymath}
\{ (x_{\alpha}), (\phi_{\alpha \beta}) \}, \:
\{ (y_{\alpha}), (\phi'_{\alpha \beta}) \} \: \in \:
\mbox{Desc}_U( {\cal C}, \{ U_{\alpha} \} )
\end{displaymath}
and a morphism
\begin{displaymath}
\{ f_{\alpha}: x_{\alpha} \rightarrow y_{\alpha} \}: \:
\{ (x_{\alpha}), (\phi_{\alpha \beta}) \} \: \longrightarrow \:
\{ (y_{\alpha}), (\phi'_{\alpha \beta}) \}
\end{displaymath}
between these two objects,
we define the action of $\Phi(U, \{ U_{\alpha} \})$ on this
morphism as,
\begin{displaymath}
f_{\alpha} \: \mapsto \: \Phi(U_{\alpha})(f_{\alpha})
\end{displaymath}
It is straightforward to check that this map is well-defined,
and moreover that
\begin{displaymath}
\Phi(U, \{ U_{\alpha} \} ): \: \mbox{Desc}_U( {\cal C}, \{ U_{\alpha} \} )
\: \longrightarrow \: \mbox{Desc}_U( {\cal D}, \{ U_{\alpha} \} )
\end{displaymath}
is a well-defined functor.

In passing, note that if $\Lambda: {\cal D} \rightarrow {\cal E}$ is
another Cartesian functor between presheaves of categories ${\cal D}$,
${\cal E}$ on $X$, then we have the relation
\begin{displaymath}
(\Lambda \Phi)(U, \{ U_{\alpha} \} ) \: = \:
\Lambda( U, \{ U_{\alpha} \} ) \circ \Phi( U, \{ U_{\alpha} \} )
\end{displaymath}

Now, let $\{ U^2_i \}$ be a refinement of $\{ U^1_{\alpha} \}$,
both open covers of the open set $U$, and let $\rho: 
\{ U^2_i \} \rightarrow \{ U^1_{\alpha} \}$ be the set of inclusions.
We shall define an invertible natural transformation
\begin{displaymath}
\Xi_{\rho}: \: \rho^*_{{\cal D}} \circ \Phi( U, \{ U^1_{\alpha} \})
\: \Longrightarrow \: \Phi(U, \{U^2_i\} ) \circ \rho^*_{{\cal C}}
\end{displaymath}
We define this natural transformation as follows.
To each object
\begin{displaymath}
\{ (x_{\alpha}), (\phi_{\alpha \beta}) \} \: \in \: \mbox{Ob }
\mbox{Desc}_U({\cal C}, \{ U^1_{\alpha} \})
\end{displaymath}
the natural transformation $\Xi_{\rho}$ associates
the morphism
\begin{displaymath}
\left\{ \: \chi_{\alpha(i), i}( x_{\alpha(i)} ):
\: \Phi( U^1_{\alpha(i)} )( x_{\alpha(i)} ) |_{U^2_i}
\: \longrightarrow \: 
\Phi( U^2_i )( x_{\alpha(i)} |_{U^2_i} ) \: \right\}
\end{displaymath}
where the $\chi$ are the natural transformations defining $\Phi$
as a Cartesian functor.
It is straightforward to check that this is a well-defined morphism,
and moreover it defines a natural transformation, in that it commutes
with morphisms $\{ (x_{\alpha}), (\phi_{\alpha \beta}) \}
\rightarrow \{ (y_{\alpha}), (\phi_{\alpha \beta}) \}$
between objects in $\mbox{Desc}_U( {\cal C}, \{ U^1_{\alpha} \})$.

If $\rho_1: \{ U^2_i \} \rightarrow \{ U^1_{\alpha} \}$
and $\rho_2: \{ U^3_a \} \rightarrow \{ U^2_i \}$ are two sets of
inclusions from refinements of open covers of $U$,
then the following diagram (closely analogous to diagram~(\ref{cartdef}))
commutes:
\begin{equation}  \label{Xiprop}
\begin{array}{ccccc}
\rho_{2 {\cal D}}^* \circ \rho_{1 {\cal D}}^* \circ
\Phi(U, \{ U^1_{\alpha} \} ) & \stackrel{ \Xi_1 }{ \Longrightarrow }
& \rho_{2 {\cal D}}^* \circ \Phi(U, \{ U^2_i \} ) \circ \rho_{1 {\cal C}}^*
& \stackrel{ \Xi_2 }{ \Longrightarrow } &
\Phi(U, \{ U^3_a \}) \circ \rho_{2 {\cal C}}^* \circ \rho_{1 {\cal C}}^* \\
\makebox[0pt][r]{ $\scriptstyle{ \lambda^{{\cal D}}_{12} }$ } \Uparrow &
& & & \Uparrow \makebox[0pt][l]{ $\scriptstyle{ \lambda^{{\cal C}}_{12} }$ } \\
(\rho_1 \rho_2)_{{\cal D}}^* \circ \Phi(U, \{ U^1_{\alpha} \}) &
& \stackrel{ \Xi_{12} }{ \Longrightarrow } & &
\Phi(U, \{ U^3_a \} ) \circ (\rho_1 \rho_2)^*_{{\cal C}}
\end{array}
\end{equation}
where the $\lambda$ are the natural transformations defined in the section
on descent categories.

We are finally ready to start defining a Cartesian functor
$\tilde{\Phi}: \tilde{{\cal C}} \rightarrow \tilde{{\cal D}}$.
Let $U \subseteq X$ be an open set, and define a functor
$\tilde{\Phi}(U): \tilde{{\cal C}}(U) \rightarrow \tilde{{\cal D}}(U)$
as follows.

Let $P$ be an object of $\tilde{{\cal C}}(U)$, that is,
$P \in \mbox{Ob } \mbox{Desc}_U({\cal C}, \{ U^1_{\alpha} \})$ for
some open cover $\{ U^1_{\alpha} \}$ of $U$.
Define 
\begin{displaymath}
\tilde{\Phi}(U)(P) \: \equiv \: \Phi(U, \{U^1_{\alpha} \})(P)
\: \in \: \mbox{Ob } \mbox{Desc}_U( {\cal D}, \{ U^1_{\alpha} \} )
\: \subseteq \: \mbox{Ob } \tilde{{\cal D}}(U)
\end{displaymath}

Let $P_1$ and $P_2$ be objects of $\tilde{{\cal C}}(U)$, meaning
\begin{displaymath}
P_i \: \in \: \mbox{Ob } \mbox{Desc}_U( {\cal C}, \{ U^i_{\alpha} \})
\end{displaymath}
for open covers $\{ U^i_{\alpha} \}$ of $U$, and $i \in \{ 1, 2 \}$.
Let $\beta \in \mbox{Hom}_{ \tilde{{\cal C}}(U) }( P_1, P_2 )$,
which means that for some refinement $\{ U^3_{\alpha} \}$ of both
$\{ U^1_{\alpha} \}$ and $\{ U^2_{\alpha} \}$,
\begin{displaymath}
\beta \: \in \: \mbox{Hom}_{ {\it Desc}( {\cal C}, \{ U^3_{\alpha} \} ) }
\left( \rho_{13}^* P_1, \rho_{23}^* P_2 \right)
\end{displaymath}
Then define 
\begin{eqnarray*}
\tilde{\Phi}(U)(\beta) & \equiv & \Xi_{23}^{-1} \circ
\Phi(U, \{ U^3_{\alpha} \} )(\beta) \circ \Xi_{13} \\
& \in & \mbox{Hom}_{ {\it Desc}({\cal D}, \{ U^3_{\alpha} \}) }
\left( \rho_{13}^* \Phi(U, \{ U^1_{\alpha} \})(P_1) ,
\rho_{23}^* \Phi(U, \{ U^2_{\alpha} \})(P_2) \right)
\end{eqnarray*}

With the definitions above, it can be shown that 
\begin{displaymath}
\tilde{\Phi}(U): \: \tilde{{\cal C}}(U) \: \longrightarrow \:
\tilde{{\cal D}}(U)
\end{displaymath}
is a well-defined functor between the categories $\tilde{{\cal C}}(U)$
and $\tilde{{\cal D}}(U)$.

In order to define a Cartesian functor $\tilde{\Phi}: \tilde{{\cal C}}
\rightarrow \tilde{{\cal D}}$, we need to specify
more than the functors $\tilde{\Phi}(U)$.  Specifically,
for any inclusion $\rho: V \hookrightarrow U$, we need to specify
an invertible natural transformation
\begin{displaymath}
\tilde{\chi}_{\rho}: \: \rho_{ \tilde{{\cal D}} }^* \circ
\tilde{\Phi}(U) \: \Longrightarrow \: \tilde{\Phi}(V) \circ
\rho_{ \tilde{{\cal C}} }^*
\end{displaymath}
satisfying certain identities.
This natural transformation is defined as follows.
To each object
\begin{displaymath}
\{ (x_{\alpha}), (\phi_{\alpha \beta}) \} \: \in \: \mbox{Ob }
\mbox{Desc}_U( {\cal C}, \{ U_{\alpha} \}) \: \subseteq \:
\tilde{{\cal C}}(U)
\end{displaymath}
(defined with respect to some open cover $\{ U_{\alpha} \}$ of $U$)
the natural transformation $\tilde{\chi}_{\rho}$ assigns the
following morphism:
\begin{displaymath}
\tilde{\chi}_{\rho} \left( \{ (x_{\alpha}), ( \phi_{\alpha \beta} ) \} 
\right) \: \equiv \: \left\{ \, \chi_{ U_{\alpha}, U_{\alpha} \cap V }
( x_{\alpha} ): \: \Phi(U_{\alpha})(x_{\alpha}) |_{ U_{\alpha} \cap V }
\: \longrightarrow \: \Phi( U_{\alpha} \cap V )( x_{\alpha}
|_{U_{\alpha} \cap V} ) \, \right\} 
\end{displaymath}
where the $\chi$ are the invertible natural transformations defining
$\Phi$ as a Cartesian functor.

It is straightforward to check that the definition given above for 
$\tilde{\chi}_{\rho}$ does in fact yield a well-defined natural
transformation.  Moreover, $\tilde{ \chi }_{\rho}$ satisfies
the usual pentagonal identity~(\ref{cartdef}).  Specifically, if $\rho_1: V
\hookrightarrow U$ and $\rho_2: W \hookrightarrow V$ are inclusions
of open sets, then the following diagram commutes:
\begin{equation}
\begin{array}{ccccc}
\rho_{2 \tilde{{\cal D}} }^* \circ \rho_{1 \tilde{{\cal D}} }^* \circ
\tilde{\Phi}(U) & \stackrel{ \tilde{\chi}_1 }{ \Longrightarrow } &
\rho_{2 \tilde{{\cal D}} }^* \circ \tilde{\Phi}(V) \circ
\rho_{1 \tilde{{\cal C}} }^* & \stackrel{ \tilde{\chi}_2 }{ \Longrightarrow }
& \tilde{ \Phi }(W) \circ \rho_{2 \tilde{{\cal C}} }^* \circ
\rho_{ 1 \tilde{{\cal C}} }^* \\
\makebox[0pt][r]{ $\scriptstyle{ \tilde{\varphi}_{12}^{ \tilde{{\cal D}} } }$ }
\Uparrow & & & & \Uparrow \makebox[0pt][l]{ $\scriptstyle{
\tilde{\varphi}_{12}^{ \tilde{{\cal C}} } }$ } \\
( \rho_1 \rho_2 )^*_{ \tilde{{\cal D}} } \circ \tilde{\Phi}(U) &
& \stackrel{ \tilde{\chi}_{12} }{ \Longrightarrow } & &
\tilde{\Phi}(W) \circ ( \rho_1 \rho_2 )^*_{ \tilde{{\cal C}} }
\end{array}
\end{equation}
where the $\tilde{\varphi}$ are the natural transformations 
defined in the previous subsection.

Thus, we have now defined a Cartesian functor $\tilde{\Phi}:
\tilde{{\cal C}} \rightarrow \tilde{{\cal D}}$.
Put another way, given a Cartesian functor $\Phi: {\cal C}
\rightarrow {\cal D}$ between two presheaves of categories
${\cal C}$, ${\cal D}$, we have now constructed a lift of $\Phi$
to a Cartesian functor between the sheafifications
$\tilde{{\cal C}}$ and $\tilde{{\cal D}}$.

In passing, we shall mention that it is straightforward to check that
if $\Lambda: {\cal D} \rightarrow {\cal E}$ is another Cartesian
functor between presheaves of categories ${\cal D}$, ${\cal E}$,
then the lift is compatible with composition -- in other words,
\begin{displaymath}
\widetilde{ ( \Lambda \circ \Phi ) } \: = \: \tilde{\Lambda} \circ
\tilde{\Phi}
\end{displaymath}

\subsubsection{Lifts of 2-arrows}
 
In this section, we shall demonstrate that 2-arrows can also be lifted to
sheafifications.  Let $\Phi_1, \Phi_2: {\cal C} \rightarrow {\cal D}$
be a pair of Cartesian functors between presheaves of categories
${\cal C}$, ${\cal D}$, and let $\psi: \Phi_1 \Rightarrow \Phi_2$
be a 2-arrow.  We shall now define a lift $\tilde{\psi}:
\tilde{\Phi}_1 \Rightarrow \tilde{\Phi}_2$ of the 2-arrow $\psi$
to a 2-arrow between the sheafifications of the Cartesian functors.

First, for any open set $U$ and any open cover $\{ U_{\alpha} \}$ of
$U$, we shall define a natural transformation
\begin{displaymath}
\psi(U, \{ U_{\alpha} \}): \: \Phi_1(U, \{ U_{\alpha} \}) \: 
\Longrightarrow \: \Phi_2(U, \{ U_{\alpha} \})
\end{displaymath}
between functors from $\mbox{Desc}_U({\cal C}, \{ U_{\alpha} \})$
to $\mbox{Desc}_U({\cal D}, \{ U_{\alpha} \})$, as follows.
Let $\{ (x_{\alpha} ), ( \phi_{\alpha \beta} ) \}$ be an object
in $\mbox{Desc}_U({\cal C}, \{ U_{\alpha} \})$.
Define a morphism 
\begin{displaymath}
\psi(U, \{ U_{\alpha} \})\left( \{ ( x_{\alpha} ), ( \phi_{\alpha \beta} ) \} 
\right): \: \Phi_1(U, \{ U_{\alpha} \}) \left( \{ ( x_{\alpha} ), ( \phi_{\alpha \beta} ) \} \right) \: \longrightarrow \:
\Phi_2(U, \{ U_{\alpha} \}) \left( \{ ( x_{\alpha} ), 
( \phi_{\alpha \beta} ) \} \right) 
\end{displaymath}
by,
\begin{displaymath}
\psi(U, \{ U_{\alpha} \}) \left( \{ ( x_{\alpha} ), ( \phi_{\alpha \beta} ) \}
\right) \: \equiv \: \left\{ \, \psi(U_{\alpha})(x_{\alpha}): \: 
\Phi_1(U_{\alpha})(x_{\alpha}) \: \longrightarrow \: 
\Phi_2(U_{\alpha})(x_{\alpha}) \, \right\}
\end{displaymath}
It is easy to check that this is a well-defined morphism in the
category $\mbox{Desc}_U({\cal D}, \{ U_{\alpha} \})$, and moreover
that this defines a natural transformation
$\Phi_1(U, \{ U_{\alpha} \}) \Rightarrow \Phi_2(U, \{ U_{\alpha} \})$.

We should mention that if $\{ U^1_{\alpha} \}$ and $\{ U^2_{\alpha} \}$
are both open covers of $U$, with $\{ U^2_{\alpha} \}$ a refinement
of $\{ U^1_{\alpha} \}$ and $\rho: \{ U^2_{\alpha} \} \rightarrow
\{ U^1_{\alpha} \}$ the set of inclusion maps, then the following
diagram commutes:
\begin{equation}
\begin{array}{ccc}
\rho^* \circ \Phi_1(U, \{ U^1_{\alpha} \}) & \stackrel{ \Xi_{12} }{
\Longrightarrow } & \Phi_1(U, \{ U^2_{\alpha} \}) \circ \rho^* \\
\makebox[0pt][r]{ $\scriptstyle{ \psi(U, \{ U^1_{\alpha} \}) }$ }
\Downarrow & & \Downarrow
\makebox[0pt][l]{ $\scriptstyle{ \psi(U, \{ U^2_{\alpha} \}) }$ } \\
\rho^* \circ \Phi_2(U, \{ U^1_{\alpha} \}) & 
\stackrel{ \Xi_{12} }{ \Longrightarrow } & \Phi_2(U, \{ U^2_{\alpha} \})
\circ \rho^*
\end{array}
\end{equation}
where $\Xi$ is the natural transformation defined earlier, relating
$\Phi(U, \{U_{\alpha} \})$ defined with respect to distinct open covers.

Now, we shall define the 2-arrow
$\tilde{\psi}: \tilde{\Phi}_1 \Rightarrow \tilde{\Phi}_2$,
lifting $\psi$ to the sheafifications.
For each open set $U$, let $P \in \mbox{Ob } \tilde{ {\cal C}}(U)$,
that is, $P \in \mbox{Ob } \mbox{Desc}_U({\cal C}, \{ U_{\alpha} \})$
for some open cover $\{ U_{\alpha} \}$ of $U$.
Define a morphism
\begin{displaymath}
\tilde{\psi}(U)(P): \: \tilde{\Phi}_1(U)(P) \: \longrightarrow
\: \tilde{\Phi}_2(U)(P)
\end{displaymath}
by,
\begin{displaymath}
\tilde{\psi}(U)(P) \: \equiv \: \psi(U, \{ U_{\alpha} \})(P)
\end{displaymath}
It is straightforward to check that $\tilde{\psi}(U)$ is a natural
transformation for each open $U$, and moreover satisfies
the usual compatibility relation, so $\tilde{\psi}$ is a 2-arrow.

In passing, we shall mention that it is easy to check that if
$\Phi_3: {\cal C} \rightarrow {\cal D}$ is another Cartesian functor
between presheaves of categories ${\cal C}$, ${\cal D}$,
and if $\psi_1: \Phi_1 \Rightarrow \Phi_2$ and $\psi_2:
\Phi_2 \Rightarrow \Phi_3$ are a pair of 2-arrows between the Cartesian
functors, then the sheafification is compatible with composition:
\begin{displaymath}
\widetilde{ ( \psi_2 \circ \psi_1) } \: = \:
\widetilde{ \psi_2 } \circ \widetilde{ \psi_1 }
\end{displaymath}

\subsection{Pullbacks of stacks}    \label{pullback}

In this section we shall define pullbacks of stacks and some
associated technology.  We shall begin by defining pullbacks
of stacks themselves, then go on to describe pullbacks of 
Cartesian functors, analogues of natural transformations between
composed pullbacks of stacks, and more.

\subsubsection{Pullbacks of stacks}

Let $f: X \rightarrow Y$ be a continuous map, and let ${\cal C}$
be a presheaf of categories on $Y$.  In this subsection we shall describe
how to construct a stack $f^* {\cal C}$ on $X$.

We should warn the reader that this section is extremely technical;
the reader should probably skip this subsection on a first reading.

Before we describe the construction of pullbacks of stacks,
we shall take a moment to review the definition of pullbacks
of sheaves of sets, which are closely analogous (and much simpler
technically).  If ${\cal F}$ is a sheaf of sets on a space $Y$
and $f: X \rightarrow Y$ is a continuous map, then
we define the presheaf of sets $f^{-1} {\cal F}$ to be the direct
limit
\begin{displaymath}
f^{-1} {\cal F}(U) \: = \: \lim_{ \stackrel{ \longrightarrow }
{ V \supseteq f(U) } } \, {\cal F}(V)
\end{displaymath}
over open subsets $V \subseteq Y$ containing $f(U)$, for any
open set $U$.  Restriction maps are defined in a straightforward manner.
To recover a sheaf, we sheafify $f^{-1} {\cal F}$.
We shall follow a very closely analogous procedure in defining
pullbacks of stacks.

In order to construct the stack $f^* {\cal C}$, 
we shall first construct a presheaf of categories which we shall
denote $f^{-1} {\cal C}$.  Once we have constructed the presheaf
$f^{-1} {\cal C}$ on $X$, we shall sheafify $f^{-1} {\cal C}$ to
recover a sheaf of categories we shall denote $f^* {\cal C}$.
The construction we will give for
$f^{-1} {\cal C}$ will work for ${\cal C}$ a presheaf of categories,
not necessarily a stack.  Readers following \cite[section 5]{brylinski}
will note that our usage of the notation
$f^*$ and $f^{-1}$ differs slightly from that reference.

In order to define $f^{-1} {\cal C}$ for a presheaf of categories ${\cal C}$
on $Y$,
one first has to
describe how to construct categories $f^* {\cal C}(U)$ associated
to each open set $U \subseteq X$, then how to build restriction functors
and invertible natural transformations.

For each open set $U \subseteq X$, the category $f^{-1} {\cal C}(U)$
is defined to be the direct limit
\begin{displaymath}
f^{-1} {\cal C}(U) \: = \: \lim_{ \stackrel{ \longrightarrow }{ f(U) 
\subseteq V}
 } \, 
{\cal C}(V)
\end{displaymath}
where the direct limit is over open sets $V \subseteq Y$ such that
$f(U) \subseteq V$.

We shall now describe precisely how one defines a direct
limit of categories.  We shall closely follow the prescription of
\cite[section 5.2]{brylinski}.

Take as objects in $f^{-1} {\cal C}(U)$,
the disjoint union of all objects in all the categories ${\cal C}(V)$.
In other words,
\begin{displaymath}
\mbox{Ob } \lim_{\stackrel{ \longrightarrow }{ f(U) \subseteq V } } \, 
{\cal C}(V) \: = \:
\coprod \, \mbox{Ob } {\cal C}(V)
\end{displaymath}

It remains to define the set of morphisms in the category
$f^{-1} {\cal C}(U)$.
Let $P_1 \in \mbox{Ob } {\cal C}(V_1)$, $P_2 \in \mbox{Ob }{\cal C}(V_2)$.
For any diagram 
\begin{equation}   \label{Wdiag}
V_1 \: \stackrel{ \rho_1 }{ \hookleftarrow } \: W \: \stackrel{
\rho_2 }{ \hookrightarrow } \: V_2
\end{equation}
for $W$ an open set in $Y$ containing $f(U)$, define $S_W$ to be
the set of morphisms $\beta: \rho_1^* P_1 \rightarrow \rho_2^* P_2$
in ${\cal C}(W)$.
Now, define an equivalence relation $\sim$ on the disjoint union
$\coprod_W S_W$ as follows: 
for any diagram~(\ref{Wdiag}) and a similar diagram
\begin{displaymath}
V_1 \: \stackrel{ \rho'_1 }{\hookleftarrow} \: W' \: \stackrel{
\rho'_2 }{\hookrightarrow} \: V_2
\end{displaymath}
($W'$ open, $f(U) \subseteq W'$)
we say the morphism $\beta: \rho_1^* P_1 \rightarrow \rho_2^* P_2$ is equivalent
to the morphism $\beta': \rho'^*_1 P_1 \rightarrow \rho'^*_2 P_2$ 
if and only if there exists a diagram
\begin{displaymath}
W \: \stackrel{ \gamma }{ \hookleftarrow } \: Z \: \stackrel{
\gamma' }{ \hookrightarrow } \: W'
\end{displaymath}
($Z$ open, $f(U) \subseteq Z$)
such that 
\begin{enumerate}
\item $\rho_i \gamma = \rho'_i \gamma'$ for $i \in \{ 1, 2 \}$,
\item there is a commutative diagram
\begin{equation}   
\begin{array}{ccccc}
(\rho_1 \gamma)^* P_1 & \stackrel{ \varphi_{\gamma, 1}(P_1) }
{ \longrightarrow } & \gamma^* \rho_1^* P_1 &
\stackrel{ \gamma^*(\beta) }{ \longrightarrow } & \gamma^* \rho_2^* P_2 \\
\makebox[0pt][r]{ $\scriptstyle{ \varphi_{\gamma, 1}(P_1) }$ }
\downarrow & & & & \uparrow \makebox[0pt][l]{ $\scriptstyle{
\varphi_{\gamma, 2}(P_2) }$ } \\
\gamma'^* \rho'^*_1 P_1 & \stackrel{ \gamma'^*(\beta') }{ \longrightarrow } &
\gamma'^* \rho'^*_2 P_2 & \stackrel{ \varphi_{\gamma, 2}(P_2) }
{ \longleftarrow } & ( \rho_2 \gamma )^* P_2
\end{array}
\end{equation}
where the $\varphi$ are the natural transformations appearing in the
definition of ${\cal C}$ as a presheaf of categories.
\end{enumerate}

The set of morphisms $\mbox{Hom}(P_1, P_2)$ in the category
$f^{-1} {\cal C}(U)$ is defined to be
the disjoint union $\coprod_W S_W$, modulo the equivalence relation
$\sim$ described above.

So far we have defined objects and morphisms in the direct
limit category $f^{-1} {\cal C}(U)$.  We shall now take a moment
to describe composition of morphisms, as the correct definition might
not be obvious to the reader.  Let $P_1, P_2, P_3 \in
\mbox{Ob } (f^{-1} {\cal C})(U)$, i.e., $P_i \in
\mbox{Ob } {\cal C}(V_i)$ for some open sets $V_i \supseteq f(U)$.
Let $\beta \in \mbox{Hom}_{ f^{-1} {\cal C}(U) }( P_1, P_2)$,
and $\alpha \in \mbox{Hom}_{ f^{-1} {\cal C}(U) }( P_2, P_3 )$.
In other words, there exists open sets $V_{\alpha}$, $V_{\beta}$,
such that $f(U) \subseteq V_{\beta} \subseteq V_1 \cap V_2$
and $f(U) \subseteq V_{\alpha} \subseteq V_2 \cap V_3$,
with
\begin{eqnarray*}
\beta & \in & \mbox{Hom}_{ {\cal C}(V_{\beta}) }( P_1 |_{ V_{\beta} } ,
P_2 |_{ V_{\beta} } ) \\
\alpha & \in & \mbox{Hom}_{ {\cal C}(V_{\alpha}) }( P_2 |_{ V_{\alpha} },
P_3 |_{ V_{\alpha} } )
\end{eqnarray*}
We define $\alpha \circ \beta$ as follows.
Let $V$ be an open set such that $f(U) \subseteq V \subseteq V_{\alpha}
\cap V_{\beta}$.  Then, define
\begin{eqnarray*}
\alpha \circ \beta & \equiv & \varphi_{ V_{\alpha} V }^{-1} \circ
\alpha |_V \circ \varphi_{ V_{\alpha} V } \circ \varphi_{
V_{\beta} V }^{-1} \circ \beta |_V \circ \varphi_{
V_{\beta} V } \\
& \in & \mbox{Hom}_{ {\cal C}(V) }\left( P_1 |_V, P_3 |_V \right)
\end{eqnarray*}
It can be shown that this definition is well-defined,
i.e., if $\alpha \sim \alpha'$ and $\beta \sim \beta'$,
then $\alpha \circ \beta \sim \alpha' \circ \beta'$.

So far we have described the categories $f^{-1} {\cal C}(U)$ associated
to any open set $U \subseteq X$.  In order to define $f^{-1} {\cal C}$
as a presheaf of categories, we still need to define restriction
functors and appropriate natural transformations.

Before we describe restriction functors, however, we shall take a moment
to reflect on the meaning of the definition of 
the categories $f^{-1} {\cal C}(U)$ given above.
Consider, for example, the special case of the identity
map $\mbox{Id}: X \rightarrow X$.  
It can be shown that $\mbox{Id}^{-1} {\cal C}(U)$ is equivalent
(as a category) to ${\cal C}(U)$.
Naively, the reader might find this result quite surprising -- the
direct limit defining $\mbox{Id}^{-1} {\cal C}(U)$ contains more objects
than ${\cal C}(U)$.  However, although it contains more objects,
there are also more isomorphisms, and in fact the number of isomorphism
classes of objects in both categories is the same.
For example, if $P_1, P_2 \in \mbox{Ob } \mbox{Id}^{-1} {\cal C}(U)$ are
two objects, then it can be shown that $P_1$ is isomorphic to $P_2$
if and only if $P_1 |_U$ is isomorphic to $P_2 |_U$.

We shall not work through the details of proving that these
are equivalent categories, though we shall take a moment to outline
the general idea.  Define a functor $F: {\cal C}(U) \rightarrow
\mbox{Id}^{-1} {\cal C}(U)$ by, $F$ maps an object $P \mapsto P$,
and $F$ maps a morphism $\beta \mapsto \beta$.
Define a functor $G: \mbox{Id}^* {\cal C}(U) \rightarrow {\cal C}(U)$
as follows.  $G$ is defined to map an object $P \mapsto P |_U$.
Suppose $P_1 \in \mbox{Ob } {\cal C}(V_1)$ and $P_2 \in
\mbox{Ob } {\cal C}(V_2)$ for open sets $V_1$, $V_2$ such that
$U \subseteq V_1 \cap V_2$ -- in other words, let $P_1$ and
$P_2$ be objects of the category $\mbox{Id}^{-1} {\cal C}(U)$.
Let $\beta \in \mbox{Hom}_{{\cal C}(V)}\left( P_1|_V, P_2|_V\right)$
for some open set $V$, $U \subseteq V \subseteq V_1 \cap V_2$,
i.e., $\beta \in \mbox{Hom}_{{\it Id}^{-1} {\cal C}(U)}(P_1, P_2)$.
Then define $G(\beta) = \varphi_{UV}^{-1} \circ \beta |_U \circ
\varphi_{UV}$.  It can be shown that $G$ is a well-defined functor
(meaning, for example, that $\beta \sim \beta'$ implies
$G(\beta) = G(\beta')$), and that $F$ and $G$ are inverses to one
another, in the sense that there exist invertible natural transformations
$F \circ G \Rightarrow \mbox{Id}_{ {\it Id}^{-1} {\cal C}(U) }$
and $G \circ F \Rightarrow \mbox{Id}_{ {\cal C}(U) }$.

In fact, more generally it can be shown that if $f: X \rightarrow Y$
is any open map (meaning, the image of any open set is open),
then for all open $U \subseteq X$,
the category $f^{-1} {\cal C}(U)$ is equivalent to the category
${\cal C}( f(U) )$.

Now that we have given some intuition for the meaning of the
direct limits used above, we shall describe the restriction
functors and natural transformations needed to describe
$f^{-1} {\cal C}$ as a presheaf of categories.

Given the categories $f^{-1} {\cal C}(U)$,
we shall now define the pullback functors $\rho^*: f^{-1} {\cal C}(U_1)
\rightarrow f^{-1} {\cal C}(U_2)$ for $\rho: U_2 \hookrightarrow U_1$.
In fact, these restriction functors are straightforward to define.
Note that $\rho$ induces a map 
\begin{displaymath}
\{ \, V \, | \, f(U_1) \subseteq V \, \} \: \longrightarrow \:
\{ \, V \, | \, f(U_2) \subseteq V \, \}
\end{displaymath}
given by $V \mapsto V$
(any open set containing $f(U_1)$, also contains $f(U_2)$).
We define $\rho^*$ to act on objects $P \in f^{-1} {\cal C}(U_1)$
as, $P \mapsto P$, and on morphisms $\beta$ as, $\beta \mapsto \beta$.
It should be clear that this map yields a well-defined\footnote{For example,
if $\beta \sim \beta'$, then it should be clear that
$\rho^*(\beta) \sim \rho^*(\beta')$.}
functor $f^{-1} {\cal C}(U_1)
\rightarrow f^{-1} {\cal C}(U_2)$.

It should be clear that the invertible natural 
transformations $( \rho_1 \rho_2 )^*
\Rightarrow \rho_2^* \circ \rho_1^*$ needed to define a presheaf of
categories are trivial.  In other words, if $P \in \mbox{Ob }
f^{-1} {\cal C}(U)$, meaning $P \in \mbox{Ob } {\cal C}(V)$ for some
open $V \supseteq f(U)$, then both the functors $(\rho_1 \rho_2)^*$
and $\rho_2^* \circ \rho_1^*$, for any pair of composable inclusions
$\rho_1$, $\rho_2$, map $P \mapsto P$, and so the morphism
that the requisite natural transformation should assign to $P$ is given
by the identity morphism.

So far we have defined a presheaf of categories $f^{-1} {\cal C}$.
Even if ${\cal C}$ is a sheaf, the presheaf $f^{-1} {\cal C}$ is, in general,
not itself a sheaf.  In order to get a sheaf from the presheaf
$f^{-1} {\cal C}$, we must sheafify the presheaf.
We shall denote the result of the sheafification by $f^* {\cal C}$.

As an illuminating example, consider $\mbox{Id}^* {\cal C}$,
where $\mbox{Id}: X \rightarrow X$ is the identity map
and ${\cal C}$ is a stack.
We pointed out earlier that each category $\mbox{Id}^* {\cal C}(U)$
is equivalent to the category ${\cal C}(U)$, for any open set $U$.
It should now be clear that $\mbox{Id}^* {\cal C}$ is equivalent
to ${\cal C}$ as a stack.

If $f$ is a homeomorphism, for example, and ${\cal C}$ is a stack,
then $f^{-1} {\cal C}$ is already a stack, not just a presheaf
of categories.  Stronger statements can be made, but this is
all we need for later use.

\subsubsection{Pullbacks of Cartesian functors}

Let $f: X \rightarrow Y$ be a continuous map,
and let $\Phi: {\cal C} \rightarrow {\cal D}$ be a Cartesian
functor between presheaves of categories ${\cal C}$,
${\cal D}$ on $Y$.  We shall construct a Cartesian functor
\begin{displaymath}
f^* \Phi: \: f^* {\cal C} \: \longrightarrow \: f^* {\cal D}
\end{displaymath}
by first constructing a Cartesian functor 
\begin{displaymath}
f^{-1} \Phi: \: f^{-1} {\cal C} \: \longrightarrow \: f^{-1} {\cal D}
\end{displaymath}
between the presheaf pullbacks $f^{-1} {\cal C}$, $f^{-1} {\cal D}$ on $X$,
and using the fact that a Cartesian functor between presheaves of
categories lifts to a Cartesian functor between the sheafifications.

We shall first define functors 
\begin{displaymath}
f^{-1} \Phi(U): \: f^{-1} {\cal C}(U) \: \rightarrow \: f^{-1} {\cal D}(U)
\end{displaymath}
for $U \subseteq X$ an open set.

We define $f^{-1} \Phi(U)$ on objects as follows.
Recall 
\begin{displaymath}
\mbox{Ob } f^{-1} {\cal C}(U) \: = \: \coprod_{ f(U) \subseteq V }
\mbox{Ob } {\cal C}(V)
\end{displaymath}
so $P \in \mbox{Ob } f^{-1} {\cal C}(U)$ means,
$P \in \mbox{Ob } {\cal C}(V)$ for some open $V \supseteq f(U)$.
Define the action of $f^{-1} \Phi(U)$ on $P$ as,
\begin{displaymath}
P \: \mapsto \: \Phi(V)(P) \: \in \: \mbox{Ob } {\cal D}(V)
\: \subseteq \: \mbox{Ob } f^{-1} {\cal D}(U)
\end{displaymath}

Now we shall define $f^{-1} \Phi(U)$ on morphisms.
Let $P_1, P_2 \in \mbox{Ob } f^{-1} {\cal C}(U)$,
meaning, $P_i \in \mbox{Ob } {\cal C}(V_i)$ for open $V_i \supseteq
f(U)$ and $i \in \{ 1, 2 \}$.
Let $\beta \in \mbox{Hom}_{ f^{-1} {\cal C}(U) }( P_1, P_2 )$.
In other words, there exists open $V_{\beta}$, $f(U) \subseteq
V_{\beta} \subseteq V_1 \cap V_2$, such that
$\beta \in \mbox{Hom}_{ {\cal C}(V_{\beta}) }( P_1 |_{ V_{\beta} },
P_2 |_{ V_{\beta} })$.
Define $f^{-1} \Phi(U)$ on $\beta$ as,
\begin{eqnarray*}
\beta & \mapsto & \chi_{V_{\beta} V_2}^{-1}(P_2) \circ
\Phi(V_{\beta})(\beta) \circ \chi_{V_{\beta}V_1}(P_1) \\
&  & \: \in \: \mbox{Hom}_{ {\cal D}(V_{\beta}) } \left( \Phi(V_1)(P_1) |_{
V_{\beta} }, \Phi(V_2)(P_2) |_{ V_{\beta} } \right)
\end{eqnarray*}
where $\chi$ is the invertible natural transformation appearing
in the definition of $\Phi: {\cal C} \rightarrow {\cal D}$ as a
Cartesian functor.

It is straightforward to check that $f^{-1} \Phi(U)$, as given above,
yields a well-defined functor $f^{-1} {\cal C}(U) \rightarrow
f^{-1} {\cal D}(U)$.

In order to create a Cartesian functor $f^{-1} \Phi$,
it remains to specify invertible natural transformations
\begin{displaymath}
f^{-1} \chi_{\rho}: \: \rho_{ f^{-1} {\cal D} }^* \circ
f^{-1} \Phi(U_1) \: \Longrightarrow \: f^{-1} \Phi(U_2) \circ
\rho_{ f^{-1} {\cal C} }^*
\end{displaymath}
for inclusions $\rho: U_2 \hookrightarrow U_1$.

In fact, the requisite natural transformations are trivial.
Let $P \in \mbox{Ob } f^{-1} {\cal C}(U_1)$,
meaning $P \in \mbox{Ob } {\cal C}(V)$ for some open $V \supseteq f(U_1)$.
First, note $P |_{ U_2 }$ is given by the same $P \in \mbox{Ob }
{\cal C}(V)$, from the definition of restriction functor for pullbacks.
Moreover,
\begin{displaymath}
f^{-1} \Phi(U_1) (P) |_{ U_2 } \: = \: f^{-1} \Phi(U_2)( P|_{U_2} )
\: = \: \Phi(V)(P) \: \in \: \mbox{Ob } {\cal D}(V) \:
\subseteq \mbox{Ob } f^{-1} {\cal D}(U_2)
\end{displaymath}
Clearly, the morphism that the natural transformation $f^{-1} \chi_{\rho}$
assigns
to the object $P$ is the identity morphism on $\Phi(V)(P)$.
This yields a well-defined natural transformation, satisfying the
usual pentagonal identity.

Thus, we have defined
a Cartesian functor $f^{-1} \Phi: f^{-1} {\cal C} \rightarrow
f^{-1} {\cal D}$.

Now that we have defined a Cartesian functor between the presheaves
$f^{-1} {\cal C}$ and $f^{-1} {\cal D}$, we can use the fact that
Cartesian functors between presheaves lift to Cartesian functors
between sheafifications to immediately recover a Cartesian
functor $f^* \Phi$:
\begin{displaymath}
f^* \Phi: \: f^* {\cal C} \: \rightarrow \: f^* {\cal D}
\end{displaymath}

In passing, we shall mention that if we have another Cartesian functor
$\Lambda: {\cal D} \rightarrow {\cal E}$ between presheaves of categories
on $X$, then it is straightforward to check that pullbacks of Cartesian
functors are compatible with composition of Cartesian functors.
In other words,
\begin{displaymath}
f^{-1} ( \Lambda \circ \Phi ) \: = \: f^{-1} \Lambda \circ f^{-1} \Phi
\end{displaymath}
and so, using an analogous result for sheafifications, we find
\begin{displaymath}
f^* ( \Lambda \circ \Phi ) \: = \: f^* \Lambda \circ f^* \Phi
\end{displaymath}

\subsubsection{Pullbacks of 2-arrows}

Let $f: X \rightarrow Y$ be a continuous map, and let $\Phi_1,
\Phi_2: {\cal C} \rightarrow {\cal D}$ be Cartesian functors
between presheaves of categories ${\cal C}$, ${\cal D}$.
Let $\psi: \Phi_1 \Rightarrow \Phi_2$ be a 2-arrow
between the Cartesian functors.
In this section we shall construct a 2-arrow
\begin{displaymath}
f^* \psi: \: f^* \Phi_1 \: \Longrightarrow \: f^* \Phi_2
\end{displaymath}
by first constructing a 2-arrow
\begin{displaymath}
f^{-1} \psi: \: f^{-1} \Phi_1 \: \Longrightarrow \:
f^{-1} \Phi_2
\end{displaymath}
between the pullbacks $f^{-1} \Phi_1$, $f^{-1} \Phi_2$,
and using the fact that a 2-arrow lifts to a 2-arrow between
sheafifications.

We shall construct $f^{-1} \psi$ as follows.
Let $U \subseteq X$ be an open set, and
let $P \in \mbox{Ob } f^{-1} {\cal C}(U)$, that is,
$P \in \mbox{Ob } {\cal C}(V)$ for some open $V \supseteq f(U)$.
Define a morphism
\begin{displaymath}
(f^{-1} \psi)(U)(P): \: (f^{-1} \Phi_1)(U)(P) \: \longrightarrow
\: (f^{-1} \Phi_2)(U)(P)
\end{displaymath}
by,
\begin{displaymath}
(f^{-1} \psi)(U)(P) \: \equiv \: \psi(V)(P): \: \Phi_1(V)(P) \:
\longrightarrow \: \Phi_2(V)(P)
\end{displaymath}
It is easy to check that this defines a natural transformation
\begin{displaymath}
(f^{-1} \psi)(U): \: (f^{-1} \Phi_1)(U) \: \Longrightarrow \:
(f^{-1} \Phi_2)(U)
\end{displaymath}
and moreover, for any inclusion $\rho: U_2 \hookrightarrow U_1$
of open subsets of $X$, these natural transformations are compatible
with the natural transformations $f^{-1} \chi_{\rho}$ defining
$f^{-1} \Phi_1$ and $f^{-1} \Phi_2$ as Cartesian functors.

Thus, we have defined a 2-arrow $f^{-1} \psi: f^{-1} \Phi_1 
\Rightarrow f^{-1} \Phi_2$.

Now that we have defined a 2-arrow between the Cartesian functors
$f^{-1} \Phi_1$ and $f^{-1} \Phi_2$, we can use the fact that
2-arrows lift to 2-arrows between sheafifications to immediately
recover a 2-arrow $f^* \psi$:
\begin{displaymath}
f^* \psi: \: f^* \Phi_1 \: \Longrightarrow \: f^* \Phi_2
\end{displaymath}

In passing, we shall mention that if we have another
Cartesian functor $\Phi_3: {\cal C} \rightarrow {\cal D}$
and a pair of 2-arrows $\psi_1: \Phi_1 \Rightarrow \Phi_2$,
$\psi_2: \Phi_2 \Rightarrow \Phi_3$, then pullback is compatible
with composition.  In other words,
\begin{displaymath}
f^{-1} ( \psi_2 \circ \psi_1 ) \: = \: (f^{-1} \psi_2) \circ (f^{-1}
\psi_1)
\end{displaymath}
By using the analogous result for sheafifications, we find
\begin{displaymath}
f^* ( \psi_2 \circ \psi_1 ) \: = \: (f^* \psi_2) \circ (f^* \psi_1)
\end{displaymath}

\subsubsection{Analogues of natural transformations} 
\label{analoguenattranssec}

In this section we shall define an analogue of natural transformation
for compositions of pullbacks.
More precisely, let $f: X \rightarrow Y$ and $g: Y \rightarrow Z$
be continuous maps between topological spaces.
For any stack ${\cal C}$ on $Z$, we shall define a Cartesian
functor 
\begin{displaymath}
\Psi_{gf}^{{\cal C}}: \: (gf)^* {\cal C} \: \longrightarrow
f^* g^* {\cal C}
\end{displaymath}
with the properties
\begin{enumerate}
\item For any Cartesian functor $\Phi: {\cal C} \rightarrow {\cal D}$
between stacks on $Z$, the following diagram commutes:
\begin{equation}    \label{nattransanalog1}
\begin{array}{ccc}
(gf)^* {\cal C} & \stackrel{ (gf)^* \Phi }{ \longrightarrow } &
(gf)^* {\cal D} \\
\makebox[0pt][r]{ $\scriptstyle{ \Psi_{gf}^{{\cal C}} }$ } \downarrow
& & \downarrow \makebox[0pt][l]{ $\scriptstyle{ \Psi_{gf}^{{\cal D}} }$ } \\
f^* g^* {\cal C} & \stackrel{ f^* g^* \Phi }{ \longrightarrow } &
f^* g^* {\cal D} 
\end{array}
\end{equation}
\item If $f: X \rightarrow Y$, $g: Y \rightarrow Z$, $h: Z \rightarrow
W$ are three continuous maps, then for any stack ${\cal C}$ on $W$
we have the commuting diagram
\begin{equation}    \label{nattransanalog2}
\begin{array}{ccc}
(hgf)^* {\cal C} & \stackrel{ \Psi_{ hg, f} }{\longrightarrow} &
\left( f^* \circ (hg)^* \right) {\cal C} \\
\makebox[0pt][r]{ $\scriptstyle{ \Psi_{ h, gf} }$ } \downarrow & &
\downarrow \makebox[0pt][l]{ $\scriptstyle{ \Psi_{h, g} }$ } \\
\left( (gf)^* \circ h^* \right) {\cal C} & \stackrel{ \Psi_{g, f} }
{\longrightarrow }
& \left( f^* \circ g^* \circ h^* \right) {\cal C}
\end{array}
\end{equation}
which the reader should immediately recognize as being analogous
to diagram~(\ref{trans}).
\end{enumerate}
Moreover, the Cartesian functor $\Psi_{gf}: (gf)^* {\cal C}
\rightarrow f^* g^* {\cal C}$ will be invertible, in the sense
that there exists a Cartesian functor $\Psi_{gf}^{-1}: f^* g^* {\cal C}
\rightarrow (gf)^* {\cal C}$ and invertible 2-arrows $\Psi \circ \Psi^{-1} 
\Rightarrow \mbox{Id}$
and $\Psi^{-1} \circ \Psi \Rightarrow \mbox{Id}$.

In order to define $\Psi$, we shall work at the level of presheaves
of categories and define a Cartesian functor
\begin{displaymath}
\Psi_{gf}: \: (gf)^{-1} {\cal C} \: \longrightarrow \: f^{-1} g^{-1} {\cal C}
\end{displaymath}

Furthermore, we shall implicitly assume that the map $f$ is open (i.e.,
images of open sets are open), and that $g$ is such that
$g^{-1} {\cal C}$ is a stack, not just a presheaf of categories.
These restrictions could almost certainly be weakened; however, in this
paper we shall only be interested in cases in which both $f$ and
$g$ are homeomorphisms, so we shall not investigate these conditions further.

To define the Cartesian functor $\Psi_{gf}: (gf)^{-1} {\cal C}
\rightarrow f^{-1} g^{-1} {\cal C}$, we shall first define functors
\begin{displaymath}
\Psi_{gf}(U): \: (gf)^{-1} {\cal C}(U) \: \longrightarrow \: f^{-1} g^{-1} 
{\cal C}(U)
\end{displaymath}
for open sets $U \subseteq X$.

We define the functor $\Psi_{gf}(U)$ on objects as follows.
Let $P$ be an object in $(gf)^{-1} {\cal C}(U)$,
which is to say, $P \in \mbox{Ob } {\cal C}(V)$ for some
open $V \supseteq (gf)(U)$.
The functor $\Psi_{gf}(U)$ acts as
\begin{eqnarray*}
P \: \mapsto \: \Psi_{gf}(U)(P) \: \equiv \: P & \in & \coprod_{ V \supseteq
(gf)(U) } \, \mbox{Ob } {\cal C}(V) \\
& & \: \subseteq \:
\coprod_{V' \supseteq f(U) } \coprod_{ V \supseteq g(V') }
\mbox{Ob } {\cal C}(V) \: = \: \mbox{Ob } f^{-1} g^{-1} {\cal C}(U)
\end{eqnarray*}

We define the functor $\Psi_{gf}(U)$ on morphisms as follows.
Let $P_1$, $P_2$ be objects in $(gf)^{-1} {\cal C}(U)$,
which is to say, $P_i \in \mbox{Ob } {\cal C}(V_i)$ for
open $V_i \supseteq (gf)(U)$, $i \in \{ 1, 2 \}$.
Let $\beta \in \mbox{Hom}_{ (gf)^{-1} {\cal C}(U) }(P_1, P_2)$,
which means that for some open $V_{\beta}$, $(gf)(U) \subseteq
V_{\beta} \subseteq V_1 \cap V_2$,
$\beta \in \mbox{Hom}_{{\cal C}(V_{\beta})}( P_1 |_{V_{\beta}}, 
P_2 |_{V_{\beta}} )$.  The functor $\Psi_{gf}(U)$ acts on $\beta$ as,
\begin{displaymath}
\beta \: \mapsto \: \Psi_{gf}(U)(\beta) \: \equiv \: \beta
\end{displaymath}
In terms of equivalence classes, denoted by brackets $[ \: ]$,
the functor $\Psi_{gf}(U)$ maps $[ \beta ]_{gf}$ to
$[ [\beta]_g ]_f$.

In order to define a Cartesian functor $\Psi_{gf}: (gf)^{-1} {\cal C}
\rightarrow f^{-1} g^{-1} {\cal C}$, we must specify an invertible
natural transformation
\begin{displaymath}
\chi_{gf}: \: \rho_{f^{-1} g^{-1} }^* \circ \Psi_{gf}(U) \:
\Longrightarrow \: \Psi_{gf}(V) \circ \rho_{ (gf)^{-1} }^*
\end{displaymath}
for every inclusion $\rho: V \hookrightarrow U$.
We define this natural transformation to be the trivial one.
In other words, given an object $P$ in $(gf)^{-1}{\cal C}(U)$,
which is to say, $P \in \mbox{Ob } {\cal C}(V)$ for some
open $V \supseteq (gf)(U)$, we define $\chi_{gf}(P) = \mbox{Id}_{
{\cal C}(V) }(P)$.  This assignment of morphisms to objects clearly
defines a natural transformation, and moreover it is easy to see
that $\chi$ satisfies the pentagonal identity for natural transformations
in Cartesian functors.

Thus, we have now defined a Cartesian functor $\Psi_{gf}:
(gf)^{-1} {\cal C} \rightarrow f^{-1} g^{-1} {\cal C}$.
It is straightforward to check that this definition satisfies the two
properties listed at the beginning of this section, and moreover that
this Cartesian functor admits an inverse Cartesian functor
$\Psi_{gf}^{-1}$, such that the composite functors
$\Psi \circ \Psi^{-1}$ and $\Psi^{-1} \circ \Psi$ can be identified with
identity functors, up to invertible 2-arrows.

Finally, we can lift this Cartesian functor to a Cartesian functor
between sheafifications
\begin{displaymath}
\Psi_{gf}: \: (gf)^* {\cal C} \: \longrightarrow \: f^* g^* {\cal C}
\end{displaymath}
satisfying the properties listed at the beginning of this section.

\subsection{Stalks of stacks}

For a presheaf of sets ${\cal F}$ on a space $X$,
recall that one can define a stalk of the sheaf ${\cal F}$ at
the point $x \in X$ to be the direct limit
\begin{displaymath}
{\cal F}_x \: \equiv \: \lim_{ \stackrel{ \longrightarrow }{ U \ni x } }
\, {\cal F}(U)
\end{displaymath}
over open sets $U \subseteq X$ containing the point $x \in X$.

We can perform the analogous construction for presheaves of
categories.  Define the stalk of a presheaf of categories
${\cal C}$ at a point $x \in X$ to be the direct limit
\begin{displaymath}
{\cal C}_x \: \equiv \: \lim_{ \stackrel{ \longrightarrow }{ U \ni x } }
\, {\cal C}(U)
\end{displaymath}
over open sets $U \subseteq X$ containing the point $x \in X$.
This direct limit is defined in precise analogy with the direct limits
defined in the previous two sections.
Note that just as the stalk of a presheaf of sets is a set,
the stalk of a presheaf of categories is a category.

This notion of stalk may give the reader some degree of 
intuition for stacks.  Furthermore, one ought to be able
to work out many stack-variants of other concepts from ordinary 
sheaves.
As we shall not use stalks of presheaves of categories in this paper, 
or other stack-theoretic versions of other sheaf theory concepts,
we shall
not speak about such matters any further.


\section{Technical notes on gerbes}    \label{advgerbesec}

In this section we shall make some highly technical remarks on gerbes.
In particular, after discussing pullbacks of gerbes and defining
torsors, we shall speak in detail about gauge transformations of gerbes,
and use such ideas to derive some basic facts about gerbes
which were mentioned earlier in this paper.
Many of the basic ideas are taken from \cite[section 5]{brylinski}.

Readers studying this paper for the first time are urged to skip
this section.

\subsection{Pullbacks of gerbes}

Let $f: X \rightarrow Y$ be a continuous map, and
${\cal C}$ a gerbe on $Y$ with band ${\cal A}$.
It is easy to check that $f^* {\cal C}$
is a gerbe on $X$, with band $f^* {\cal A}$ \cite[prop. 5.2.6]{brylinski}.
Moreover, the element of the sheaf cohomology group
$H^2(X, f^* {\cal A})$ characterizing $f^* {\cal C}$ is precisely
the pullback of the element of the sheaf cohomology group
$H^2(Y, {\cal A})$ characterizing ${\cal C}$ \cite[section 5.2]{brylinski},
as the reader might have guessed.

If ${\cal C}$ has a connective structure, then
(at least for $f$ a diffeomorphism) one naturally obtains a connective
structure on $f^* {\cal C}$.  Let $\mbox{Co}$ denote the connective
structure on ${\cal C}$, meaning that $\mbox{Co}$ is a Cartesian
functor
\begin{displaymath}
\mbox{Co}: \: {\cal C} \: \longrightarrow \: \mbox{Tors}_Y(\Omega^1)
\end{displaymath}
which is compatible with the bands of either gerbe.
This Cartesian functor can be lifted to a Cartesian functor
\begin{displaymath}
f^* \mbox{Co}:  \: f^* {\cal C} \: \longrightarrow \: f^* 
\mbox{Tors}_Y(\Omega^1)
\end{displaymath}
and at least in the special case that $f$ is a diffeomorphism
(the only case we shall need), $f^* \mbox{Tors}_Y(\Omega^1) \cong
\mbox{Tors}_X(\Omega^1)$.  Thus, at least for $f$ a diffeomorphism
we find that a connective structure on ${\cal C}$ naturally
defines a connective structure on $f^* {\cal C}$.

Let $K$ denote a curving on $( {\cal C}, \mbox{Co})$.  At least in the
case that $f$ is a diffeomorphism, we can pullback $K$ to a curving
$f^* K$ on $( f^* {\cal C}, f^* \mbox{Co})$.
We shall outline the details here.  Since $f$ is assumed to be a 
diffeomorphism, we know that $f^* {\cal C} \cong f^{-1} {\cal C}$,
and so it suffices to define $f^* K$ at the level of $f^{-1} {\cal C}$
and $f^{-1} \mbox{Co}$.  Let $U \subseteq X$ be open, and
let $P \in \mbox{Ob } (f^{-1} {\cal C})(U)$, i.e.,
$P \in \mbox{Ob } {\cal C}(V)$ for some open $V \supseteq f(U)$.
Recall that by definition,
\begin{displaymath}
(f^{-1} \mbox{Co})(U)(P) \: = \: \mbox{Co}(V)(P)
\end{displaymath}
Let $\nabla$ be a section of $(f^{-1} \mbox{Co})(U)(P)$, which is to say,
a section of $\mbox{Co}(V)(P)$.
We are now finally ready to define $f^* K$.
Define
\begin{displaymath}
(f^* K)(\nabla) \: \equiv \: (f|_U)^* \left[ K(\nabla) \right]
\end{displaymath}
It is straightforward to check that this definition satisfies the
defining axioms for a curving.

To summarize, there exist natural notions of pullback for both
gerbes and connections on gerbes.

\subsection{Torsors}    \label{torsdef}

Most mathematical descriptions of gerbes rely heavily on
torsors.  For the most part, we have strenuously avoided
speaking of torsors in this text, but at times their use
is unavoidable.  In this section we shall define torsors.
Our discussion will largely follow \cite[section 5.1]{brylinski}
and \cite{iversen}.

A torsor with respect to a group $G$ is a set
with an action of $G$ that is free and transitive.
An example of a $G$-torsor for a topological group $G$
is the group $G$ itself (though when
described as a torsor, one implicitly drops the group structure).
Let $C^{\infty}(G)$ denote the group of smooth maps from a manifold
$X$ into a Lie group $G$, then an example of a $C^{\infty}(G)$-torsor
is any principal $G$-bundle.  (Some authors abuse notation and refer
to a principal $G$-bundle as a $G$-torsor, rather than
a $C^{\infty}(G)$-torsor; we shall specifically
avoid such mangled usage.)
Let $\Omega^1(X)$ denote the abelian group consisting
of 1-forms on a manifold $X$, then an example of an $\Omega^1(X)$-torsor
is the space of connections on any principal $U(1)$-bundle.
(Any two such connections differ by a 1-form, which can be seen as follows:
Let $A^{\alpha}_{\mu}$ and $A^{' \alpha}_{\mu}$ denote two locally-defined
connections on the bundle with respect to a good open cover $\{ U_{\alpha} \}$.
Then $A^{\alpha} - A^{\beta} = A^{' \alpha} - A^{' \beta} =
d \, \mbox{ln } g_{\alpha \beta}$ on overlaps, so in particular
$A^{\alpha} - A^{' \alpha}$ is a globally-defined 1-form.)

The torsors that we use in this paper are torsors with respect
to a sheaf of abelian groups, not just a group.
A torsor with respect to a sheaf of groups (say, ${\cal A}$)
is a sheaf of sets, say, ${\cal F}$, such that the set
${\cal F}(U)$ associated to any open set $U$ is a torsor
with respect to the group ${\cal A}(U)$,
and such that the action of ${\cal A}$ commutes with restriction maps.

A morphism of ${\cal A}$-torsors $\phi: {\cal F} \rightarrow {\cal G}$
is a morphism of sheaves of sets, such that the action of
${\cal A}$ commutes with $\phi$.

An example of a torsor with respect to the sheaf\footnote{Note our
notation is slightly ambiguous:  we use $C^{\infty}(G)$ to denote
both the group of smooth maps into $G$, and the sheaf of smooth
local maps into $G$.
The correct interpretation should be clear from context.} 
${\cal A} = 
C^{\infty}(G)$ of $C^{\infty}$ maps into $G$ is the sheaf of
local sections of a smooth principal $G$-bundle.

The set of isomorphism classes of ${\cal A}$-torsors has an
(abelian) group structure.
Let ${\cal F}$, ${\cal G}$ be a pair of ${\cal A}$-torsors.
The product of sheaves ${\cal F} \times {\cal G}$ is an
$({\cal A} \times {\cal A})$-torsor.
We define the ${\cal A}$-torsor ${\cal F} \cdot {\cal G}$
(the result of the group operation) to be the sheaf associated
to the presheaf 
\begin{displaymath}
\frac{ ( {\cal F} \times {\cal G} ) \times {\cal A} }{ {\cal A} \times
{\cal A} }
\end{displaymath}
where ${\cal A} \times {\cal A}$ acts on ${\cal A}$ via the product
map ${\cal A} \times {\cal A} \rightarrow {\cal A}$.

Given any ${\cal A}$-torsor ${\cal F}$, there is a natural definition
of ${\cal F}^{-1}$.  Specifically, the ${\cal A}$-torsor
${\cal F}^{-1}$ is defined by the property that for any open
$U \subseteq X$, the set ${\cal F}^{-1}(U)$ is the set of
torsor isomorphisms ${\cal F}(U) \stackrel{ \sim}{ \longrightarrow}
{\cal A}(U)$.

It can be shown that the group of isomorphism classes of
${\cal A}$-torsors, over a space $X$, is in natural bijection
with the group $H^1(X, {\cal A})$.

It should be clear that
that any morphism of ${\cal A}$-torsors, over the same
space, is necessarily an isomorphism.  This is a generalization of a
similar result for principal $G$-bundles \cite[section 4.3]{husemoller},
namely that any morphism of principal $G$-bundles for fixed $G$ over
the same base space is necessarily an isomorphism.

We should also mention a technical lemma which we shall use
in what follows.  Let $I$, $K$ be a pair of ${\cal A}$-torsors
over a space $X$, for some sheaf of abelian groups ${\cal A}$.
We shall show that in order to define an 
isomorphism $I \rightarrow K$ of ${\cal A}$-torsors $I$, $K$,
it suffices to show how a set of local sections $\{ s_{\alpha} \}$
of $I$, defined with respect to an open cover $\{ U_{\alpha} \}$ of $X$,
are mapped.  To define the isomorphism for other elements of
the sets $I(U_{\alpha})$, $K(U_{\alpha})$, use the action of
the group ${\cal A}(U_{\alpha})$ -- in other words, any other
element of $I(U_{\alpha})$ will differ from $s_{\alpha}$ by
an element of the group ${\cal A}(U_{\alpha})$, so we can define
its image to be the image of $s_{\alpha}$ modulo the same group element,
thus explicitly recovering an isomorphism of sets which, by construction,
commutes with the action of ${\cal A}(U_{\alpha})$.  We can construct
maps $I(W) \rightarrow K(W)$ for $W \subseteq U_{\alpha}$ for some
$U_{\alpha}$ by using restriction in the obvious way, and we can
construct maps for $W \supseteq U_{\alpha}$ by performing all possible
gluings of elements of $\{ I(U_{\alpha}) \}$.

More information on torsors can be found in,
for example, \cite{iversen}, \cite[section VIII.2]{maclane},
or \cite[section 5.1]{brylinski}.

\subsection{Gauge transformations of gerbes}    \label{gtap}

Just as maps $X \rightarrow G$ define gauge transformations
of principal $G$-bundles on a space $X$, we will see
explicitly in this section that (equivalence classes of) bundles define gauge
transformations of gerbes of band $C^{\infty}(G)$.

To fully explain these ideas will take some time.  
Much of the material we present in subsections~\ref{gtobj}
and \ref{morphtors} is taken
from \cite[section 5.2]{brylinski}.

\subsubsection{Gauge transformations of objects}   \label{gtobj}

Let ${\cal C}$ denote a gerbe on a manifold $X$, with band ${\cal A}$.
We shall assume ${\cal A} = C^{\infty}(G)$ for some abelian Lie
group $G$, for simplicity.
Let $U \subseteq X$ be an open subset of a space $X$, and let $I$ be
a principal $G$-bundle 
on $U$.
Strictly speaking, we should take $I$ to be a ${\cal A} |_U$-torsor,
not a principal $G$-bundle;
however, where possible, we are trying to avoid the language of torsors.

Given any object $P \in \mbox{Ob } {\cal C}(U)$, we shall show
how to use $I$ to construct another object we shall denote
$P \times I \in \mbox{Ob } {\cal C}(U)$.
Let $\{ U_{\alpha} \}$ be a good open cover of $U$,
so that $I(U_{\alpha}) \neq \emptyset$ for all $\alpha$.
Let $\{ s_{\alpha} \in I(U_{\alpha}) \}$ be a set of local sections
of $I$, over the open subsets $U_{\alpha}$.
Define $g_{\alpha \beta}$ to be the (unique)
element of ${\cal A}(U_{\alpha \beta})$
that transports $s_{\beta} |_{U_{\alpha \beta}}$ to $s_{\alpha} |_{
U_{\alpha \beta} }$.  (Recall that strictly speaking, we should
interpret $I$ as a torsor, so that there is no natural group law
on the $\{ s_{\alpha} \}$ {\it per se}.)

We can now define the object $P \times I$, using the gluing law
for objects, as follows.  Define a set of isomorphisms
\begin{displaymath}
\phi_{\alpha \beta} : \: P |_{U_{\beta}} |_{U_{\alpha \beta}}
\: \longrightarrow \: P |_{U_{\alpha}} |_{U_{\alpha \beta}}
\end{displaymath}
by,
\begin{displaymath}
\phi_{\alpha \beta} \: \equiv \: \varphi_{\alpha, \alpha \beta}
\circ g_{\alpha \beta} \circ \varphi_{\beta, \alpha \beta}^{-1}
\end{displaymath}
It is straightforward to check that these satisfy the axioms for
the gluing law for objects, and so from said gluing law
we recover a new object in ${\cal C}(U)$ which we shall denote
$P \times I$.

In passing, we should mention a minor technical problem with
the description above, namely that the object $P \times I$ is
almost, but not quite, uniquely specified by a set of local
sections of $I$.  Recall that the gluing law for objects
yields objects that are unique only up to unique isomorphism commuting
with the gluing maps $\psi_{\alpha}$.
Thus, the object $P \times I$ is not uniquely defined -- however,
there exists a unique isomorphism (commuting with the
gluing maps $\psi_{\alpha}$) between any two objects that one
might label ``$P \times I$.''  Thus, in order to uniquely specify
an action of a torsor $I$ together with a set of local sections of $I$
on objects, we must choose specific examples of $P \times I$ for
each object $P$.  In the next subsection we shall derive a functor
describing the action of the torsor $I$, and it is straightforward
to check that for any two such functors differing only in the choices
made of objects ``$P \times I$,'' the isomorphisms between the choices
define a 2-arrow between functors.  We shall not speak further about
this issue, except when absolutely necessary.

In defining the action of a torsor $I$ on objects $P$,
we referred to a specific choice of a set of local sections of $I$.
What happens if we choose a distinct set of local sections of $I$?
The answer is that any two sets of local sections of $I$ will 
define isomorphic objects
$P \times I$.  We shall outline how this is proven in the special
case that the two sets of local sections in question are defined
with respect to the same open cover $\{ U_{\alpha} \}$.
(It is straightforward to check that the same result also holds
for local sections defined with respect to distinct open covers,
but the details are more cumbersome.)
Let $U \subseteq X$ be an open set, $I$ an ${\cal A}|_U$-torsor,
and $\{ s_{\alpha} \}$, $\{ s'_{\alpha} \}$ two sets of local sections
of $I$, both defined over the same open cover $\{ U_{\alpha} \}$ of $U$.
Let $P \times I$ and $(P \times I)'$ denote the two objects
obtained from gluing via the local sections $\{ s_{\alpha} \}$,
$\{ s'_{\alpha} \}$, and let
\begin{eqnarray*}
\psi_{\alpha}(P): \: (P \times I)|_{U_{\alpha}} & \longrightarrow &
P |_{U_{\alpha}} \\
\psi'_{\alpha}(P): \: (P \times I)'|_{U_{\alpha}} & \longrightarrow &
P |_{U_{\alpha}}
\end{eqnarray*}
denote the corresponding isomorphisms.
Define $f_{\alpha}: (P \times I)|_{U_{\alpha}} \rightarrow
(P \times I)'|_{U_{\alpha}}$ by,
\begin{displaymath}
f_{\alpha} \: \equiv \: ( \psi'_{\alpha}(P) )^{-1} \circ (
s'_{\alpha} - s_{\alpha} ) \circ \psi_{\alpha}(P)
\end{displaymath}
where we have used $(s'_{\alpha} - s_{\alpha})$ to denote the element
of the band mapping $s_{\alpha} \mapsto s'_{\alpha}$.
It is straightforward to check that the $\{ f_{\alpha} \}$ satisfy
the gluing axiom for morphisms, and so there exists a (unique)
morphism $P \times I \rightarrow (P \times I)'$. 
Thus, distinct local sections of $I$ define isomorphic objects $P \times I$.

\subsubsection{Induced equivalences of categories}

A ${\cal A}|_U$-torsor $I$ not only induces a map between objects
of the category ${\cal C}(U)$, but it also induces a self-equivalence
of the category ${\cal C}(U)$, which we shall now explain.

We shall denote the proposed functor by $I(U)$.
We explained the action of $I(U)$ on objects, namely
$P \mapsto P \times I$, in the previous subsection.
We define the action of $I(U)$ on morphisms as follows.

Let $P_1$, $P_2$ be objects of ${\cal C}(U)$, and let
$\beta \in \mbox{Hom}_{ {\cal C}(U) }(P_1, P_2)$.
Let $\psi_{\alpha}(P_i)$ denote the isomorphisms
\begin{displaymath}
\psi_{\alpha}(P_i): \: ( P_i \times I ) |_{U_{\alpha}} \: \stackrel{ \sim }{
\longrightarrow } \: P_i |_{U_{\alpha}}
\end{displaymath}
($i \in \{ 1, 2 \}$)
constructed at the same time as the $P_i \times I$, in the gluing law for
objects, for $\{ U_{\alpha} \}$ an open cover of $U$.  Define morphisms
\begin{displaymath}
( \beta \times I ) |_{U_{\alpha}} : \: ( P_1 \times I ) |_{U_{\alpha}}
\: \longrightarrow \: ( P_2 \times I )|_{U_{\alpha}}
\end{displaymath}
by
\begin{displaymath}
( \beta \times I ) |_{U_{\alpha}} \: \equiv \:
( \psi_{\alpha}(P_2) )^{-1} \circ \beta |_{U_{\alpha}} \circ
\psi_{\alpha}(P_1)
\end{displaymath}
One can then use the gluing law for morphisms to glue together
the $ ( \beta \times I ) |_{ U_{\alpha}}$ to form a (unique)
morphism $\beta \times I: P_1 \times I \rightarrow P_2 \times I$.

It is straightforward to check that the map we have just defined,
namely $I(U): \beta \mapsto \beta \times I$, completes the definition
of a functor $I(U): {\cal C}(U) \rightarrow {\cal C}(U)$.

In passing, note that the maps $\psi_{\alpha}$ defined above,
define a natural transformation $\rho^* \circ I_{{\cal C}}(U)
\Rightarrow \rho^* \circ \mbox{Id}_{{\cal C}}(U)$, for
$\rho: U_{\alpha} \hookrightarrow U$ inclusion.

It is easy to check that there exist invertible natural
transformations
\begin{eqnarray*}
I^{-1}(U) \circ I(U) & \Longrightarrow & \mbox{Id}_{ {\cal C}(U) } \\
I(U) \circ I^{-1}(U) & \Longrightarrow & \mbox{Id}_{ {\cal C}(U) }
\end{eqnarray*}
where the $I^{-1}(U)$ are the functors associated to the dual torsors,
so each functor $I(U)$ defines a self-equivalence of the category
${\cal C}(U)$.

So far we have shown how an ${\cal A}|_U$-torsor $I$
(together with a specific choice of local sections of $I$) defines
an equivalence of categories $I(U): {\cal C}(U) \rightarrow {\cal C}(U)$.
We shall now use $I$ to define a Cartesian functor.

Let $I$ be an ${\cal A}$-torsor, let $\{ U_{\alpha} \}$ be a good
open cover of $X$, and let $\{ s_{\alpha} \}$ be a set of local
sections of $I$, defined with respect to $\{ U_{\alpha} \}$.
We have already demonstrated how to define a family of functors
$I(U): {\cal C}(U) \rightarrow {\cal C}(U)$, each defined
by the ${\cal A}|_U$-torsor $I|_U$ and local sections $\{ s_{\alpha} |_U \}$
of $I|_U$, defined with respect to the open cover $\{ U_{\alpha} \cap U \}$
of $U$.  In order to define a Cartesian functor $I: {\cal C} \rightarrow
{\cal C}$, it remains to define invertible natural transformations
$\chi_{\rho}: \rho^* \circ I(U) \Rightarrow I(V) \circ \rho^*$
for each inclusion $\rho: V \hookrightarrow U$, obeying the usual constraint.

Let $\rho: U_2 \hookrightarrow U_1$ be an inclusion of open sets,
and let $P \in \mbox{Ob } {\cal C}(U_1)$.
Let
\begin{eqnarray*}
\psi^1_{\alpha}(P): & (P \times I|_{U_1} ) |_{U_1 \cap U_{\alpha}} 
\: (\, = \, I(U_1)(P)|_{U_1 \cap U_{\alpha}} \, ) &
\longrightarrow \: P|_{ U_1 \cap U_{\alpha} } \\
\psi^2_{\alpha}(P): & (P|_{U_2} \times I|_{U_2} ) |_{U_2 \cap U_{\alpha} } 
\: (\, = \, I(U_2)(P|_{U_2})|_{U_2 \cap U_{\alpha}} \, ) &
\longrightarrow \: P|_{U_2} |_{U_2 \cap U_{\alpha} }
\end{eqnarray*}
be the isomorphisms appearing in the gluing law for objects.
Define $f_{\alpha}: I(U_1)(P)|_{U_2}|_{U_2 \cap U_{\alpha} } \rightarrow
I(U_2)(P|_{U_2})|_{U_2 \cap U_{\alpha}}$ by,
\begin{displaymath}
f_{\alpha} \: \equiv \: (\psi^2_{\alpha}(P))^{-1} \circ \varphi_{2, \alpha 2}
\circ \varphi^{-1}_{\alpha 1, \alpha 2} \circ \psi^1_{\alpha}(P) |_{U_2 \cap
U_{\alpha} } \circ \varphi_{\alpha 1, \alpha 2} \circ \varphi^{-1}_{
2, \alpha 2}
\end{displaymath}
where $\varphi$ are the natural transformations defining ${\cal C}$ as a
presheaf of categories, and we have implicitly used the notation
$U_{\alpha i} = U_i \cap U_{\alpha}$ ($i \in \{ 1, 2 \}$).
It is straightforward to check that the $f_{\alpha}$ satisfy the gluing
axiom for morphisms, and so there exists a unique morphism
\begin{displaymath}
\chi_{\rho}(P): \: I(U_1)(P)|_{U_2} \: \longrightarrow \:
I(U_2)(P|_{U_2})
\end{displaymath}
such that $\chi_{\rho}(P) |_{U_2 \cap U_{\alpha}} = f_{\alpha}$.
Moreover, it is straightforward to check that the $\chi_{\rho}$
define a natural transformation $\rho^* \circ I(U_1) \Rightarrow
I(U_2) \circ \rho^*$, and moreover that this set of natural transformations
makes diagram~(\ref{cartdef}) commute.

Thus, we have just defined the natural transformations needed to
describe $I: {\cal C} \rightarrow {\cal C}$ as a Cartesian functor.
Moreover, it should be clear that $I$ defines a map of gerbes, i.e.,
commutes with the action of the band.

The Cartesian functor $I: {\cal C} \rightarrow {\cal C}$ we defined above
depends explicitly upon a choice of local sections $\{ s_{\alpha} \}$
of $I$, with respect to some open cover $\{ U_{\alpha} \}$ of $X$.
How do two Cartesian functors defined by distinct choices of local
sections of the same ${\cal A}$-torsor $I$ differ?
It is straightforward to check that any two such 
Cartesian functors $I: {\cal C} \rightarrow {\cal C}$ differ by a 2-arrow.
In other words, if $I$ and $I'$ are two Cartesian functors ${\cal C}
\rightarrow {\cal C}$, both associated to the same ${\cal A}$-torsor
$I$ but differing in the choice of local sections, then there
exists an invertible 2-arrow $\eta: I \Rightarrow I'$.
The natural transformations over each open set $U$ are defined
by the morphisms $P \times I \rightarrow (P \times I)'$ we defined
earlier in section~\ref{gtobj}, in studying this same issue in the context
of gauge transformations of individual objects.

We should also mention an interesting special case of the
matter above.  Let $I^1_{{\cal C}}$ denote an automorphism of the
gerbe ${\cal C}$, associated to the ${\cal A}$-torsor $I$
and defined by local sections $\{ s^1_{\alpha} \}$ defined
over an open cover $\{ U^1_{\alpha} \}$ of $X$.  Let
$\{ U^2_i \}$ be a refinement of $\{ U^1_{\alpha} \}$,
and consider the automorphism $I^2_{{\cal C}}$ defined by
the local sections $\{ s^2_i \equiv s^1_{\alpha(i)}|_{U^2_i} \}$
of $I$, defined with respect to the cover $\{ U^2_i \}$.
It is straightforward to check that, in this special case,
the distinction between
$I^1_{{\cal C}}$ and $I^2_{{\cal C}}$ is identical to the ambiguity
in defining $I_{{\cal C}}$ on objects, for fixed choices of local sections.
Thus, for judicious choices in the definition of $I^1_{{\cal C}}$,
$I^1_{{\cal C}}$ is the same functor as $I^2_{{\cal C}}$.

Suppose $I^1$ and $I^2$ are a pair of ${\cal A}$-torsors.
Let $I^1_{{\cal C}}$ and $I^2_{{\cal C}}$ be corresponding
automorphisms of the gerbe ${\cal C}$.  We shall now show
that (under the correct circumstances) the gerbe automorphism
$( I^1 \cdot I^2 )_{{\cal C}}$ is identical to the automorphism
$I^1_{{\cal C}} \circ I^2_{{\cal C}}$.  Assume (without loss of
generality) that the local
sections defining $I^1_{{\cal C}}$ and $I^2_{{\cal C}}$
are defined with respect to the same open cover $\{ U_{\alpha} \}$
of $X$ (if not, restrict to a mutual refinement); let $\{ s^1_{\alpha} \}$
and $\{ s^2_{\alpha} \}$ denote the local sections of $I^1$, $I^2$,
respectively, defining $I^1_{{\cal C}}$ and $I^2_{{\cal C}}$.
Then it is straightforward to check that (with the usual
judicious choices) the gerbe automorphism $( I^1 \cdot I^2 )_{{\cal C}}$
defined with respect to the local sections\footnote{Our notation
is slightly sloppy, in that ${\cal A}(U_{\alpha})$ is an abelian
group, not a ring, and $I^i(U_{\alpha})$ is a set, not a module.}
$\{ s^1_{\alpha} \otimes_{{\cal A}(U_{\alpha})}
s^2_{\alpha} \}$ of the torsor $I^1 \cdot I^2$ is identical
to the composition $I^1_{{\cal C}} \circ I^2_{{\cal C}}$,
i.e.,
\begin{displaymath}
\left( \, I^1 \cdot I^2 \, \right)_{{\cal C}} \: = \: I^1_{{\cal C}} \circ
I^2_{{\cal C}}
\end{displaymath}

Suppose $I^1$ and $I^2$ are a pair of ${\cal A}$-torsors.
We shall show here that a torsor isomorphism
$\omega: I^1 \rightarrow I^2$ is equivalent to a 2-arrow
$\overline{\omega}: I^1_{{\cal C}} \Rightarrow I^2_{{\cal C}}$.
First, we shall define some notation.  Assume $\{ s^1_{\alpha} \}$
and $\{ s^2_{\alpha} \}$ are local sections of $I^1$, $I^2$ defining
the gerbe automorphisms $I^1_{{\cal C}}$, $I^2_{{\cal C}}$, assumed
to both\footnote{If they are not both defined over the same open
cover, then restrict to a mutual refinement.} 
be defined over an open cover $\{ U_{\alpha} \}$ of $X$.
Assume first that we are given $\omega: I^1 \rightarrow I^2$;
we shall describe how to construct $\overline{\omega}:
I^1_{{\cal C}} \Rightarrow I^2_{{\cal C}}$.
Let $U$ be an open set, and denote the natural transformations associated
to $I^1_{{\cal C}}$, $I^2_{{\cal C}}$ by $\psi_{\alpha}$, i.e.,
\begin{displaymath}
\psi^i_{\alpha}(P): \: I^i_{{\cal C}}(U)(P)|_{U \cap U_{\alpha}}
\: \longrightarrow \: P|_{U \cap U_{\alpha}}
\end{displaymath}
for $P \in \mbox{Ob } {\cal C}(U)$, and $i \in \{ 1, 2 \}$.
Then we define 
\begin{displaymath}
\overline{\omega}_{\alpha}(U)(P): \: I^1_{{\cal C}}(U)(P)|_{U \cap U_{\alpha} }
 \: \longrightarrow
\: I^2_{{\cal C}}(U)(P)|_{U \cap U_{\alpha}}
\end{displaymath}
by,
\begin{displaymath}
\overline{\omega}_{\alpha}(U)(P) \: \equiv \: \left( \psi^2_{\alpha}(P) 
\right)^{-1} \circ \left( s^2_{\alpha} \, - \, \omega(s^1_{\alpha}) \right)
\circ \psi^1_{\alpha}(P)
\end{displaymath}
It is easy to check that the $\overline{\omega}_{\alpha}(U)(P)$ satisfy
the gluing axiom for morphisms, and so can be glued together to form
a unique morphism
\begin{displaymath}
\overline{\omega}(U)(P): \: I^1_{{\cal C}}(U)(P) \: \longrightarrow \:
I^2_{{\cal C}}(U)(P)
\end{displaymath}
whose restriction to $U \cap U_{\alpha}$ is given by
$\overline{\omega}_{\alpha}(U)(P)$.  Moreover, it is straightforward to
check that $\overline{\omega}(U): I^1_{{\cal C}}(U) \Rightarrow
I^2_{{\cal C}}(U)$ is a natural transformation, and finally that
$\overline{\omega}: I^1_{{\cal C}} \Rightarrow I^2_{{\cal C}}$ is a
2-arrow.  Thus, given a torsor isomorphism $\omega: I^1 \rightarrow I^2$,
we have constructed a 2-arrow $\overline{\omega}: I^1_{{\cal C}}
\Rightarrow I^2_{{\cal C}}$.

Conversely, given a 2-arrow $\overline{\omega}: I^1_{{\cal C}}
\Rightarrow I^2_{{\cal C}}$, we shall now construct a torsor isomorphism
$\omega: I^1 \rightarrow I^2$.  We shall define $\omega$ by describing
the action of $\omega$ on the local sections $\{ s^1_{\alpha} \}$ of
$I^1_{{\cal C}}$.  Using the same conventions as in the paragraph above,
for any open $U$ and $P \in \mbox{Ob } {\cal C}(U)$, define
$g(U \cap U_{\alpha} )(P)$ to be the element of the abelian group
${\cal A}(U \cap U_{\alpha})$ associated to the automorphism
\begin{displaymath}
\psi^2_{\alpha}(P) \circ \overline{\omega}(U)(P) \circ
\left( \psi^1_{\alpha}(P) \right)^{-1}: \: P|_{U_{\alpha}}
\: \longrightarrow \: P|_{U_{\alpha}}
\end{displaymath}
Assume without loss of generality\footnote{If not, restrict to a suitable
refinement.} that $\{ U_{\alpha} \}$ is a good cover, and so each
category ${\cal C}(U_{\alpha})$ contains a single isomorphism class of
objects\footnote{We shall prove later that isomorphism classes of
objects in each category ${\cal C}(U)$ are in one-to-one correspondence
with elements of $H^1(U, {\cal A}|_U)$.}.  Then, for $U = U_{\alpha}$,
the band element $g(U_{\alpha})$ defined above is independent of the
choice of $P$.  Define
\begin{displaymath}
\omega(s^1_{\alpha}) \: \equiv \: g(U_{\alpha})^{-1} \cdot s^2_{\alpha}
\end{displaymath}
To define a torsor isomorphism, it suffices to describe the action
on a set of local sections associated to a cover, which we have just
done.  Thus, we have just associated a torsor isomorphism $\omega: I^1 
\rightarrow I^2$ to the 2-arrow $\overline{\omega}: I^1_{{\cal C}}
\Rightarrow I^2_{{\cal C}}$.  To summarize the results of this paragraph
and the last, a torsor isomorphism $\omega: I^1 \rightarrow I^2$ is
equivalent to a 2-arrow $\overline{\omega}: I^1_{{\cal C}} \Rightarrow
I^2_{{\cal C}}$.

As a consequence of the results in the last paragraph,
we see that isomorphic torsors define isomorphic gerbe automorphisms.
Thus, distinct gerbe automorphisms are defined by equivalence classes
of torsors, not individual torsors {\it per se}.
This fact will be quite important when discussing how the
group cohomology group $H^2(\Gamma, U(1))$ appears when describing
equivariant structures on $B$ fields -- modding out by group coboundaries
ultimately comes from the fact that only equivalence classes of torsors
define distinct gerbe automorphisms.

In passing, we should also speak briefly on pullbacks, and how
the pullback of an automorphism associated to a torsor $I$ is
related to the automorphism associated to the pullback of the torsor.
Let $f: X \rightarrow Y$ be a continuous map, and let ${\cal C}$
be a gerbe on $Y$, with band ${\cal A}$.  Let $I$ be an ${\cal A}$-torsor,
and let $I_{{\cal C}}$ denote an associated automorphism of ${\cal C}$
(defined with respect to local sections $\{ s_{\alpha} \}$,
over an open cover $\{ U_{\alpha} \}$ of $Y$).
Let $(f^* I)_{{\cal C}}$ denote the automorphism of the gerbe
$f^* {\cal C}$ defined by the $f^* {\cal A}$-torsor $f^* I$,
with respect to the same open sections $\{ s_{\alpha} \}$ of
$f^* I$, over the open cover $\{ f^{-1}( U_{\alpha} ) \}$ of $X$.
It is straightforward to check that, at least in the case that
$f$ is a homeomorphism (so that $f^{-1} {\cal C} \cong f^* {\cal C}$
and $f^{-1} I \cong f^* I$), the gerbe automorphisms
$f^*( I_{{\cal C}} )$ and $( f^* I )_{{\cal C}}$ coincide\footnote{In
fact, we are being slightly sloppy.  In defining the action of $I$,
recall there was a minor ambiguity in the definition of the action
on the objects of ${\cal C}$.  Distinct choices differ by unique
morphisms, so we assumed such choices were made, and thereafter
ignored the point.  Essentially the same problem arises here.
The correct statement is that $f^*( I_{{\cal C}} )$ and
$(f^* I)_{{\cal C}}$ differ by at most a unique 2-arrow;
but by making judicious choices, they can be assumed to coincide.}.

\subsubsection{Action on connections}   \label{gtconnec}

In this section we shall describe how gauge transformations
of gerbes act on connective structures and curvings.
We briefly touched on these matters in section~\ref{gerbeconnecintro};
here we re-examine them in more detail.

To fix notation, let ${\cal C}$ be a gerbe with
connective structure $\mbox{Co}$.
let $U \subseteq X$ be open, and $P \in \mbox{Ob } {\cal C}(U)$.
Let $I$ be an ${\cal A}|_U$-torsor, let $\{ U_{\alpha} \}$
be a good open cover of $U$, and let $\{ s_{\alpha} \}$
be a set of local sections of $I$, defined with respect to
$\{ U_{\alpha} \}$.  Let $I_{{\cal C}}: {\cal C}|_U
\rightarrow {\cal C}|_U$ denote the gerbe automorphism corresponding
to $I$ with sections $\{ s_{\alpha} \}$.

Consider the action of $I_{{\cal C}}$ on $P$.  By definition
of $P \times I = I_{{\cal C}}(U)(P)$, we have morphisms
$\psi_{\alpha}(P): (P \times I)|_{U_{\alpha}}$ such that the
following diagram commutes:
\begin{displaymath}
\begin{array}{ccccccc}
(P \times I)|_{U_{\beta}} |_{U_{\alpha \beta}} & \multicolumn{2}{c}{ \stackrel{
\varphi_{\beta, \alpha \beta} }{ \longleftarrow } } &
(P \times I)|_{U_{\alpha \beta}} & \multicolumn{2}{c}{
\stackrel{ \varphi_{\alpha, \alpha \beta} }{ \longrightarrow } }
& (P \times I)|_{U_{\alpha}}|_{U_{\alpha \beta}} \\
\makebox[0pt][r]{ $\scriptstyle{ \psi_{\beta}(P)|_{U_{\alpha \beta}} }$ }
\downarrow & & & & & & \downarrow
\makebox[0pt][l]{ $\scriptstyle{ \psi_{\alpha}(P)|_{U_{\alpha \beta}} }$ } \\
P|_{U_{\beta}} |_{U_{\alpha \beta}} &
\stackrel{ \varphi_{\beta, \alpha \beta} }{ \longleftarrow } &
P|_{U_{\alpha \beta}} & \stackrel{ s_{\alpha} - s_{\beta} }{ \longrightarrow }
& P|_{U_{\alpha \beta}} & \stackrel{ \varphi_{\alpha, \alpha \beta} }{
\longrightarrow } & P|_{U_{\alpha}}|_{U_{\alpha \beta}}
\end{array}
\end{displaymath}
where the $\varphi$ are the natural transformations defining ${\cal C}$
as a presheaf of categories, and $s_{\alpha} - s_{\beta}$ denotes
the band element mapping $s_{\beta} \rightarrow s_{\alpha}$.

By applying the Cartesian functor $\mbox{Co}$, we can (with a bit of
work) recover the following commutative diagram:
\begin{equation}  \label{Cogauge}
\begin{array}{ccccccc}
\mbox{Co}(P \times I) |_{U_{\beta}} |_{U_{\alpha \beta}} &
\multicolumn{2}{c}{ \stackrel{ \varphi_{\beta, \alpha \beta} }{ \longleftarrow}
} & \mbox{Co}(P \times I) |_{U_{\alpha \beta}} &
\multicolumn{2}{c}{ \stackrel{ \varphi_{\alpha, \alpha \beta} }{ 
\longrightarrow } } &
\mbox{Co}(P \times I) |_{U_{\alpha}} |_{U_{\alpha \beta}} \\
\downarrow & & & & & & \downarrow \\
\mbox{Co}(P) |_{U_{\beta}} |_{U_{\alpha \beta}} &
\stackrel{ \varphi_{\beta, \alpha \beta} }{ \longleftarrow } &
\mbox{Co}(P) |_{U_{\alpha \beta}} &
\stackrel{ - d \, {\it ln} \: (s_{\alpha} - s_{\beta} ) }{ 
\longrightarrow } &
\mbox{Co}(P)|_{U_{\alpha \beta}} &
\stackrel{ \varphi_{\alpha, \alpha \beta} }{ \longrightarrow } &
\mbox{Co}(P)|_{U_{\alpha}} |_{U_{\alpha \beta}}
\end{array}
\end{equation}
where we have abbreviated $\mbox{Co}(U)(P)$ by $\mbox{Co}(P)$, for example,
and where the map
\begin{displaymath}
\mbox{Co}(U)(P \times I)|_{U_{\alpha}} |_{U_{\alpha \beta}} \: \longrightarrow
\: \mbox{Co}(U)(P) |_{U_{\alpha}} |_{U_{\alpha \beta}}
\end{displaymath}
is given by the composition
\begin{displaymath}
(\chi_{\alpha})^{-1} \circ ( \chi_{\alpha \beta} )^{-1} \circ
\mbox{Co}(U_{\alpha \beta})(\psi_{\alpha}(P)|_{U_{\alpha \beta}}) \circ
\chi_{\alpha \beta} \circ \chi_{\alpha}
\end{displaymath}
The $\chi$ are the natural transformations defining $\mbox{Co}$ as a 
Cartesian functor.

From diagram~(\ref{Cogauge}), it should be clear that
$\mbox{Co}(U)(P \times I)$ and $\mbox{Co}(U)(P)$ differ by the
$\Omega^1|_U$-torsor of connections on $I$.  Moreover, 
a precise specification of how a particular section $\nabla
\in \Gamma(U, \mbox{Co}(U)(P))$ is mapped is determined by the
local sections $\{ s_{\alpha} \}$ determining $I_{{\cal C}}$,
and is equivalent to a choice of connection on $I$.
(Compare, for example, \cite[section 5.3, equ'n~(5-11)]{brylinski}.)

In other words, let $\{ A^{\alpha} \}$ be a family of 1-forms
on $\{ U_{\alpha} \}$ defining a connection on $I$.  Assume
the action of $I_{{\cal C}}$ on sections of $\mbox{Co}(U)(P)$
is determined by $\{ A^{\alpha} \}$.
Then for any $\nabla_{\alpha} \in \Gamma(U_{\alpha},
\mbox{Co}(U_{\alpha})(P|_{U_{\alpha}})$, 
$\nabla_{\alpha} \mapsto \nabla_{\alpha} + A^{\alpha}$
under the action of $I_{{\cal C}}$.

Phrased more simply still, a gauge transformation of a gerbe
with connective structure is defined by an equivalence class of
bundles with connection.  (After all, a 2-arrow between gerbe automorphisms
lifts to a ``connection-preserving'' map between the connective structures,
as noted earlier.)

How does $I_{{\cal C}}$ act on a curving $K$ on $( {\cal C},
\mbox{Co})$?  The answer should be immediately clear from the
definition of curving:  
\begin{displaymath}
K(\nabla_{\alpha} + A^{\alpha})
\: = \: K(\nabla_{\alpha}) \: + \: d A^{\alpha}
\end{displaymath}
Much earlier in this paper we remarked that an automorphism of a gerbe
with connection is defined by an equivalence class of bundles
with {\it flat} connection.  Here, we can see that more explicitly.
If the connection is flat, then the curving $K$ is invariant under
the gauge transformation.  Put another way, 
just as constant gauge transformations
define bundle automorphisms that preserve the connection,
a gauge transformation of a gerbe that preserves the connection
is an equivalence class of bundles with flat connection.

\subsubsection{Gauge transformations commute with gerbe maps}

Let $F: {\cal C} \rightarrow {\cal D}$ be a map between gerbes
${\cal C}$, ${\cal D}$, that is, a Cartesian functor commuting
with the action of the band.  Assume ${\cal C}$ and ${\cal D}$
both have band ${\cal A}$, and are both defined over a space $X$.

In this section, we shall show that if $I$ is any ${\cal A}$-torsor,
then there exists an invertible 2-arrow 
\begin{displaymath}
\kappa: \: F \circ I_{{\cal C}} \: \Longrightarrow \:
I_{{\cal D}} \circ F
\end{displaymath}
where $I_{{\cal C}}: {\cal C} \rightarrow {\cal C}$ and
$I_{{\cal D}}: {\cal D} \rightarrow {\cal D}$ are gerbe maps
induced by the ${\cal A}$-torsor $I$.

In fact, this is quite straightforward.
Without loss of generality\footnote{As the gerbe automorphisms defined
by distinct choices of local sections of $I$ differ by an invertible
2-arrow, clearly we are free to choose any convenient sets of local sections
without changing the result.}, assume that the local sections of $I$
defining $I_{{\cal C}}$ and $I_{{\cal D}}$ are identical.
Let $\{ U_{\alpha} \}$ be a good open cover of $X$, and let
$\{ s_{\alpha} \}$ be the local sections of $I$ over $\{ U_{\alpha} \}$
which define $I_{{\cal C}}$ and $I_{{\cal D}}$.
Let
\begin{eqnarray*}
\psi^{{\cal C}}_{\alpha}(P): & I_{{\cal C}}(U)(P)|_{ U \cap U_{\alpha} }
& \longrightarrow \: P|_{ U \cap U_{\alpha} } \: \:
\mbox{ for } P \in \mbox{Ob } {\cal C}(U) \\
\psi^{{\cal D}}_{\alpha}(P'): & I_{{\cal D}}(U)(P')|_{ U \cap U_{\alpha} }
& \longrightarrow \: P'|_{ U \cap U_{\alpha} } \: \:
\mbox{ for } P' \in \mbox{Ob } {\cal D}(U) 
\end{eqnarray*}
(for open $U \subseteq X$)
denote the isomorphisms obtained in the definition of gauge transformation.
Define 
\begin{displaymath}
\kappa_{\alpha}(U)(P): \: ( F \circ I_{{\cal C}} )(U)(P) |_{ U \cap U_{\alpha} }
\: \longrightarrow \: ( I_{{\cal D}} \circ F )(U)(P) |_{ U \cap U_{\alpha} }
\end{displaymath}
for open $U \subseteq X$ and $P \in \mbox{Ob }{\cal C}(U)$ by,
\begin{displaymath}
\kappa_{\alpha}(U)(P) \: \equiv \: \psi_{\alpha}^{{\cal D}} \left(
F(U)(P) \right)^{-1} \circ \left( \chi_{\alpha}^F \right)^{-1} \circ 
F(U \cap U_{\alpha} )\left( \psi^{{\cal C}}_{\alpha}(P) \right)
\circ \chi_{\alpha}^F
\end{displaymath}
where the $\chi^F$ are the natural transformations defining $F$ as a 
Cartesian functor.
It is straightforward to check that the $\kappa_{\alpha}(U)(P)$ satisfy
the gluing axiom for morphisms, so they can be glued together to form
a unique morphism 
\begin{displaymath}
\kappa(U)(P): \: (F \circ I_{{\cal C}} )(U)(P) \: \longrightarrow
\: ( I_{{\cal D}} \circ F )(U)(P)
\end{displaymath}
such that $\kappa(U)(P) |_{ U \cap U_{\alpha} } = \kappa_{\alpha}(U)(P)$.

It is straightforward to check that $\kappa(U)$ defines a natural
transformation $(F \circ I_{{\cal C}})(U) \Rightarrow (I_{{\cal D}} 
\circ F)(U)$,
and moreover that $\kappa$ defines an invertible 2-arrow
$F \circ I_{{\cal C}} \Rightarrow I_{{\cal D}} \circ F$.

Thus, loosely speaking, gauge transformations commute with gerbe maps.

\subsubsection{Sheaves of morphisms as torsors}   \label{morphtors}

If $P_1$ and $P_2$ are any two objects of ${\cal C}(U)$,
then the sheaf of (iso)morphisms $\underline{\mbox{Hom}}_U(P_1, P_2)$ is
an ${\cal A} |_U$-torsor\footnote{Those readers also studying
\cite[section 5]{brylinski} will note that in that reference,
$\underline{\mbox{Hom}}(P_1, P_2)$ is instead denoted 
$\underline{\mbox{Isom}}(P_1, P_2)$.}.  
In particular, ${\cal A}|_U \cong \underline{\mbox{Aut}}(P_1)$
acts on the left, and it should be clear that this action is
free and transitive -- for any open $V \subseteq U$, the set
$\mbox{Hom}(P_1|_V, P_2|_V)$ is either empty (and so trivially an
${\cal A}(V)$-torsor), or is nonempty and is manifestly a torsor
under the group
${\cal A}(V)$.  Thus, as a sheaf of abelian groups, ${\cal A} |_U$ acts
freely and transitively on the sheaf $\underline{\mbox{Hom}}_U(P_1, P_2)$,
and moreover its action commutes with restriction maps, 
so $\underline{\mbox{Hom}}_U(P_1,P_2)$ is an ${\cal A}|_U$-torsor.

One can easily check that if $F: {\cal C} \rightarrow {\cal D}$ is
a map of gerbes of band ${\cal A}$ over the same space $X$, 
then for any open $U$
and for any two objects $P_a, P_b \in \mbox{Ob } {\cal C}(U)$,
the induced map
\begin{displaymath}
\underline{ \mbox{Hom} }_U(P_a, P_b) \: \longrightarrow \:
\underline{ \mbox{Hom} }_U( F(U)(P_a), F(U)(P_b) )
\end{displaymath}
is a morphism of torsors.
As any morphism of ${\cal A}$-torsors is an isomorphism,
this means that the ${\cal A}$-torsors $\underline{\mbox{Hom}}_U
(P_a, P_b)$ and $\underline{\mbox{Hom}}_U( F(U)(P_a), F(U)(P_b))$
are isomorphic.

It is straightforward to check that the torsor 
$\underline{\mbox{Hom}}_U(P_1, P_2)$
induces the gauge transformation on ${\cal C}(U)$ that
maps the object $P_1$ to the object $P_2$, as the reader
has probably guessed.

Moreover, it is straightforward to see that
$\underline{\mbox{Hom}}_U(P, P \times I) \cong I$ as ${\cal A}|_U$-torsors,
for any ${\cal A}|_U$-torsor $I$.
Let $\{ U_{\alpha} \}$ be a good open cover of $U$,
and let $\{ s_{\alpha} \in I(U_{\alpha}) \}$ be the set
of local sections of $I$ defining the map $P \mapsto P \times I$.
Recall that we constructed $P \times I$ using the gluing law for
objects, and the same gluing law also yields a set of
isomorphisms
\begin{displaymath}
\psi_{\alpha}(P): \: (P \times I) |_{U_{\alpha}} \: \longrightarrow \:
P |_{U_{\alpha}}
\end{displaymath}
We can define an isomorphism of ${\cal A} |_U$-torsors
by specifying that the sections $\{ s_{\alpha} \}$ should
map to the isomorphisms $\{ \psi^{-1}_{\alpha} \}$.
As mentioned in the section on torsors,
to define an isomorphism between two torsors, it suffices to describe
how a set of local sections (defined with respect to an open cover)
are mapped, thus we have now defined an isomorphism 
$I \stackrel{ \sim }{\longrightarrow} \underline{\mbox{Hom}}_U(P,
P \times I)$ of ${\cal A}|_U$-torsors.

One implication of the fact that $\underline{\mbox{Hom}}_U(P, P \times I)
\cong I$ is that the action of
$I$ is free on equivalence classes of objects in ${\cal C}(U)$:
$P \times I$ is isomorphic to $P$ if and only if
$I$ is trivializable, i.e., only if $I$ has a global section does there
exist a morphism $P \rightarrow P \times I$, not just between their
restrictions to open subsets.

Moreover, the action of torsors on objects is not only free, but
transitive:  any two objects $P_1, P_2 \in \mbox{Ob } {\cal C}(U)$ 
can be related by
a set of local sections of the ${\cal A}|_U$-torsor 
$\underline{\mbox{Hom}}_U(P_1,
P_2)$.

Thus, the set of equivalence classes of objects of ${\cal C}(U)$
is a torsor under the action of the abelian group $H^1(U, {\cal A} |_U)$.

This means that there exists a (noncanonical) one-to-one correspondence between
equivalence classes of objects of ${\cal C}(U)$ and
equivalence classes of ${\cal A}|_U$-torsors.
In the case that ${\cal A} = C^{\infty}(G)$ for some abelian $G$,
we can rephrase this by saying that equivalence classes of objects
of ${\cal C}(U)$, for any gerbe ${\cal C}$ with band ${\cal A}$,
are in one-to-one correspondence with equivalence classes of
principal $G$-bundles on $U$.
This implies that all gerbes with band ${\cal A} = C^{\infty}(G)$
look locally like the trivial gerbe $\mbox{Tors}(G)$,
just as all bundles look locally like a trivial bundle.

This last result gives a great deal of insight into gerbes,
and so is worth repeating.  Just as all bundles can be locally
trivialized, all gerbes can be locally trivialized.
Gerbes with band ${\cal A} = C^{\infty}(G)$ for abelian $G$
are locally isomorphic to the stack of principal $G$-bundles,
a trivial gerbe. 

We shall conclude this section by explicitly demonstrating such
local trivializations.  Local trivializations are not needed
to define a sheaf -- however, as they can be used to give insight
into local structure, we shall work them out explicitly.
More precisely, for any open set $U$ such that ${\cal C}(U) \neq
\emptyset$, we shall construct functors
\begin{displaymath}
\begin{array}{rccc}
F(U): & {\cal C} & \longrightarrow & \mbox{Tors}({\cal A})(U) \\
F^*(U): & \mbox{Tors}({\cal A})(U) & \longrightarrow & {\cal C}(U)
\end{array}
\end{displaymath}
such that there exist invertible natural transformations
$F^*(U) \circ F(U) \Rightarrow \mbox{Id}$ and 
$F(U) \circ F^*(U) \Rightarrow \mbox{Id}$.

Fix an object $P_0 \in \mbox{Ob } {\cal C}(U)$,
and a torsor $I_0 \in \mbox{Ob } \mbox{Tors}({\cal A})(U)$.
We define the functor $F(U)$ as follows.
Let $P \in \mbox{Ob } {\cal C}(U)$ be any object,
and define the ${\cal A}|_U$-torsor 
\begin{displaymath}
I \: \equiv \: \underline{ \mbox{Hom} }_U( P_0, P)
\end{displaymath}
Define 
\begin{displaymath}
F(U)(P) \: \equiv \: I \cdot I_0
\end{displaymath}
Next, let $P_a, P_b \in \mbox{Ob } {\cal C}(U)$ be objects,
and $\beta: P_a \rightarrow P_b$ be a morphism.
Define the ${\cal A}|_U$-torsors
\begin{eqnarray*}
I^a & \equiv & \underline{ \mbox{Hom} }_U(P_0, P_a) \\
I^b & \equiv & \underline{ \mbox{Hom} }_U(P_0, P_b)
\end{eqnarray*}
The morphism $\beta$ induces a morphism of torsors
\begin{displaymath}
\beta^{\#}: \: I^a \: \longrightarrow \: I^b
\end{displaymath}
and so we define
\begin{displaymath}
F(U)(\beta) \: \equiv \: \beta^{\#}: \: I^a \cdot I_0
\: \longrightarrow \: I^b \cdot I_0
\end{displaymath}
With these definitions, $F(U)$ is a well-defined functor
from ${\cal C}(U)$ to $\mbox{Tors}({\cal A})(U)$.

Next, we shall define the functor $F^*(U)$.
Let $I$ be a ${\cal A}|_U$-torsor in $\mbox{Ob } \mbox{Tors}({\cal A})(U)$,
and define
\begin{displaymath}
F^*(U)(I) \: \equiv \: \left( I \cdot I_0^{-1} \right)_{{\cal C}}(U)(P_0)
\end{displaymath}
Next, let $I^a$, $I^b$ be ${\cal A}|_U$-torsors in $\mbox{Ob }
\mbox{Tors}({\cal A})(U)$, and let $\beta: I^a \rightarrow I^b$
be a morphism of ${\cal A}|_U$-torsors.  The morphism $\beta$
defines a 2-arrow
\begin{displaymath}
\beta^*: \: \left( I^a \cdot I_0^{-1} \right)_{{\cal C}} \:
\Longrightarrow \: \left( I^b \cdot I_0^{-1} \right)_{{\cal C}}
\end{displaymath}
Define
\begin{displaymath}
F^*(U)(\beta) \: \equiv \: \beta^*(U)(P_0)
\end{displaymath}
With these definitions, $F^*(U)$ is a well-defined functor
from $\mbox{Tors}({\cal A})(U)$ to ${\cal C}(U)$.

Finally, it is straightforward to check that there exist
invertible natural transformations 
\begin{eqnarray*}
F^*(U) \circ F(U) & \Longrightarrow & \mbox{Id}_{ {\cal C}(U) } \\
F(U) \circ F^*(U) & \Longrightarrow & \mbox{Id}_{ {\it Tors}({\cal A})(U) }
\end{eqnarray*}

Thus, we have now finished demonstrating explicitly that for
any $U$ such that ${\cal C}(U) \neq \emptyset$,
the category ${\cal C}(U)$ is equivalent to the category
$\mbox{Tors}({\cal A})(U)$.

\subsubsection{Sheaves of natural transformations as torsors}

We shall now show that sheaves of natural transformations 
are ${\cal A}$-torsors.
More specifically,
let $F, G: {\cal C} \rightarrow {\cal D}$ be a pair of Cartesian
functors defining maps of gerbes, between the gerbes ${\cal C}$ and
${\cal D}$, assumed to both have band ${\cal A}$ and both be defined over
the same space $X$.  We shall show that the sheaf of
local 2-arrows $\underline{2R}_U(F,G)$  
is an ${\cal A}|_U$-torsor, for any open $U \subseteq X$.
We shall also derive some useful properties of these torsors.

We shall now argue that the sheaf of sets $\underline{2R}_U(F,G)$
is an ${\cal A}|_U$-torsor.
We define the action of ${\cal A}|_U$ as follows.
Let $V \subseteq U$ be open, and let $g \in {\cal A}(V)$.
Let $\psi$ be an element of the set $\underline{2R}_U(F,G)(V)$,
which is to say, $\psi$ is a collection of natural transformations
$F(W) \Rightarrow G(W)$, one for each open $W \subseteq V$.
The element $g \in {\cal A}(V)$ 
acts on each natural transformation $\psi(W)$ as,
\begin{displaymath}
\psi(W) \: \mapsto \: \psi(W) \circ ( \rho^*_W g ) 
\end{displaymath}
where $\rho^*$ is the restriction map in the sheaf ${\cal A}$,
and for each $P \in \mbox{Ob }{\cal C}(W)$, $\rho^* g$ is
interpreted as an element of $\mbox{Aut}( F(W)(P) )$.

It is easy to see that this action of ${\cal A}$ is free,
and commutes with the restriction map.
To check that it is transitive requires more work, which we
shall outline here.  Let $\psi_a$, $\psi_b$ be two elements of
the set $\underline{2R}_U(F,G)(V)$.  For any $P \in \mbox{Ob }
{\cal C}(V)$, $\psi_b(V)(P)^{-1} \circ \psi_a(V)(P)$ is an
automorphism of $F(V)(P)$, which we can identify with an element
$g(P) \in {\cal A}(V)$.  Any two objects $P_1$, $P_2$ which
are related by a morphism will clearly define isomorphic
group elements: $g(P_1) = g(P_2)$.  More generally, one
can show that $g(P_1) = g(P_2)$ even if $P_1$, $P_2$ are not
related by a morphism, by restricting to an open subset $W \subseteq V$
such that there exists a morphism $\beta: P_1 |_W \rightarrow P_2 |_W$.
Define a morphism 
\begin{displaymath}
\lambda: \: F(V)(P_1)|_W \: \longrightarrow \: F(V)(P_2)|_W
\end{displaymath}
by,
\begin{displaymath}
\lambda \equiv ( \chi^F_W )^{-1} \circ F(W)(\beta) \circ \chi^F_W
\end{displaymath}
then it is straightforward to check that
\begin{displaymath}
\left( \psi_b(V)(P_2)^{-1} \circ \psi_a(V)(P_2) \right) |_W \: = \:
\lambda \circ \left( \psi_b(V)(P_1)^{-1} \circ \psi_a(V)(P_1) \right)|_W
\circ \lambda^{-1}
\end{displaymath}
which implies that $g(P_1) = g(P_2)$.
Thus, for any pair of 2-arrows $\psi_a, \psi_b \in
\underline{2R}_U(F,G)(V)$, there exists $g \in {\cal A}(V)$ such that
$\psi_a = \psi_b \circ g$.  This means that the action of ${\cal A}|_U$ on
the sheaf of sets $\underline{2R}_U(F,G)$ is transitive, and so we
have finished demonstrating that $\underline{2R}_U(F,G)$ is an
${\cal A}|_U$-torsor.

We shall now argue that if $V$ is an open subset of $U$ such that
the categories ${\cal C}(V)$ and ${\cal D}(V)$ each only contain
a single isomorphism class of objects, then the set of
local 2-arrows $\underline{2R}_U(F,G)(V)$ is nonempty.
First, we shall construct natural transformations $F(V) \Rightarrow
G(V)$.  Fix some $P_0 \in \mbox{Ob } {\cal C}(V)$, and
a morphism $\psi(V)(P_0): F(V)(P_0) \rightarrow G(V)(P_0)$.
For any other $P \in \mbox{Ob } {\cal C}(V)$, let $\beta: P \rightarrow
P_0$ be a morphism, and define $\psi(V)(P): F(V)(P) \rightarrow G(V)(P)$
by, $\psi(V)(P) \equiv G(V)(\beta)^{-1} \circ \psi(V)(P_0) \circ
F(V)(\beta)$.  It is straightforward to check that
$\psi(V)(P)$ is independent of the choice\footnote{In other words,
if $\psi_{\alpha}(V)(P)$ and $\psi_{\beta}(V)(P)$ denote the
two isomorphisms defined by any two morphisms $\alpha, \beta: P \rightarrow
P_0$, then $\psi_{\alpha}(V)(P) = \psi_{\beta}(V)(P)$.} of $\beta$,
and moreover that this defines a natural transformation $\psi(V):
F(V) \Rightarrow G(V)$.  In fact, we can generate an entire
${\cal A}(V)$-torsor of natural transformations in this fashion,
by choosing different $P_0$ and different $\psi(V)(P_0)$.
In passing, note that this also implies that a natural transformation
$F(V) \circ |_V \Rightarrow G(V) \circ |_V$, for example,
uniquely determines a natural transformation $F(V) \Rightarrow G(V)$
(assuming that, for $|_V$ denoting a restriction functor associated
to the inclusion $V \hookrightarrow U$, ${\cal C}(U)$ and ${\cal D}(U)$
are both nonempty).

So far we have constructed a natural transformation $\psi(V): F(V)
\Rightarrow G(V)$; we shall now show that any such natural transformation
defines a local 2-arrow $F|_V \Rightarrow G|_V$, i.e., an element of
$\underline{2R}_U(F,G)(V)$.
In order to construct an element of $\underline{2R}_U(F,G)(V)$,
we need to specify (suitably compatible) natural transformations
$F(W) \Rightarrow G(W)$ for all open $W \subseteq V$.  For any
$\psi(V): F(V) \Rightarrow G(V)$ and any open $W \subseteq V$,
define $\psi(W)$ to be the natural transformation generated by
the composition
\begin{displaymath}
F(W) \circ \rho^* \: \stackrel{ \chi^F_{\rho} }{ \Longleftarrow }
\: \rho^* \circ F(V) \:
\stackrel{ \psi(V) }{ \Longrightarrow }
\: \rho^* \circ G(V) \: \stackrel{ \chi^G_{\rho} }{ \Longrightarrow }
\: G(W) \circ \rho^*
\end{displaymath}
where $\rho: W \hookrightarrow V$ is inclusion.  The composition above
only defines a natural transformation $F(W) \circ \rho^* \Rightarrow G(W) \circ
\rho^*$, not $F(W) \Rightarrow G(W)$;
however using the ideas in the previous paragraph, it should be clear
that the composition can be used to
determine a natural transformation $F(W) \Rightarrow G(W)$ (uniquely,
in fact, if every object of ${\cal D}(W)$ is isomorphic to the restriction
of an object of ${\cal D}(V)$).
Moreover, it is straightforward to check that this collection of
natural transformations has the property that for any inclusion of
open sets $\rho: W_2 \hookrightarrow W_1$, the following
diagram commutes:
\begin{equation}
\begin{array}{ccc}
\rho^* \circ F(W_1) & \stackrel{ \psi(W_1) }{ \Longrightarrow } &
\rho^* \circ G(W_1) \\
\makebox[0pt][r]{ $\scriptstyle{ \chi^F_{\rho} }$ } \Downarrow & &
\Downarrow \makebox[0pt][l]{ $\scriptstyle{ \chi^G_{\rho} }$ } \\
F(W_2) \circ \rho^* & \stackrel{ \psi(W_2) }{ \Longrightarrow } &
G(W_2) \circ \rho^*
\end{array}
\end{equation}
Thus, given any natural transformation $\psi(V): F(V) \Rightarrow G(V)$,
we can construct a local 2-arrow $\psi: F|_V \Rightarrow G|_V$.
In other words, we have constructed an element of $\underline{2R}_U(F,G)(V)$,
and so if $V \subseteq U$ is an open set such that ${\cal C}(V)$ and
${\cal D}(V)$ each only contain a single isomorphism class of objects,
then the set of local 2-arrows $\underline{2R}_U(F,G)(V)$ is nonempty.

We shall now argue that 
for any ${\cal A}|_U$-torsor $I$,
there is an isomorphism
\begin{displaymath}
I \: \stackrel{ \sim }{ \longrightarrow } \:
\underline{2R}_U(F, I_{{\cal D}} \circ F)
\end{displaymath}
of ${\cal A}|_U$-torsors.  We shall define this isomorphism by
describing how a distinguished set of local sections $\{ s_{\alpha} \}$
of $I$,
defined with respect to an open cover $\{ U_{\alpha} \}$ of $U$,
are mapped into local sections of $\underline{2R}_U(F, I_{{\cal D}}\circ F)$.
Denote the local sections of $I$ defining
the gerbe map $I_{{\cal D}}: {\cal D} \rightarrow {\cal D}$ by
$\{ s_{\alpha} \}$, and assume (without loss of generality\footnote{If 
this is not true, then simply restrict the $\{ s_{\alpha} \}$ to
elements of a suitable refinement.}) that
they are defined with respect to an open cover $\{ U_{\alpha} \}$
of $U$ such that ${\cal C}(U_{\alpha})$ and ${\cal D}(U_{\alpha})$ each
only contain a single isomorphism class of objects, for all $\alpha$.
By definition of $I_{{\cal D}}$, there exist natural transformations
\begin{displaymath}
\psi'_{\alpha}(U_{\alpha}): \: |_{U_{\alpha}} \circ I_{{\cal D}}(U)
\: \Longrightarrow \: |_{U_{\alpha}} \circ \mbox{Id}_{\cal D}(U)
\end{displaymath}
where the $|_{U_{\alpha}}$ denote restriction functors associated
to the inclusion $U_{\alpha} \hookrightarrow U$.
Define new natural transformations 
\begin{displaymath}
\psi_{\alpha}(U_{\alpha}): \: F( U_{\alpha} ) \: \Longrightarrow \:
(I_{{\cal D}} \circ F)(U_{\alpha})
\end{displaymath}
as the natural transformations associated to the composition
\begin{displaymath}
F( U_{\alpha} ) \circ |_{U_{\alpha}} \: \stackrel{ \chi^F_{\alpha} }
{ \Longleftarrow } \: |_{U_{\alpha}} \circ F( U ) \:
\stackrel{ \psi'_{\alpha}(U_{\alpha}) }{ \Longleftarrow }
\: |_{U_{\alpha}} \circ (I_{{\cal D}} \circ F)(U) \:
\stackrel{ \chi^{IF}_{\alpha} }{ \Longrightarrow } \:
(I_{{\cal D}} \circ F)(U_{\alpha}) \circ |_{U_{\alpha}}
\end{displaymath}
as in the previous paragraphs.  Furthermore, proceeding
again as in the previous few paragraphs, extend
$\psi_{\alpha}(U_{\alpha})$ to an element $\psi_{\alpha} \in
\underline{2R}_U(F,G)( 
U_{\alpha})$.  Finally, define the isomorphism
$I \rightarrow \underline{2R}_U(F, I_{{\cal D}} \circ F)$ of
${\cal A}|_U$-torsors by mapping $s_{\alpha} \mapsto
\psi_{\alpha}$.  As described in the section on torsors, to define
an isomorphism it suffices to describe how a set of local sections,
defined with respect to an open cover, is mapped. 
Thus, $I \cong \underline{2R}_U(F, I_{{\cal D}} \circ F)$.

We have just shown that $I \cong \underline{2R}_U(F, I \circ F)$
as ${\cal A}|_U$-torsors.  One important consequence which is immediately
derived from this fact is that if there exists a 2-arrow
$F \Rightarrow I \circ F$ for some ${\cal A}$-torsor $I$,
then $I$ must have a global section, and so is trivial.

\subsubsection{Gerbe maps differ by gauge transformations}  
\label{gerbemapsdiffgauge}

In this subsection we shall show that any two gerbe maps differ
by a gauge transformation of the gerbe (i.e., a bundle, modulo equivalence).
There is a precisely analogous notion for bundles:  any two
morphisms $F, G: P_1 \rightarrow P_2$ between principal $G$-bundles
differ by a gauge transformation (namely, $F \circ G^{-1}$, which is
a bundle automorphism, and hence a gauge transformation).

Let $F, G: {\cal C} 
\rightarrow {\cal D}$ be any two gerbe maps, and open $U \subseteq X$ open.
We shall first show that
there exists an ${\cal A}|_U$-torsor $I$ and a local 2-arrow 
$\psi: G |_U \Rightarrow (I_{{\cal D}} \circ F)|_U$ (i.e.,
a global section of the sheaf $\underline{2R}_U(G, I_{{\cal D}} \circ F)$).
Given $F$ and $G$, define $I$ to be the ${\cal A}|_U$-torsor
$\underline{2R}_U(F,G)$.  Let $\{ U_{\alpha} \}$ be a good
open cover of $U$, and let $\{ s_{\alpha} \}$ be a set of local
sections of $I$ defining the gerbe automorphism $I_{{\cal D}}$.
Let 
\begin{displaymath}
\psi^I_{\alpha}(U): \: |_{U_{\alpha}} \circ I_{{\cal D}}(U)  
\: \Longrightarrow \: |_{U_{\alpha}} \circ \mbox{Id}_{{\cal D}}(U)
\end{displaymath}
(where $|_{U_{\alpha}}$ denotes the restriction functors associated
to the inclusions $U_{\alpha} \hookrightarrow U$)
denote the natural transformations associated to the $\{ s_{\alpha} \}$,
appearing in the definition of $I_{{\cal D}}$. 
Let $\psi^s_{\alpha}: F|_{U_{\alpha}} \Rightarrow G|_{U_{\alpha}}$
denote the $s_{\alpha}$, interpreted explicitly as local 2-arrows.
Now, we shall define a natural transformation $G(U) \Rightarrow 
(I_{{\cal D}} \circ F)(U)$, by using the gluing law.
For any object $P \in \mbox{Ob }{\cal C}(U)$, define a morphism
\begin{displaymath}
f_{\alpha}(P): \: G(U)(P) |_{U_{\alpha}} \: \longrightarrow 
\: (I_{{\cal D}} \circ F)(U)(P)|_{U_{\alpha}}
\end{displaymath} 
by,
\begin{displaymath}
f_{\alpha}(P) \: \equiv \: \psi^I_{\alpha} \left( F(U)(P) \right)^{-1} \circ
( \chi^F_{\alpha} )^{-1} \circ 
\psi^s_{\alpha}(U_{\alpha})(P|_{U_{\alpha}})^{-1}
\circ \chi^G_{\alpha}
\end{displaymath}
where the $\chi$ denote natural transformations defining $F$, $G$
as Cartesian functors.  It is straightforward to check that
the $f_{\alpha}(P)$ satisfy the gluing axiom for morphisms, and so
define a morphism $\psi(U)(P): G(U)(P) \rightarrow (I_{{\cal D}} 
\circ F)(U)(P)$.  Moreover, it is straightforward to check that
the $\psi(U)(P)$ define a natural transformation $\psi(U):
G(U) \Rightarrow (I_{{\cal D}} \circ F)(U)$.  Finally, a few paragraphs
earlier we argued that a natural transformation $\psi(U): G(U)
\Rightarrow (I_{{\cal D}} \circ F)(U)$ defines a local
2-arrow $G|_U \Rightarrow (I_{{\cal D}} \circ F)|_U$.
Thus, we have explicitly constructed a global section of the sheaf
$\underline{2R}_U(G, I_{{\cal D}} \circ F)$.

In other words, we have just shown that for any two gerbe maps $F$,
$G$, and for any open $U \subseteq X$, there exists a ${\cal A}|_U$-torsor
$I$ and a local 2-arrow $\psi: G|_U \Rightarrow (I_{{\cal D}} \circ F)|_U$.
(Of course, $I$ is only defined up to isomorphism, as always.)

Note that the result above implies that if $\Phi: {\cal C} \rightarrow 
{\cal C}$ is any automorphism of the gerbe ${\cal C}$,
then there exists an ${\cal A}$-torsor $I$ such that $\Phi$ is equivalent
to $I_{{\cal C}}: {\cal C} \rightarrow {\cal C}$,
i.e., there exists an invertible 2-arrow $\Phi \Rightarrow I_{{\cal C}}$.

Next, suppose that $F$ and $G$ define equivalences of gerbes with
connective structure.  This means that we also have invertible 2-arrows
\begin{eqnarray*}
\Psi^F: \: \mbox{Co}_{{\cal C}} & \Longrightarrow & 
\mbox{Co}_{{\cal D}} \circ F \\
\Psi^g: \: \mbox{Co}_{{\cal C}} & \Longrightarrow &
\mbox{Co}_{{\cal D}} \circ G
\end{eqnarray*}
It is straightforward to check that the difference between these
2-arrows is defined by a choice of connection on $I$
(up to equivalence, as always).

Finally, suppose that $F$ and $G$ define equivalences of gerbes with
connective structure and curving.  This means that in addition to also
having invertible 2-arrows $\Psi^F$, $\Psi^G$ as above, the curving $K$
is invariant under the action of the 2-arrows.  The difference between
such data is again defined by a bundle $I$ with connection, and the
constraint that the curving is invariant becomes the constraint that
the connection on $I$ is flat.
So, the difference between two equivalences of gerbes with connection
(connective structure and curving) is defined by a bundle $I$
with flat connection.

In passing, we should mention that there is a closely analogous notion
for bundles.  Given two morphisms $F, G: P_1 \rightarrow P_2$
between, say, principal $U(1)$-bundles $P_1$, $P_2$ with connection,
the difference $F \circ G^{-1}$ is a gauge transformation which
preserves the connection -- in other words, a constant\footnote{Assuming
the base space is connected.  Locally constant, more generally.}
gauge transformation.

\subsubsection{Maps of gerbes are equivalences of gerbes}  \label{gmapequiv}

In this subsection we shall argue that any map between
gerbes with the same band, over the same space, is necessarily
an isomorphism.  This is a direct analogue of the statement
that any morphism of principal $G$-bundles, for fixed $G$ and
over a fixed space, is necessarily an isomorphism
\cite[section 4.3]{husemoller}.

Let $F: {\cal C} \rightarrow {\cal D}$ be a map between
the gerbes ${\cal C}$, ${\cal D}$, both assumed to have band
${\cal A}$ and both be defined over a fixed space $X$.
We shall argue that $F$ is an equivalence of gerbes.
More precisely, we shall show that each functor $F(U):
{\cal C}(U) \rightarrow {\cal D}(U)$ is an equivalence
of categories, in that there exists a functor
$F^*(U): {\cal D}(U) \rightarrow {\cal C}(U)$ and
invertible natural transformations
\begin{eqnarray*}
\psi_1: & F(U) \circ F^*(U) & \Longrightarrow \: \mbox{Id}_{ {\cal D}(U) } \\
\psi_2: & F^*(U) \circ F(U) & \Longrightarrow \: \mbox{Id}_{ {\cal C}(U) }
\end{eqnarray*}

We define the functor $F^*(U): {\cal D}(U) \rightarrow {\cal C}(U)$
on objects as follows.  
Fix some arbitrary object $P_0 \in \mbox{Ob }{\cal C}(U)$.
For any object $P \in \mbox{Ob } {\cal D}(U)$,
define an ${\cal A}|_U$-torsor 
\begin{displaymath}
I \: = \: \underline{ \mbox{Hom} }_U( F(U)(P_0), P )
\end{displaymath}
Define $F^*(U)(P) \equiv I_{{\cal C}}(U)(P_0)$.

We define $F^*(U)$ on morphisms as follows.
Let $\beta: P_a \rightarrow P_b$ be a morphism between objects
$P_a, P_b \in \mbox{Ob }{\cal D}(U)$.
Define ${\cal A}|_U$-torsors
\begin{eqnarray*}
I^a & \equiv & \underline{ \mbox{Hom} }_U( F(U)(P_0), P_a ) \\
I^b & \equiv & \underline{ \mbox{Hom} }_U( F(U)(P_0), P_b ) 
\end{eqnarray*}
The morphism $\beta$ defines a morphism of torsors $\beta^{\#}:
I^a \rightarrow I^b$, and thus a 2-arrow $\beta^*: I^a_{{\cal C}}
\Rightarrow I^b_{{\cal C}}$.
Define 
\begin{displaymath}
F^*(U)(\beta) \: \equiv \: \beta^*(U)(P_0)
\end{displaymath}
It is straightforward to check that with these definitions,
$F^*(U)$ is a well-defined functor ${\cal D}(U) \rightarrow {\cal C}(U)$.

Before proving that $F^*(U)$ defines an inverse to $F(U)$,
we shall briefly attempt to provide some intuition for this
result.  First, note that for any object $P \in \mbox{Ob } {\cal D}(U)$,
there exists a 2-arrow such that
\begin{displaymath}
(F F^*)(U)(P) \: = \: (F I_{{\cal C}})(U)(P_0) \: \Longrightarrow \:
(I_{{\cal D}} F)(U)(P_0) \: = \: P
\end{displaymath}
This is not quite sufficient to prove that $F$ is an equivalence of
categories, because the 2-arrow above depends upon $P$,
whereas we need to find a single natural transformation.
Similarly, using the fact that
\begin{displaymath}
\underline{ \mbox{Hom} }_U \left( F(U)(P_0), F(U)(P) \right)
\: \cong \: \underline{\mbox{Hom}}_U( P_0, P)
\end{displaymath}
for any object $P \in \mbox{Ob }{\cal C}(U)$, we see that
\begin{displaymath}
(F^* F)(U)(P) \: = \: I_{{\cal C}}(U)(P_0) \: \cong \: P
\end{displaymath}
Again, the remarks above are not intended to be proofs, but are intended
merely to give the reader some intuition as to why our definition
of $F^*(U)$ is a correct one.

We shall now construct invertible natural transformations
\begin{eqnarray*}
\psi_1: & F(U) \circ F^*(U) & \Longrightarrow \: \mbox{Id}_{ {\cal D}(U) } \\
\psi_2: & F^*(U) \circ F(U) & \Longrightarrow \: \mbox{Id}_{ {\cal C}(U) }
\end{eqnarray*}
Existence of these natural transformations, together with $F^*(U)$,
will suffice to prove that $F(U)$ is an equivalence of categories,
and that $F: {\cal C} \rightarrow {\cal D}$ is an equivalence of gerbes.

We shall define $\psi_1: (F F^*)(U) \Rightarrow \mbox{Id}$ as follows.
Let $P_i$ be a family of objects of ${\cal D}(U)$, one for each
equivalence class of objects in ${\cal D}(U)$.
Define a family of ${\cal A}|_U$-torsors
\begin{displaymath}
I^i \: \equiv \: \underline{ \mbox{Hom} }_U \left( F(U)(P_0), P_i \right)
\end{displaymath}
Fix a family of 2-arrows $\Psi_i$:
\begin{displaymath}
\Psi_i: \: ( F I^i_{{\cal C}} ) \: \Longrightarrow \: 
(I^i_{{\cal D}} F)
\end{displaymath}
Now, for any object $P \in \mbox{Ob } {\cal D}(U)$,
let $f: P \rightarrow P_i$ be a morphism from $P$ to some $P_i$,
and define $\psi_1(P): (F F^*)(U)(P) \rightarrow P$ by,
\begin{displaymath}
\psi_1(P) \: \equiv \: f^{-1} \circ \Psi_i(U)(P_0) \circ
(F F^*)(U)(f)
\end{displaymath}
It is straightforward to check that $\psi_1(P)$ is independent of
the choice of $f$, and moreover that $\psi_1(P)$ defines
a natural transformation
\begin{displaymath}
\psi_1: \: F(U) \circ F^*(U) \: \Longrightarrow \: \mbox{Id}_{ {\cal D}(U) }
\end{displaymath}

We shall define $\psi_2: (F^* F)(U) \Rightarrow \mbox{Id}$ as follows.
Let $P'_i$ be a family of objects of ${\cal C}(U)$,
one for each equivalence class of objects in ${\cal C}(U)$.
Define families of ${\cal A}|_U$-torsors
\begin{eqnarray*}
I^i & \equiv & \underline{\mbox{Hom}}_U(P_0, P_i) \\
I'^i & \equiv & \underline{\mbox{Hom}}_U\left(
F(U)(P_0), F(U)(P_i) \right)
\end{eqnarray*}
and fix a family of 2-arrows
\begin{displaymath}
\Psi'_i: \: I'^i_{{\cal C}} \: \Longrightarrow I^i_{{\cal C}}
\end{displaymath}
Now, for any object $P \in \mbox{Ob } {\cal C}(U)$, let
$f: P \rightarrow P'_i$ for a morphism from $P$ to some $P'_i$,
and define $\psi_2(P): (F^* F)(U)(P) \rightarrow P$ by,
\begin{displaymath}
\psi_2(P) \: \equiv \:f^{-1} \circ \Psi'_i(U)(P_0) \circ 
(F^* F)(U)(f)
\end{displaymath}
It is straightforward to check that $\psi_2(P)$ is independent of
the choice of $f$, and moreover that $\psi_2(P)$ defines
a natural transformation
\begin{displaymath}
\psi_2: \: F^*(U) \circ F(U) \: \Longrightarrow \:
\mbox{Id}_{ {\cal C}(U) }
\end{displaymath}

Thus, any gerbe map $F: {\cal C} \rightarrow {\cal D}$ between gerbes
of the same band, over the same space, is necessarily an equivalence
of gerbes.

\section{Equivariant gerbes}

In this section we shall define the notion of equivariance under
a group $\Gamma$ acting on a space $X$ for gerbes defined on the
space $X$.  We shall (loosely) follow \cite[section 7.3]{brylinski}.
We shall assume that $X$ is connected.
We should also mention that we shall often refer to pullbacks of stacks
and gerbes, concepts which were defined in section~\ref{pullback}.
More generally, in both this section and the next,
we shall often make use of results proven in
sections~\ref{advstacksec} and \ref{advgerbesec}, without
specific attribution.

First, however, we shall define the notion of an equivariant stack.
Let $\Gamma$ be a group acting on a topological space $X$
by homeomorphisms, and let ${\cal C}$ be a stack on $X$.
Loosely, for ${\cal C}$ to be equivariant under $\Gamma$ means
that $g^* {\cal C}$ should be isomorphic (in the appropriate sense)
to ${\cal C}$ for all $g \in \Gamma$.
Technically, an equivariant structure on a stack ${\cal C}$ consists
of the following data:
\begin{enumerate}
\item  A Cartesian functor $\Phi_g: g^* {\cal C} \rightarrow {\cal C}$
defining an equivalence of stacks, for each $g \in \Gamma$.
\item For each pair $(g_1, g_2) \in \Gamma \times \Gamma$,
an invertible 2-arrow
\begin{displaymath}
\psi_{g_1, g_2}: \: \Phi_{g_1 g_2} \: \Longrightarrow
\: \Phi_{g_2} \circ g_2^* \Phi_{g_1} \circ \Psi_{g_1,g_2}^{{\cal C}}
\end{displaymath}
between Cartesian functors $(g_1 g_2)^* {\cal C} \rightarrow {\cal C}$,
where $\Psi_{g_1,g_2}^{{\cal C}}: (g_1 g_2)^* {\cal C}
\rightarrow g_2^* g_1^* {\cal C}$ is the analogue of a natural
transformation defined in 
section~\ref{analoguenattranssec}.
\end{enumerate}
Moreover, the 2-arrows $\psi_{g_1, g_2}$ are required to make
the following diagram commute:
\begin{equation}   \label{equivst2ar}
\begin{array}{ccc}
\Phi_{g_1 g_2 g_3} & \stackrel{ \psi_{g_1 g_2, g_3} }
{\Longrightarrow} & \Phi_{g_3} \circ g_3^* \Phi_{g_1 g_2} \circ
\Psi_{g_1 g_2, g_3} \\
\makebox[0pt][r]{ $\scriptstyle{ \psi_{g_1, g_2 g_3} }$ } \Downarrow &
& \Downarrow \makebox[0pt][l]{ $\scriptstyle{ \psi_{g_1, g_2} }$ } \\
\Phi_{g_2 g_3} \circ (g_2 g_3)^* \Phi_{g_1} \circ \Psi_{g_1, g_2 g_3} &
\stackrel{ \psi_{g_2, g_3} }{ \Longrightarrow } &
\Phi_{g_3} \circ g_3^* \left( \Phi_{g_2} \circ g_2^* \Phi_{g_1} 
\circ \Psi_{g_1, g_2} \right) \circ \Psi_{g_1 g_2, g_3}
\end{array}
\end{equation}
Note that in order to make sense out of the diagram above,
we are using the fact that the $\Psi$ obey 
\begin{displaymath}
\Psi_{g_1, g_2} \circ \Psi_{g_1 g_2, g_3} \: = \:
\Psi_{g_2, g_3} \circ \Psi_{g_1, g_2 g_3}
\end{displaymath}
and also that
\begin{displaymath}
\Psi_{g_2, g_3} \circ (g_2 g_3)^* \Phi_{g_1} \: = \:
g_3^* g_2^* \Phi_{g_1} \circ \Psi_{g_2, g_3}
\end{displaymath}
as is discussed in section~\ref{analoguenattranssec}.

An equivariant structure on a gerbe is defined to be an equivariant structure
on the underlying stack such that
each $\Phi_g: g^* {\cal C} \rightarrow {\cal C}$ defines an equivalence
of gerbes.

We shall now argue that any two distinct equivariant structures
on a gerbe differ by a choice of principal $G$-bundles $T_g$ (for
${\cal A} = C^{\infty}(G)$), one for each $g \in \Gamma$,
and a set of isomorphisms of principal $G$-bundles
\begin{displaymath}
\omega_{g_1, g_2}: \: T_{g_1 g_2} \: \longrightarrow \: T_{g_2}
\cdot g_2^* T_{g_1}
\end{displaymath}
such that the following diagram commutes:
\begin{equation}  \label{Tequiv}
\begin{array}{ccc}
T_{g_1 g_2 g_3} & \stackrel{ \omega_{g_1 g_2, g_3} }{ \longrightarrow } &
T_{g_3} \cdot g_3^* T_{g_1 g_2} \\
\makebox[0pt][r]{ $\scriptstyle{ \omega_{ g_1, g_2 g_3 } }$ } \downarrow &
& \downarrow \makebox[0pt][l]{ $\scriptstyle{ \omega_{g_1, g_2} }$ } \\
T_{g_2 g_3} \cdot (g_2 g_3)^* T_{g_1} & \stackrel{
\omega_{g_2, g_3} }{ \longrightarrow } &
T_{g_3} \cdot g_3^* \left( T_{g_2} \cdot g_2^* T_{g_1} \right)
\end{array}
\end{equation}
modulo equivalences of bundles.

In passing, we should point out the close formal resemblance
between diagram~(\ref{Tequiv}) and structures appearing
in \cite[section 1]{freed1}.

Suppose we have two distinct equivariant structures on a gerbe,
that is, two sets of gerbe maps
$\Phi_g, \Phi'_g: g^*{\cal C} \rightarrow {\cal C}$
and corresponding 2-arrows $\psi_{g_1, g_2}$, $\psi'_{g_1, g_2}$.
First, from section~\ref{gerbemapsdiffgauge},
any two gerbe maps $\Phi_g$, $\Phi'_g$ differ
by a gauge transformation, that is, (for band ${\cal A} = 
C^{\infty}(G)$) there exists a principal $G$-bundle we shall denote $T_g$
and a 2-arrow $\lambda_g: \Phi_g \Rightarrow T_g \circ \Phi'_g$.
Note we have used $T_g$ to denote both a principal $G$-bundle and
an associated gerbe automorphism.  

Note that we have been slightly
sloppy -- to completely specify a gerbe automorphism associated to the
bundle $T_g$, we would also need to specify a choice of local sections.
Any two choices of local sections define automorphisms that differ by a 2-arrow,
so changing the choice of local sections merely corresponds to changing
$\lambda_g$.  Thus, we shall not belabor the choice of local sections
any further.

Next, let $\omega_{g_1, g_2}: T_{g_1 g_2} \Rightarrow T_{g_2} \circ
g_2^* T_{g_1}$ be a 2-arrow such that the following diagram commutes:
\begin{equation}  \label{omegadef}
\begin{array}{ccccc}
\Phi_{g_1 g_2} & & \stackrel{ \psi_{g_1, g_2} }{ \Longrightarrow }
 & & \Phi_{g_2} \circ g_2^* \Phi_{g_1} \\
\makebox[0pt][r]{ $\scriptstyle{ \lambda_{g_1 g_2} }$ } \Downarrow &
& & & \Downarrow \makebox[0pt][l]{ $\scriptstyle{ \lambda_{g_2} \circ
g_2^* \lambda_{g_1} }$ } \\
T_{g_1 g_2} \circ \Phi'_{g_1 g_2} & \stackrel{ \psi'_{g_1, g_2} }{
\Longrightarrow } & T_{g_1 g_2} \circ \left( \Phi'_{g_2} \circ
g_2^* \Phi'_{g_1} \right) &
\stackrel{ \Upsilon_{1,2} \circ \omega_{g_1, g_2} }{ \Longrightarrow } &
( T_{g_2} \circ \Phi'_{g_2} ) \circ g_2^* (
T_{g_1} \circ \Phi'_{g_1} )
\end{array}
\end{equation}
where $\Psi$ are implicit (omitted for aesthetic reasons), and
where we have used $\Upsilon_{1,2}$ to denote a 2-arrow
\begin{displaymath}
\Upsilon_{1,2}: \: ( T_{g_2} \circ g_2^* T_{g_1} ) \circ \left( \Phi'_{g_2} 
\circ g_2^*
\Phi'_{g_1} \circ \Psi_{g_1, g_2} \right) \: 
\Longrightarrow \: ( T_{g_2} \circ \Phi'_{g_2} ) \circ g_2^* (
T_{g_1} \circ \Phi'_{g_1} ) \circ \Psi_{g_1, g_2}
\end{displaymath}
describing how we commute the gauge transformations past gerbe maps.
The $\Upsilon$ should not be assumed to be completely arbitrary;
we shall assume that the following diagram of $\Upsilon$ commutes:
\begin{displaymath}
\begin{array}{ccc}
\left( T_{g_3} \cdot g_3^*( T_{g_2} \cdot g_2^* T_{g_1}) \right)
\circ \Phi'_{g_3} \circ g_3^*\left( \Phi'_{g_2} \circ g_2^* \Phi'_{g_1}
\right)  &
\Longrightarrow &
T_{g_3} \circ \Phi'_{g_3} \circ \left( T_{g_2} \cdot g_2^* T_{g_1} \right)
\circ \left( \Phi'_{g_2} \circ g_2^* \Phi_{g_1} \right)
\\
\Downarrow & & \Downarrow \\
\left( T_{g_3} \cdot g_3^* T_{g_2} \right) \circ 
\Phi'_{g_3} \circ g_3^*\left( \Phi'_{g_2} \circ g_2^* T_{g_1} \circ g_2^*
\Phi'_{g_1} \right) 
& \Longrightarrow &
T_{g_3} \circ \Phi'_{g_3} \circ g_3^* \left(
T_{g_2} \circ \Phi'_{g_2} \circ g_2^* T_{g_1} \circ g_2^* \Phi'_{g_1}
\right) 
\end{array}
\end{displaymath}
where we have omitted $\Psi_{g_1, g_2}$ and $\Psi_{g_1 g_2, g_3}$.

The specification of the 2-arrow $\omega_{g_1, g_2}: T_{g_1 g_2} \Rightarrow
T_{g_2} \circ g_2^* T_{g_1}$ is equivalent to a specification of
an isomorphism of
principal $G$-bundles 
\begin{displaymath}
\omega_{g_1, g_2}: \: T_{g_1 g_2} \: \longrightarrow \: T_{g_2}
\cdot g_2^* T_{g_1}
\end{displaymath}
and it is straightforward to check that
the requirement that diagram~(\ref{equivst2ar}) commute for
both the $\psi_{g_1, g_2}$ and the $\psi'_{g_1, g_2}$
implies that
the following diagram of 2-arrows commutes:
\begin{equation}  \label{Tequiv2}
\begin{array}{ccc}
T_{g_1 g_2 g_3} & \stackrel{ \omega_{g_1 g_2, g_3} }{ \Longrightarrow } &
T_{g_3} \circ g_3^* T_{g_1 g_2} \\
\makebox[0pt][r]{ $\scriptstyle{ \omega_{ g_1, g_2 g_3 } }$ } \Downarrow &
& \Downarrow \makebox[0pt][l]{ $\scriptstyle{ \omega_{g_1, g_2} }$ } \\
T_{g_2 g_3} \circ (g_2 g_3)^* T_{g_1} & \stackrel{ 
\omega_{g_2, g_3} }{ \Longrightarrow } &
T_{g_3} \circ g_3^* \left( T_{g_2} \circ g_2^* T_{g_1} \right)
\end{array}
\end{equation}
which implies that diagram~(\ref{Tequiv}) commutes, as claimed.

To summarize our progress so far, we have discovered that the
difference between two equivariant structures on a gerbe 
is described by the data $(T_g, \omega_{g_1, g_2})$ such that
diagram~(\ref{Tequiv}) commutes.
However, we have been a little sloppy.  We could replace any of the
bundles principal $G$-bundles $T_g$ by isomorphic bundles $T'_g$,
as only equivalence classes of principal $G$-bundles are relevant.
If $\kappa_g: T_g \rightarrow T'_g$ are isomorphisms,
then the difference between two equivariant structures can also
be described by the data $\left(T'_g, \kappa_{g_1 g_2} \circ \omega_{g_1, g_2}
\circ ( \kappa_{g_2} \otimes g_2^* \kappa_{g_1} )^{-1} \right)$.

At the end of the day, we will recover a classification of equivariant
structures preserving the gerbe connection, in which we shall
find 
$H^2(\Gamma, G)$.  Before we begin working out the details,
we should take a moment to explain the general idea.
First, since automorphisms of gerbes with connection are defined by
equivalence classes of bundles with connection,
we will also specify connections on the bundles $T_g$,
and the isomorphisms $\omega_{g_1, g_2}$ will be forced to preserve
those connections.
Then, we will demand that the connection on the gerbe be invariant
under all gerbe equivalences, in precise analogy with our strategy
for studying equivariant bundles in \cite{dt1}.  This will
imply that any two equivariant structures differ by a set of
bundles $\{ T_g \}$ with flat connection.  We find the group 
$H^2(\Gamma, U(1))$ by
taking the bundles $T_g$ to be topologically trivial, not just flat,
with gauge-trivial connections.  Cocycle representatives of
elements of $H^2(\Gamma, U(1))$ will be defined by the
isomorphisms $\omega_{g_1, g_2}$; the group cocycle condition will
come from commutivity of diagram~(\ref{Tequiv}).

\section{Equivariant gerbes with connection}

In the previous section we defined the notion of equivariant
structure for gerbes.  In this section we shall extend this notion
to define equivariant structures for gerbes with connection
(connective structure and curving).  As we have only defined
connections for gerbes with band ${\cal A} = C^{\infty}(U(1))$,
we shall assume throughout this section that all gerbes
have band $C^{\infty}(U(1))$.

Let ${\cal C}$ be a gerbe which is equivariant with respect
to the action of a group $\Gamma$ acting by diffeomorphisms.
Under what circumstances will the equivariant structure
respect the connection on ${\cal C}$ ?  
As the reader has probably guessed, the gerbe maps $\Phi_g: g^* {\cal C}
\rightarrow {\cal C}$ are required to be equivalences
of gerbes with connection, so we must also specify 2-arrows
\begin{displaymath}
\kappa_g: \: g^* \mbox{Co} \: \Longrightarrow \: \mbox{Co} \circ
\Phi_g
\end{displaymath}
(Recall that the pullback of a gerbe with
connection is another gerbe with connection -- a connection on ${\cal C}$
naturally induces a connection on $g^* {\cal C}$.
In order for the equivariant structure to respect this connection,
we must demand that the equivalence of gerbes $\Phi_g$ respect
the connection.)

So far we have defined equivariant structures on gerbes with connection.
How are these equivariant structures classified?

Suppose we have two equivariant structures on a gerbe with connection,
that is, two sets of gerbe maps $(\Phi_g, \kappa_g), (\Phi'_g, \kappa'_g)$ 
defining equivalences
of gerbes with connection, and corresponding 2-arrows
$\psi_{g_1, g_2}, \psi'_{g_1, g_2}$ satisfying the conditions above.
How are these two equivariant structures related?

In the previous section, we mentioned that for any pair
$\Phi_g$, $\Phi'_g$, there exists\footnote{Up to isomorphism.} 
a principal $U(1)$-bundle\footnote{
We get principal $U(1)$ bundles because we have assumed the
band is $C^{\infty}(U(1))$ in this section.  In the previous
section we did not have such a constraint on the band.} $T_g$ and
a 2-arrow $\lambda_g: \Phi_g \Rightarrow T_g \circ \Phi'_g$, using
results in section~\ref{gerbemapsdiffgauge}.  Here, because the gerbe
maps define equivalences of gerbes with connection, to describe the
difference between the gerbe maps we must also specify a connection
on each bundle $T_g$, and the connections are constrained to be flat.
Moreover, from commutivity of diagram~(\ref{omegadef}), as applied to the
connective structures, we find that the morphisms $\omega_{g_1, g_2}$
must preserve the connection on each bundle.

In other words, so far we have found that the difference between
two equivariant structures on a gerbe with connection is defined
by a set of principal $U(1)$-bundles $T_g$, each with a flat connection,
together with connection-preserving isomorphisms
$\omega_{g_1, g_2}: T_{g_1 g_2} \rightarrow T_{g_2} \cdot g_2^* T_{g_1}$
such that diagram~(\ref{Tequiv}) commutes.

As before, the bundles with connection are only defined up to isomorphism.
If $\kappa_g: T_g \rightarrow T'_g$ defines a set of isomorphisms of bundles
with connection, then we can replace the data $(T_g, \omega_{g_1, g_2})$
with the data  $\left(T'_g, \kappa_{g_1 g_2} \circ \omega_{g_1, g_2}
\circ ( \kappa_{g_2} \otimes g_2^* \kappa_{g_1} )^{-1} \right)$.

At the end of the day, we wish to find how the group $H^2(\Gamma, U(1))$
appears in describing the difference between two equivariant structures.
This group appears as follows.  Take all the bundles $T_g$ to be topologically
trivial, with gauge-trivial connections.  Then, we can use the fact
that the bundles $T_g$ are only defined up to isomorphism to replace
each $T_g$ with the canonical trivial bundle with identically zero connection.
As the isomorphisms $\omega_{g_1, g_2}$ are constrained to preserve
the connection, this means they must be constant (assuming the underlying
space is connected).  From diagram~(\ref{Tequiv}) we see that the
$\omega_{g_1, g_2}$ define a group 2-cocycle. 
Now, even after making this choice of $T_g$'s, there is still a residual
gauge invariant -- we can gauge-transform each $T_g$ by a constant
gauge transformation, which preserves the (identically zero) connection
on each $T_g$.  It is clear that these constant gauge transformations
of each $T_g$ change the $\omega_{g_1, g_2}$ by a group coboundary.
Thus, we have found elements of the group $H^2(\Gamma, U(1))$ lurking 
in the differences
between any two equivariant structures on a gerbe with connection.

In general, however, there will be additional possible orbifold
group actions, beyond those classified by elements of $H^2(\Gamma, U(1))$.
In retrospect, we should not be surprised -- for example\footnote{
 Lest we give the wrong impression,
classifying equivariant structures is not the same thing as
calculating cohomology, but in very special cases,
cohomology calculations can shed light.
},
the Cartan-Leray spectral sequence for $H^2(X/\Gamma, {\bf Z})$
(for $\Gamma$ freely-acting) contains contributions from more than
just $H^2(\Gamma, U(1))$.  We shall discuss this matter further
in \cite{dt3}.

\section{Check:  loop spaces}    \label{loopspaces}

In this section we shall shed some light on the methods and results
of the previous two sections by thinking about gerbes in terms
of loop spaces.

First, note that a principal $U(1)$ bundle $P$ with connection on a 
manifold $M$
determines a $U(1)$-valued function on $LM$, the loop space of $M$.
More precisely, for any loop in $M$, we can assign an element of $U(1)$
given by the value of the Wilson loop.  Thus, Wilson loops assign
elements of $U(1)$ to each loop in $M$, and so define a $U(1)$-valued
function on $LM$.

The assignment above, of $U(1)$-valued functions on $LM$ to 
principal $U(1)$ bundles with connection on $M$, is a basic example
of a more general principle.  Namely, to any $n$-gerbe with connection
on a manifold $M$, one can assign an $(n-1)$-gerbe with connection
on the loop space $LM$.  (This can be seen in terms of Deligne
cohomology; see, for example, \cite[section 6.5]{brylinski}.)
More relevantly to this paper, to any (1-)gerbe with connection 
(and band $C^{\infty}(U(1))$) on a
manifold $M$, one can assign\footnote{Strictly speaking,
to a (1-)gerbe with connective structure on $M$, we can assign
a bundle on $LM$.  A curving on that connective structure can be
used to define a connection on the bundle on $LM$.
See \cite[section 6.2]{brylinski} for more details.
In this framework, gerbe maps become morphisms of principal
bundles (more generally, torsors) on $LM$, and gerbe maps related
by invertible 2-arrows map to the same morphism of principal
bundles on $LM$.} 
a principal $U(1)$ bundle with connection
on the loop space $LM$.  (For more details, see,
for example, \cite[section 6.2]{brylinski}; a derivation of this
fact at the level of Deligne cohomology is also given in 
\cite{gaw}.)

Thus, the reader might naively be led to suspect that an equivariant
gerbe on a space $M$ is equivalent to an equivariant bundle on $LM$.
Unfortunately, this is not quite correct.  The essential difficulty
is that, in general, the map from $n$-gerbes on $M$ to $(n-1)$-gerbes
on $LM$ is a many-to-one map.  For example, consider the map from
principal $U(1)$ bundles on $M$ to $U(1)$-valued functions on $LM$
defined by Wilson loops, as described at the beginning of this section.
Specifying Wilson loops about every loop on a manifold $M$ (i.e.,
specifying a $U(1)$-valued function on $LM$) does not uniquely
determine a principal $U(1)$ bundle with connection.
Instead, such a set of Wilson loops only determines an equivalence
class of principal $U(1)$ bundles with connection 
\cite[prop. 1.12.3]{kostant}.

In the present situation, because a (1-)gerbe with connection on $M$ can not be
uniquely determined by a principal $U(1)$ bundle with connection on $LM$,
we can not completely describe equivariant gerbes with connection 
in terms of equivariant bundles with connection on $LM$.

However, it is true that an equivariant structure on a gerbe with
connection on $M$ does determine an equivariant structure on
the corresponding bundle with connection on $LM$.
We shall merely outline the results.  Recall that an equivariant
structure on a gerbe ${\cal C}$ with connection is defined by
a collection of gerbe maps
\begin{displaymath}
\Phi_g: \: g^* {\cal C} \: \longrightarrow \: {\cal C}
\end{displaymath}
and invertible 2-arrows
\begin{displaymath}
\psi_{g_1, g_2}: \: \Phi_{g_1 g_2} \: \Longrightarrow \:
\Phi_{g_2} \circ g_2^* \Phi_{g_1} \circ \Psi^{{\cal C}}_{g_1, g_2}
\end{displaymath}
subject to various constraints.
We shall let $P$ denote the bundle with connection on $LM$ corresponding
to the gerbe ${\cal C}$ on $M$.  The gerbe maps $\Phi_g$
become bundle isomorphisms $\phi_g: g^* P \rightarrow P$,
and the existence of invertible 2-arrows $\psi_{g_1, g_2}$
implies that $\phi_{g_1 g_2} = \phi_{g_1} \circ g_2^* \phi_{g_1}$,
which, recall from \cite{dt1}, defines an equivariant structure on the
bundle $P$.

Recall in \cite{dt1} we pointed out that any equivariant structure
on a bundle with connection could be obtained from any other equivariant
structure via a set of constant gauge transformations.
In the present context, however, there is a slight subtlety.
Gauge transformations of $P$ on $LM$ are determined by 
principal $U(1)$ bundles with connection on $M$, that is, 
$U(1)$-valued functions on $LM$.  However, not all $U(1)$-valued
functions on $LM$ can be understood in terms of Wilson loops on bundles
with connection on $M$.  In fact, the only constant $U(1)$-valued
function on $LM$ that can be understood in terms of Wilson loops
on bundles with connection on $M$, is the trivial constant function
that is the identity for all points on $LM$.
To see this fact, suppose that $L$ is a principal $U(1)$ bundle with
connection on $M$ with the property that all Wilson loops are 
equal to some (single) element of $U(1)$, call it $x$.
Let $W(\gamma)$ denote the value of the Wilson loop about any
loop $\gamma$, then it should be clear that 
\begin{displaymath}
W( 2 \gamma) \: = \: W(\gamma)^2
\end{displaymath}
or, in other words, $x = x^2$.  But as $x \in U(1)$, 
the only way that this can be satisfied is if $x$ is the
identity.  Thus, the only constant gauge transformations on $LM$
that can be understood as coming from bundles with connection on $M$
are those that are identically the identity.


If $LM$ is connected, then there is only one constant gauge transformation
allowed between
equivariant structures on $P \rightarrow LM$.  From this the reader might 
incorrectly conclude that this must mean that there can only
be one equivariant structure on a gerbe with connection.
The problem with this argument is that the map from
bundles (with connection) on $M$ to functions on $LM$ is a many-to-one map.
At the level of the loop space $LM$, there is only one equivariant
structure; however, there are actually multiple equivariant structures
on the gerbe on $M$, all of which map to the same equivariant
structure on the bundle $P$ on $LM$. 

To classify equivariant structures on a gerbe with connection,
we must return to the analysis of the previous section.
In passing, however, there is one slight additional bit of insight
that can be gained from thinking in terms of loop spaces.
We argued that on $LM$, for $LM$ connected,
there was only one equivariant structure
on the bundle $P$; any other differed from it by a constant
gauge transformation on $LM$.  However, specifying a constant
gauge transformation on $LM$ only determines an equivalence
class of bundles with connection on $M$.  In the previous section,
we found that equivariant structures on a gerbe with connection
on $M$ were determined by a set of isomorphisms of (trivial) bundles
with (trivial) connection; here we see that these isomorphisms must
preserve the trivial connection, and so must be constant, in order to
remain within the equivalence class determined by the trivial
constant gauge transformation on $LM$.  This sheds some
light on the constraints on these isomorphisms determined in the
previous section.

In general, $LM$ has one component for each element of $\pi_1(M)$.
Differences between equivariant structures on a bundle with connection
on a non-simply-connected space are not described by constant gauge
transformations, but rather by locally constant gauge transformations.
These degrees of freedom correspond to taking the bundles $T_g$
with connection to be flat, but nontrivial.

\section{Conclusions}

In this paper we have accomplished two things.  First, we have
given a thorough review of gerbes in terms of stacks, at a relatively
basic level (i.e., without using the language of sites).
Second, we have discussed the classification of equivariant structures
on (1-)gerbes with connection, and proven that in general
the set of such equivariant structures is a torsor under a group that
includes 
$H^2(\Gamma, U(1))$, as claimed in \cite{dt1}, providing
a simple geometric understanding of discrete torsion.



\section{Acknowledgements}

We would like to thank P.~Aspinwall, D.~Freed, A.~Knutson, D.~Morrison, and
R.~Plesser for useful conversations.

\end{document}